\documentclass[journal=jctcce,amsmath,amssymb,manuscript=article]{achemso}
\usepackage[utf8]{inputenc} 
\usepackage[T1]{fontenc}    
\usepackage{hyperref}       
\usepackage{url}            
\usepackage{booktabs}       
\usepackage{amsfonts}       
\usepackage{nicefrac}       
\usepackage{microtype}      
\usepackage{fancyhdr}       
\usepackage{graphicx}       
\usepackage{braket}
\usepackage{amsmath}
\usepackage{svg}
\usepackage{pifont}
\usepackage{ulem}
\usepackage{comment}
\usepackage{CJKutf8}
\usepackage{textcomp}
\usepackage{cleveref}
\usepackage[ruled,vlined]{algorithm2e}

\graphicspath{{Fig/}}     

\newcommand{\CY}[1]{{\color{black}{#1}}}

\title{Applying Space-Group Symmetry to Speed up Hybrid-Functional Calculations within the Framework of Numerical Atomic Orbitals
}

\author{Yu Cao}
\affiliation{HEDPS, CAPT, School of Physics and College of Engineering, Peking University, Beijing, 100871, China}
\alsoaffiliation[IOP]{Institute of Physics, Chinese Academy of Sciences, Beijing, 100190, China}
\author{Min-Ye Zhang}
\affiliation[IOP]{Institute of Physics, Chinese Academy of Sciences, Beijing, 100190, China}
\alsoaffiliation[NOMAD]{The NOMAD Laboratory at the FHI of the Max Planck Society, Faradayweg 4-6, D-14195 Berlin-Dahlem, Germany}
\author{Peize Lin}
\affiliation[IAI]{Institute of Artificial Intelligence, Hefei Comprehensive National Science Center, Hefei 230026, Anhui, China.}
\alsoaffiliation[IOP]{Institute of Physics, Chinese Academy of Sciences, Beijing, 100190, China}
\author{Mohan Chen}
\affiliation[CAPT]{CAPT, HEDPS, College of Engineering and School of Physics, Peking University, Beijing, 100871, China}
\alsoaffiliation[AISI]{AI for Science Institute, Beijing, 100080, China}
\email{mohanchen@pku.edu.cn}

\author{Xinguo Ren}
\affiliation[IOP]{Institute of Physics, Chinese Academy of Sciences, Beijing, 100190, China}
\email{renxg@iphy.ac.cn}


\begin{document}

%
%
%
%
\begin{tocentry}
	\centering
	\includegraphics[width=0.8\textwidth]{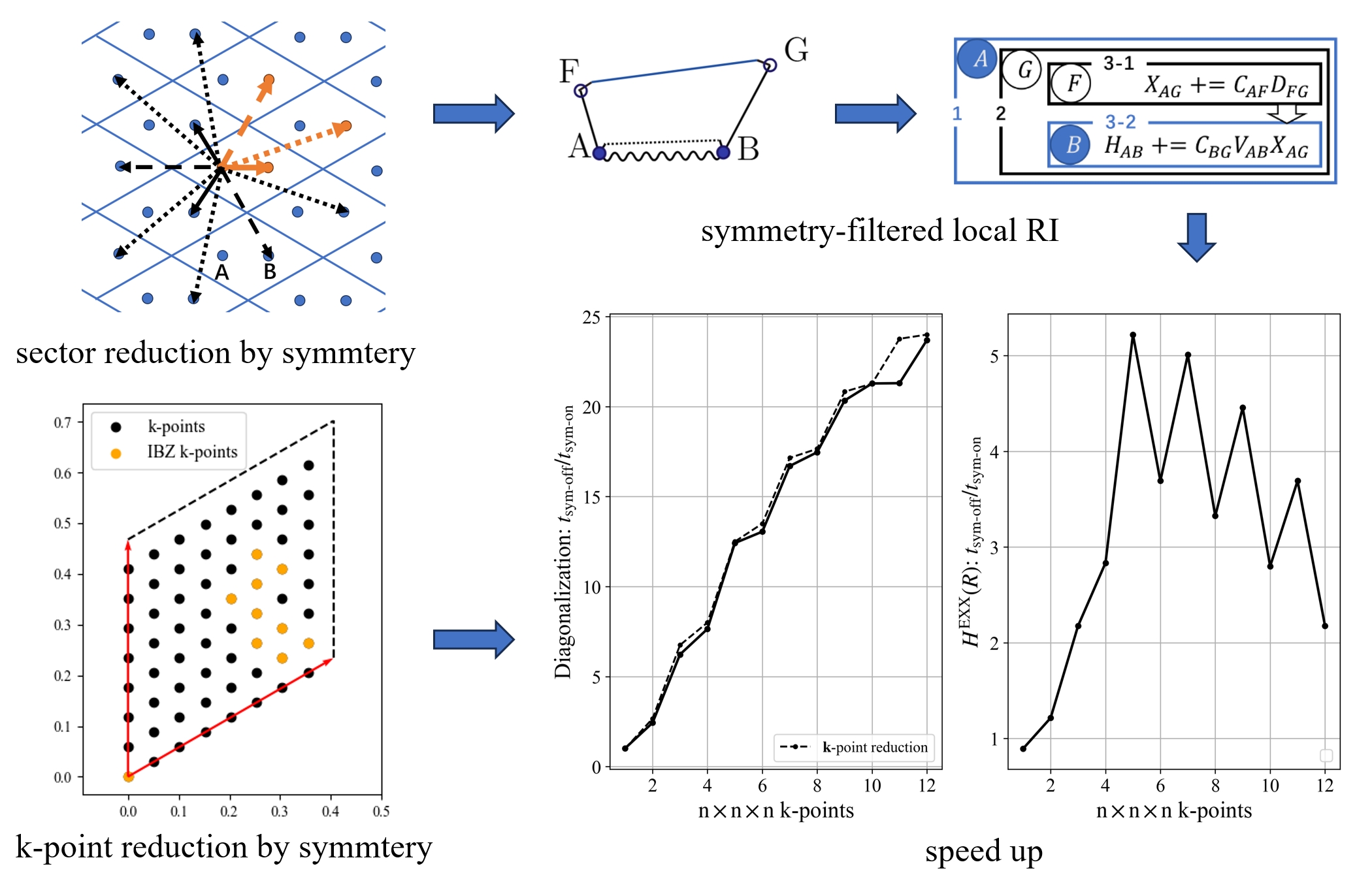}
\end{tocentry}


\begin{abstract}
Building upon the efficient implementation of hybrid density functionals (HDFs) for large-scale periodic systems within the framework of numerical atomic orbital bases using the localized resolution of identity (RI) technique, we have developed an algorithm that exploits the space group symmetry in key operation steps of HDF calculations, leading to further improvements in two ways. First, the reduction of $\mathbf{k}$-points in the Brillouin zone can reduce the number of Kohn-Sham equations to be solved. This necessitates the correct implementation of the rotation relation between the density matrices of equivalent $\mathbf{k}$-points within the representation of atomic orbitals. Second, the reduction of the real-space sector can accelerate the construction of the exact-exchange part of the Hamiltonian in real space. 

We have implemented this algorithm in the ABACUS software interfaced with LibRI, and tested its performance for several types of crystal systems with different symmetries. The expected speed-up is achieved in both aspects: the time of solving the Kohn-Sham equations decreases in proportion with the reduction of $\mathbf{k}$-points, while the construction of the Hamiltonian in real space is sped up by several times, with
the degree of acceleration depending on the size and symmetry of the system.
\end{abstract}

\section{Introduction}

Kohn-Sham density functional theory (KS-DFT)\cite{hohenberg_inhomogeneous_1964,kohn_self-consistent_1965} is a widely used computational scheme for first-principles calculations. Exact in principle in its formulation for the ground state, the exchange-correlation (XC) functional of KS-DFT has to be approximated in practice, and the different levels of approximations can be classified into five rungs according to the so-called ``Jacob's ladder'' \cite{perdew_jacobs_2001}.
Hybrid density functionals (HDFs) \cite{becke_densityfunctional_1993, zhang_hybrid_2020}, formulated within the generalized KS (GKS) framework \cite{seidl_generalized_1996},
belong to the fourth rung of the ladder and can overcome some of the drawbacks of the (semi-)local XC functionals \cite{lee_development_1988,perdew_generalized_1996,sun_strongly_2015}.
In particular, the self-interaction error\cite{perdew_self-interaction_1981} and/or the delocalization error \cite{mori-sanchez_localization_2008} are mitigated and the under-reported band gaps
are corrected. These are achieved
by employing nonlocal effective potentials derived from the Hartree-Fock-type exact exchange (EXX) with different forms of Coulomb kernels.

The high accuracy of HDFs comes with a price: With a canonical $O(N^4)$-scaling,  constructing the EXX non-local  potential matrix is computationally much more
expensive than its local or semilocal counterparts. 
To deal with this issue, various algorithms have been developed within different numerical frameworks, ranging from
quadratic \cite{almlof_principles_1982,haser_improvements_1989} or even linear-scaling \cite{burant_linear_1996,schwegler_linear_1996, schwegler_linear_1997,ochsenfeld_linear_1998, rudberg_kohnsham_2011} algorithms for finite systems and periodic systems \cite{pisani_exact-exchange_1980,pisani_hartree-fock_1988,dovesi_c_2014,guidon_ab_2008,guidon_robust_2009,paier_accurate_2009} designed for Gaussian-type orbitals (GTOs), to periodic implementations based on projector augmented wave (PAW) \cite{paier_screened_2006}, pseudopotential planewave (PW) \cite{lin_adaptively_2016}, and linearized augmented plane wave (LAPW)\cite{betzinger_hybrid_2010} schemes. Efficient PW-based algorithms were developed in terms of Wannier orbitals \cite{wu_order-n_2009}, by constructing an adaptive compressed exchange
operator \cite{lin_adaptively_2016}, or more recently based on the interpolative separable density fitting (ISDF) techniques \cite{hu_interpolative_2017,Wu/Qin/etal:2022}. 

Meanwhile, recent years have seen a surge of HDF implementations based on numerical atomic orbitals (NAOs), exploiting their favorable properties
such as compactness and strict spatial localization.
The NAO-based EXX implementations are enabled by expanding NAOs in terms of GTOs \cite{shang_implementation_2010,qin_interpolative_2020} to 
ease the calculation of two-electron Coulomb repulsion integrals (ERIs) or by invoking the resolution-of-the-identity (RI) technique \cite{feyereisen_use_1993,vahtras_integral_1993,weigend_ri-mp2_1998}, also known the variational density fitting (DF) approach \cite{whitten_coulombic_1973,dunlap_approximations_1979,dunlap_variational_2010,  qin_interpolative_2020, lee_systematically_2020, henneke_fast_2020}.
In RI/DF, the products of orbitals are expanded in terms of a set of auxiliary basis functions (ABFs), which reduces the 4-center integrals into three- and two-center ones and hence leads to computational speed-up and memory saving. Different variants have been developed to make RI/DF more efficient by going beyond the original global scheme,\cite{sodt_linear_2006,sodt_hartree-fock_2008, pisani_local-mp2_2005,pisani_periodic_2008, reine_variational_2008} including localized RI or pair-atomic RI (LRI/PARI).\cite{ren_resolution--identity_2012,merlot_attractive_2013,ihrig_accurate_2015,levchenko_hybrid_2015,wirz_resolution---identity_2017,Kokott/etal:2024}
In general, RI/DF technique can be also employed for GTOs \cite{weigend_fully_2002, eshuis_fast_2010,del_ben_electron_2013, Bussy/Hutter:2024}, or mixed basis sets \cite{sun_gaussian_2017}.
Other low-rank decompositions for electron repulsion ERIs include Cholesky decomposition (CD)\cite{beebe_simplifications_1977,koch_reduced_2003} and tensor hypercontraction (THC)\cite{hohenstein_tensor_2012,parrish_tensor_2012, lu_compression_2015}, etc. 

The LRI-based linear-scaling algorithm for evaluating the EXX contributions for periodic systems has been implemented \cite{lin_efficient_2021,lin_accuracy_2020} 
in the ABACUS code \cite{li_large-scale_2016,lin_ab_2024}, which employs systematically generated NAOs \cite{chen_systematically_2010,chen_electronic_2011,lin_strategy_2021} as
basis functions. 
However, the crystalline symmetry has not been exploited in the original EXX implementation \cite{lin_accuracy_2020,lin_efficient_2021}.
In this work, we extend the EXX implementation in ABACUS to fully account for the space-group symmetry,
which leads to significant further reduction of the computational cost, facilitating efficient HDF calculations for large systems with high symmetries and/or systems with dense ${\mathbf k}$ grids. 

The space-group symmetry can be exploited in first-principles calculations in several different ways \cite{dovesi_c_2014}:
Firstly, it leads to ${\mathbf{k}}$-point reduction. Specifically, equivalent ${\mathbf{k}}$-points connected by the rotation part of any space-group symmetry operation contribute equal band energies, and their corresponding wavefunctions differ at most by a phase factor \cite{dresselhaus_group_2007}. Therefore, only one of them needs to be solved by diagonalizing the Kohn-Sham (KS) Hamiltonian. The total charge density can be recovered from the contribution of irreducible {$\mathbf{k}$}-points by symmetrization, where rotating the density on the grid only amounts to rotating the grid coordinates in practice as $\hat{P}\rho(\mathbf{r})=\rho(\hat{P}^{-1}\mathbf{r})$. 
Secondly, real-space integrals that need to be evaluated can be significantly reduced. 
The one- and two-electron integrals linked by symmetries can be transformed into each other 
by matrix representations of the symmetry group\cite{dacre_use_1970,dupuis_molecular_1977,pisani_hartree-fock_1988}.
In the past, this kind of real-space sector reduction was used to reduce the calculation of ERIs in CRYSTAL \cite{ferrero_coupled_2008,orlando_full_2014, dovesi_c_2014, dovesi_quantum-mechanical_2018, dovesi_crystal_2020}, Exciton\cite{10Charles-IRQuads}, 
POLYATOM\cite{csizmadia_non-empirical_1966, elder_use_1973},
HONDO\cite{dupuis_molecular_1977}, 
DISCO\cite{almlof_principles_1982}, DSCF\cite{haser_improvements_1989}, \CY{Turbomole\cite{94CPL-Wullen},} and other codes \cite{rusakov_space_2013}. However, it should be mentioned that \CY{\sout{most of}all} these codes directly deal with the four-center ERIs, i.e., without using the RI/DF technique. 
\CY{Although RI is supported in Turbomole, the reported tests are done only under the $C_1$ symmetry \cite{02PCCP-Weigend, 08JCC-Weigend}. The efficacy of utilizing the space space is not obvious from their publications. }
Thirdly, the symmetry can be incorporated in terms of the so-called symmetry-adapted basis. 
The AO basis set can be shrunk to the (space-group-)symmetry-adapted crystalline orbitals (SACOs), which can speed up the diagonalization step \cite{zicovich-wilson_use_1998,zicovich-wilson_use_1998-1,haser_improvements_1989}, and also applied to beyond-DFT methods such as self-consistent GW \cite{dongxinyang-2025}.
Finally, the RI/ISDF-based calculations  
can also be accelerated by exploiting symmetry \cite{yeh_low-scaling_2024-1,haser_molecular_1991,hser_exploiting_1991,furche_turbomole_2014, sierka_fast_2003, gao_accelerating_2020} through adapted auxiliary basis functions (ABFs) \cite{yeh_low-scaling_2024-1} or symmetry-reduced interpolating points \cite{gao_accelerating_2020}.
The last two kinds of symmetry exploitation are, however, beyond the scope of this paper. \CY{In brief, although
the space group symmetry has been used to speed up the periodic EXX calculations, a detailed description of the algorithm and systematic benchmark of the efficacy of applying such symmetry in the context of LRI-based EXX calculations has not been reported before, to the best of knowledge. A thorough investigation along these lines constitute the major motivation of the present work. }


 



The space-group symmetry analysis has been previously implemented in ABACUS 
and utilized to speed up KS-DFT calculation with (semi-)local XC functionals, but it cannot be straightforwardly used functionals that depend on the density matrix, such as the EXX functional here. This is because the infrastructure for rotating
the density matrix according to symmetry operations was not implemented previously. Thus implementing
the infrastructure supporting density matrix rotations will be the first objective of the present work. Once this is done, GKS calculation can also be restricted to irreducible Brillouin-zone (BZ) ${\mathbf{k}}$-points. Furthermore, applying a similar rotation to the Hamiltonian matrix allows us to reduce the RI-based real-space integration into the irreducible sector, which leads to additional savings in the computational workload in constructing
the EXX Hamiltonian. This is the second and major part of this work.

The paper is organized as follows: In \cref{sec:theory&method}, the rotation relationships of the Hamiltonian and density matrices within the representation of atomic orbitals in both real and $\mathbf{k}$ spaces are derived. \Cref{sec:implement} presents concrete implementation details based on the local RI framework for atomic orbital basis sets. Illustrating test results of the efficacy of our algorithm and implementation are discussed in \cref{sec:result}, followed by the conclusion in \cref{sec:conclusion}.  In addition, some basic knowledge to derive the relation in \cref{sec:theory&method} is given in \cref{ap:opformulas}. In \cref{ap:Cs} 
 the rotation relation of the RI expansion coefficient tensors is derived. Certain
limitations and an in-depth analysis of the performance bottleneck of our current implementation are discussed in \Cref{ap:IRQ} and \Cref{ap:why-down}, respectively.
Additional results are presented in \cref{ap:tests}.

\section{Basic Theory and Method}
\label{sec:theory&method}

In this section, we present the theoretical formulations which allow us to make use of 
symmetries within the framework of linear combination of atomic orbitals (LCAO). We start by reviewing the workflow of SCF calculations using HDFs within the LCAO framework, whereby the steps at which the symmetry can be exploited to reduce the computation cost
are identified. Then we derive the transformations of the Hamiltonian and density matrices under symmetry operations in both real and reciprocal spaces. This provides concrete formulas for one to apply symmetries in these steps for
the Hamiltonian that depends explicitly on the density matrix (such as the EXX and/or HDFs).

\subsection{Overall workflow}
\label{sec:imp:workflow}
\begin{figure}[htbp]
    \centering
    \includegraphics[width=0.5\textwidth]{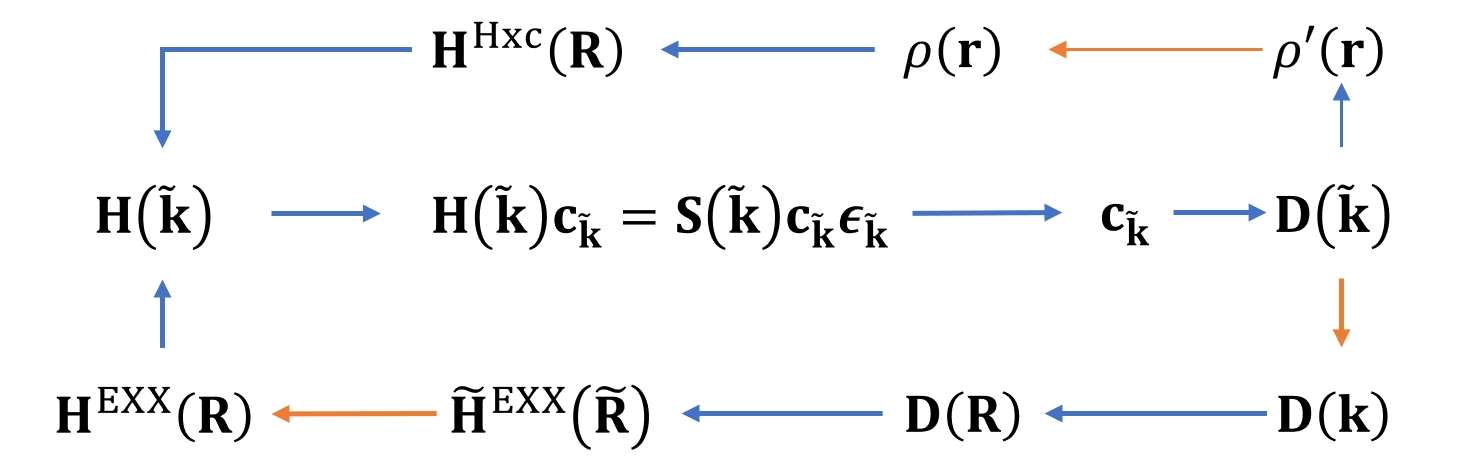}
    \caption{Flowchart of the HDF-based SCF loop that utilizes the space-group symmetry. Density-matrix-dependent Hamiltonian terms need the correct $D(\mathbf{R})$ Fourier-transformed from $D(\mathbf{k})$ in the whole BZ, while (semi-)local Hamiltonian terms depend on the symmetrized charge density $\rho(\mathbf{r})$.
    The orange arrows indicate that the symmetry relation is used in the step.}
    \label{fig:workflow}
\end{figure}

For HDF calculations in the LCAO basis, the Hamiltonian is composed of terms that depend on local densities and density gradients, and term(s) that depend on the
density matrix. This is illustrated in Fig.~\ref{fig:workflow}, which shows the major steps in a typical SCF
calculation using HDFs. Here, the full Hamiltonian $H$ consists of $H^\text{Hxc}$, which depends only on the density $\rho$ (and the density gradients which are omitted for simplicity), and $H^\text{EXX}$, which depends on the density matrix $D$. In the LCAO approach, these Hamiltonian components are first computed in real space (hence having a dependence on the lattice vector $\mathbf{R}$) and then Fourier transformed to the $\mathbf{k}$ space. For the $\mathbf{k}$-space Hamiltonian matrices, we only need them in the irreducible BZ (IBZ), denoted as $H(\tilde{\mathbf{k}})$ in Fig.~\ref{fig:workflow}. However, to perform the Fourier transformation, 
we need the Hamiltonian matrices $H(\mathbf{R})$ over the full real-space sectors (the full Born-von K\'arm\'an (BvK) supercell).
On the other hand, by diagonalizing
$H(\tilde{k})$, one can only get the KS eigenvectors and density matrices within the IBZ. Obviously, symmetrization
operations are needed to transform the density matrix (and the corresponding density) from IBZ to the full BZ, 
as indicated by the yellow arrows in \cref{fig:workflow}.

As indicated in Fig.~\ref{fig:workflow}, to compute the Hamiltonian part that depends only on the density, we just need to symmetrize the charge density (and compute the gradients of the symmetrized density), 
which is relatively straightforward.
Namely, the charge densities contributed by the wavefunctions of symmetry-equivalent $\mathbf{k}$-points ($\mathbf{k}$ and $\hat{P}\mathbf{k}$) are connected by some coordinate transformation \cite{dresselhaus_group_2007}, and can be added up via a ``charge density symmetrization'' procedure, 
thereby yielding the correct density and $H^\text{Hxc}$.
However, to obtain the right Hamiltonian containing terms that
depend on the density matrix, an extra step is to rotate the density matrix 
from the irreducible part of the BZ to the full BZ. 
For systems having time-reversal symmetry, the degeneracy of $\mathbf{k}$ and $-\mathbf{k}$ is also exploited, where
the relation $\mathbf{D}(\mathbf{k})=\mathbf{D}^*(-\mathbf{k})$ is used to restore the density matrix outside the IBZ, in addition to the space-group symmetry.


Restricting $H(\tilde{k})$ to the IBZ reduces the number of KS equations to be solved, saving time for
the diagonalization of the Hamiltonian matrices. However, as alluded to above, to perform the Fourier 
transformation, we need to have the real-space Hamiltonian $H^{\text{EXX}}(\mathbf{R})$ over the full sector,
whose evaluation represents the most significant bottleneck in HDF calculations. 
Naturally, if the real-space sector can be reduced by symmetry in a similar way as the reduction of 
the $\mathbf{k}$-points,  then we only need to calculate part of $H^{\text{EXX}}(\mathbf{R})$ in the irreducible sector (denoted as $\tilde{H}^{\text{EXX}}(\mathbf{\tilde{R}})$). 
Obviously, the rotation to the full sector (Eq.~\ref{eq:H-mat}) takes negligible time compared to the calculation of $H^{\text{EXX}}(\mathbf{R})$.

In summary, as can be seen in Fig.~\ref{fig:workflow} and discussed above, to utilize the space group symmetry in HDF calculations, we need to 1) restore the density matrices in the $\mathbf{k}$ from the IBZ to full BZ, and 2) to obtain the Hamiltonian matrices in real space from the irreducible sector to full sector. Below we shall
devote ourselves to deriving the key formula to perform such transformations within the LCAO framework. 
The specifics related to the LRI approximation for evaluating the two-electron Coulomb integrals will be 
deferred to Sec.~\ref{sec:implement}.


\subsection{Symmetry-associated Hamiltonian on atomic orbitals}
\label{ssec:hamiltonian}
If a space-group symmetry operation $\hat{P}=\{V|\mathbf{f}\}$, defined by a rotation $V$ followed by a translation $f$, does not change the system Hamiltonian, the following relationship is satisfied,
\begin{equation}
\hat{P}\hat{H}\hat{P}^{-1}=\hat{H} \, .
\end{equation}
Furthermore, since $\hat{P}$ is also a unitary transformation, inserting $\hat{P}^{-1}\hat{P}$ into the Hamilton matrix under certain basis set representation $\{\phi\}$ gives
\begin{equation}
\label{eq:hrep}
\braket{\phi_i|\hat{H}|\phi_j}=\braket{\phi_i|\hat{P}^{-1}\hat{P}\hat{H}\hat{P}^{-1}\hat{P}|\phi_j}
=\braket{\hat{P}\phi_i|\hat{H}|\hat{P}\phi_j}\, .
\end{equation}
The electronic wavefunction of the periodic systems can be expanded in terms of Bloch orbitals, 
\begin{equation}
    \label{eq:wf_expansion}
    \psi_\mathbf{k}(\mathbf{r})=\sum_{\mathcal{U}\mu}c^\mathbf{k}_{\mathcal{U}\mu}\phi^\mathbf{k}_{\mathcal{U}\mu}(\mathbf{r})\, ,
\end{equation}
which themselves can be constructed using atomic orbitals (AOs) summed over lattice vectors in the BvK supercell,
\begin{equation}
    \phi^\mathbf{k}_{\mathcal{U}\mu}(\mathbf{r})=\frac{1}{\sqrt{N_\mathbf{k}}}\sum_\mathbf{R}e^{i\mathbf{k\cdot R}}\phi^\mathbf{R}_{\mathcal{U}\mu}(\mathbf{r})\, ,
    \label{eq:Bloch_expansion}
\end{equation}
where  $\mu=\{nlm\}$ is the combined index of atomic orbitals centering on atom $\mathcal{U}$, 
$c^\mathbf{k}_{\mathcal{U}\mu}$ the expansion coefficients,
and the real-space AO basis
\begin{equation}
    \phi^\mathbf{R}_{\mathcal{U}\mu}(\mathbf{r})=\phi_{\mathcal{U}\mu}(\mathbf{r-s_\mathcal{U}-R})
    %
\end{equation}
with $s_{\mathcal{U}}$ denoting the atomic position within the unit cell $\mathbf{R}$.
$N_\mathbf{k}$ is the number of $\mathbf{k}$-points in the full first BZ, which equals to the number of unit cells in the BvK supercell.

\begin{figure}
    \centering
    \includegraphics[width=0.5\linewidth]{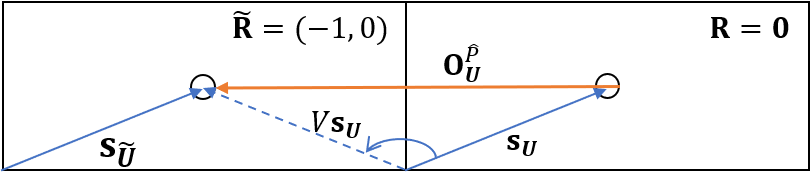}
    \caption{A two-dimensional schematic illustration of the definition of $\mathbf{O}^{\hat{P}}_\mathcal{U}$ for $\mathbf{f}=0$. }
    \label{fig:retlat}
\end{figure}

For an AO basis located at atom $\mathcal{U}$ in unit cell $\mathbf{R}$,
acting $\hat{P}$ on it yields a linear combination of the orbitals on the transformed atom $\tilde{\mathcal{U}}$ in unit cell $\tilde{\mathbf{R}}$, i.e. 
$
    \hat{P}\phi^\mathbf{R}_{\mathcal{U}\mu}(\mathbf{r})=\sum_{\mu'}\phi^{\tilde{\mathbf{R}}}_{\tilde{\mathcal{U}}\mu'}(\mathbf{r})\tilde{T}_{\mu'\mu}(V)
$,
where $\mathbf{\tilde{T}}(V)$ is the representation matrix of the rotation operator $V$ on the basis functions $\{\phi_\mu(\mathbf{r})\}$ centered on the original atom $\mathcal{U}$.
Therefore, the symmetry operator $\hat{P}=\{V|\mathbf{f}\}$ can be represented by the full basis set $\{\phi^{\mathbf{R}}_{\mathcal{U}\mu}(\mathbf{r})\}$ as
\begin{equation}
\begin{aligned}
    \label{eq:rot-atomic-orb}
    \hat{P}\phi^\mathbf{R}_{\mathcal{U}\mu}(\mathbf{r})
    &=
    \sum_{\mathcal{U}'\mu'}\phi^{\tilde{\mathbf{R}}}_{\mathcal{U}'\mu'}(\mathbf{r})\delta_{\mathcal{U}'\tilde{\mathcal{U}}}\tilde{T}_{\mathcal{U}'\mu',\mathcal{U}\mu}(V)\, .
\end{aligned}
\end{equation}
The rotation matrix $\tilde{T}_{\mu'\mu}(V)$ can in turn be constructed by the unitary transformed Wigner $D$ matrix (see \cref{ap:opformulas} for further details).
The transformed atomic position $\{\mathbf{s}_{\tilde{\mathcal{U}}},\tilde{\mathbf{R}}\}$ is related to its original position
$\{\mathbf{s}_\mathcal{U},\mathbf{R}\}$ via
\begin{equation}
  \begin{aligned}
    \tilde{\mathbf{R}} & =V\mathbf{R}+\mathbf{O}^{\hat{P}}_\mathcal{U} \\
    \mathbf{s}_{\tilde{\mathcal{U}}} & =V\mathbf{s}_\mathcal{U}+\mathbf{f}-\mathbf{O}^{\hat{P}}_\mathcal{U}
 \end{aligned}
 \label{eq:atom_position_transform}
\end{equation}
where $\mathbf{O}^{\hat{P}}_\mathcal{U}$ is the lattice vector corresponding to the unit cell into which
the image of atom $\mathcal{U}$ in the original cell $\mathbf{R}=(0,0,0)$ is transformed. Introducing $\mathbf{O}^{\hat{P}}_\mathcal{U}$ in Eq.~\eqref{eq:atom_position_transform} ensures
that $\mathbf{s}_{\tilde{\mathcal{U}}}$ is located within the unit cell at the origin $\mathbf{R}=(0,0,0)$. 
Fig.~\ref{fig:retlat} illustrates this situation for a simple case with $\mathbf{R}=(0,0,0)$ and $\mathbf{f}=0$. 

Now, applying Eqs.~\eqref{eq:hrep} and \eqref{eq:rot-atomic-orb} to the Hamiltonian matrix within the AO representation gives rise to the following transformation relationship between its matrix blocks associated with atomic pairs,
\begin{equation}
\begin{aligned}
H_{\mathcal{U}\mu,\mathcal{V}\nu}(\mathbf{R})
&=\braket{\phi^\mathbf{0}_{\mathcal{U}\mu}|\hat{H}|\phi^\mathbf{R}_{\mathcal{V}\nu}}
=\braket{\hat{P}\phi^\mathbf{0}_{\mathcal{U}\mu}|\hat{H}|\hat{P}\phi^\mathbf{R}_{\mathcal{V},\nu}}\\
&=\sum_{\mu'\nu'}\tilde{T}^*_{\mathcal{\tilde{U}}\mu',\mathcal{U}\mu}(V)\braket{\phi^\mathbf{\tilde{0}}_{\tilde{\mathcal{U}}\mu'}|\hat{H}|\phi_{\tilde{\mathcal{V}}\nu'}^\mathbf{\tilde{R}}}\tilde{T}_{\mathcal{\tilde{V}}\nu',\mathcal{V}\nu}(V)\\
&=\sum_{\mu'\nu'}\tilde{T}^*_{\mathcal{\tilde{U}}\mu',\mathcal{U}\mu}(V)
H_{\tilde{\mathcal{U}}\mu',\tilde{\mathcal{V}},\nu'}(V\mathbf{R}+\mathbf{O}_\mathcal{V}^{\hat{P}}-\mathbf{O}_\mathcal{U}^{\hat{P}})
\tilde{T}_{\mathcal{\tilde{V}}\nu',\mathcal{V}\nu}(V)\, .
\end{aligned}
\label{eq:Ham_matrix_trans}
\end{equation}
Here
\begin{equation}
    H_{\mathcal{U}\mu,\mathcal{V}\nu}(\mathbf{R})\equiv\braket{\phi^\mathbf{0}_{\mathcal{U}\mu}|\hat{H}|\phi^\mathbf{R}_{\mathcal
    V\nu}}
    =\iint{d\mathbf{r}d\mathbf{r'}\phi_{\mathcal{U}\mu}(\mathbf{r}-\mathbf{s}_{\mathcal{U}})
    H(\mathbf{r},\mathbf{r'})
    \phi_{\mathcal{V}\nu}}(\mathbf{r'}-\mathbf{s}_\mathcal{V}-\mathbf{R})
\end{equation}
is the Hamiltonian matrix between the AOs centered on two atoms located in two unit cells separated by a lattice vector $\mathbf{R}$.
In matrix form, Eq.~\ref{eq:Ham_matrix_trans} can be expressed as
\begin{equation}
\mathbf{H}_\mathcal{UV}(\mathbf{R})=\mathbf{\tilde{T}}^\dagger_{\mathcal{\tilde{U}},\mathcal{U}}(V)\mathbf{H}_\mathcal{\tilde{U}\tilde{V}}(V\mathbf{R}+\mathbf{O}_\mathcal{V}^{\hat{P}}-\mathbf{O}_\mathcal{U}^{\hat{P}})\mathbf{\tilde{T}}_{\mathcal{\tilde{V}},\mathcal{V}}(V)\, .
\label{eq:H-mat}
\end{equation}

The rule for rotating the Bloch orbitals can be derived from the rotation of the AOs, i.e., Eq.~\eqref{eq:rot-atomic-orb}.
From the invariance of the inner product, the relation between $\mathbf{k}$ and its rotated counterpart $\tilde{\mathbf{k}}=V\mathbf{k}$ can be obtained
\begin{equation}
\mathbf{k\cdot R}=(V\mathbf{k})\cdot V\mathbf{R}
=(\mathbf{\tilde{k}+K})(\mathbf{\tilde{R}}-\mathbf{O}_\mathcal{U}^{\hat{P}})
\label{eq:inner_product}
\end{equation}
where $\mathbf{K}$ is a reciprocal lattice vector. Consequently, we have 
\begin{equation}
\exp({i\mathbf{k\cdot R}})
=\exp({i\mathbf{\tilde{k}}\cdot\mathbf{\tilde{R}}-i\mathbf{\tilde{k}}\cdot\mathbf{O}_\mathcal{U}^{\hat{P}}})\, .
\label{eq:phase}
\end{equation}
Using Eqs.~\eqref{eq:Bloch_expansion}, \eqref{eq:rot-atomic-orb}, and \eqref{eq:phase}, one can show that a rotated Bloch orbital $\hat{P}\phi^\mathbf{k}(\mathbf{r})$ can be expressed by a linear combination of the Bloch orbitals associated with the rotated $\mathbf{k}$-point $\tilde{\mathbf{k}}=V\mathbf{k}$ and the rotated atom $\tilde{\mathcal{U}}$, namely,
\begin{equation}
 \label{eq:rot-bloch}
\begin{aligned}
\hat{P}\phi_{\mathcal{U}\mu}^\mathbf{k}(\mathbf{r})
&=\frac{1}{\sqrt{N_\mathbf{k}}}\sum_\mathbf{R}e^{i\mathbf{k\cdot R}}\sum_{\mathcal{U}'\mu'}\phi_{\mathcal{U}'\mu'}^{\tilde{\mathbf{R}}}(\mathbf{r})\delta_{\mathcal{U}'\tilde{\mathcal{U}}}\tilde{T}_{\mathcal{U}'\mu',\mathcal{U}\mu}(V)\\
&=\sum_{\mathcal{U}'\mu'} e^{-i\mathbf{\tilde{k}\cdot O^{\hat{P}}_\mathcal{U}}}
\frac{1}{\sqrt{N_\mathbf{k}}}\sum_\mathbf{R}e^{i\mathbf{\tilde{k}\cdot \tilde{R}}}\phi_{\mathcal{U}'\mu'}^{\tilde{\mathbf{R}}}(\mathbf{r})\delta_{\mathcal{U}'\tilde{\mathcal{U}}}\tilde{T}_{\mathcal{U}'\mu',\mathcal{U}\mu}(V)\\
&=\sum_{\mathcal{U}'\mu'}\phi_{\mathcal{U}'\mu'}^{\tilde{\mathbf{k}}}(\mathbf{r})e^{-i\mathbf{\tilde{k}}\cdot\mathbf{O}^{\hat{P}}_\mathcal{U}}\delta_{\mathcal{U}'\tilde{\mathcal{U}}}\tilde{T}_{\mathcal{U}'\mu',\mathcal{U}\mu}(V)\\
&\equiv\sum_{\mathcal{U}'\mu'}\phi_{\mathcal{U}'\mu'}^{\tilde{\mathbf{k}}}(\mathbf{r})M_{\mathcal{U}'\mu',\mathcal{U}\mu}(\hat{P};\mathbf{k})
\end{aligned}
\end{equation}
where we define matrix
\begin{equation}
    M_{\mathcal{U}'\mu',\mathcal{U}\mu}(\hat{P};\mathbf{k})=\delta_{\mathcal{U}'\tilde{\mathcal{U}}}\tilde{T}_{\mathcal{U}'\mu',\mathcal{U}\mu}(V)e^{-i\tilde{\mathbf{k}}\cdot\mathbf{O}^{\hat{P}}_\mathcal{U}}\, .
    \label{eq:T2M}
\end{equation}
which is the representation matrix of $\hat{P}$
in the basis of Bloch orbitals.
It follows that, for the Hamiltonian matrix represented in terms of the Bloch orbitals, the Hamiltonian blocks belonging to the symmetry-linked atom pairs are related by the following transformation,
\begin{equation}
\label{eq:rot-hamilt-bloch}
\begin{aligned}
H_{\mathcal{U}\mu,\mathcal{V}\nu}(\mathbf{k})
&=\braket{\hat{P}\phi^\mathbf{k}_{\mathcal{U}\mu}|\hat{H}|\hat{P}\phi^\mathbf{k}_{\mathcal{V}\nu}}\\
&=\sum_{\mathcal{U}'\mu'}\sum_{\mathcal{V}'\nu'}
M^*_{\mathcal{U}'\mu',\mathcal{U}\mu}(\hat{P};\mathbf{k})
\braket{\phi^\mathbf{\tilde{k}}_{\mathcal{U}'\mu'}|\hat{H}|\phi^\mathbf{\tilde{k}}_{\mathcal{V}'\nu'}}
M_{\mathcal{V}'\nu',\mathcal{V}\nu}(\hat{P};\mathbf{k})\\
&=\sum_{\mu'\nu'}\tilde{T}^*_{\mathcal{\tilde{U}}\mu',\mathcal{U}\mu}(V)H_{\tilde{\mathcal{U}}\tilde{\mathcal{V}}}(\tilde{\mathbf{k}})\tilde{T}_{\mathcal{\tilde{V}}\nu',\mathcal{V}\nu}(V)
e^{i\tilde{\mathbf{k}}(\mathbf{O}^{\hat{P}}_\mathcal{U}-\mathbf{O}^{\hat{P}}_\mathcal{V})} \, 
\end{aligned}
\end{equation}
or in matrix form,
\begin{equation}
\begin{aligned}
    \label{eq:rot-hamilt-bloch-matrix}
    \mathbf{H}(\mathbf{k})&=\mathbf{M}^\dagger(\hat{P};\mathbf{k})\mathbf{H}(\mathbf{\tilde{k}})\mathbf{M}(\hat{P};\mathbf{k})\, .\\
\end{aligned}
\end{equation}

\subsection{Symmetry-associated wave functions and density matrix of periodic systems}

In this subsection, we will derive how to transform the wavefunction coefficients and density matrices between the symmetry-associated $\mathbf{k}$-points, which is needed to restore the electronic structure information over the full Brillouin zone (BZ) from its irreducible wedge. 

According to Bloch's theorem, the wave function of a periodic system with wave vector $\mathbf{k}$ gets a phase factor under the pure translation of a Bravais lattice vector $\mathbf{R}$, i.e.
\begin{equation}
\label{eq:bloch-theorem}
\psi_\mathbf{k}(\mathbf{r-R})=\{E|\mathbf{R}\}\psi_\mathbf{k}(\mathbf{r})=e^{-i\mathbf{k\cdot R}}\psi_\mathbf{k}(\mathbf{r})
\end{equation}
where $E$ is the identity operation. 
To reduce the $\mathbf{k}$-points at which the Kohn-Sham equation has to be explicitly solved in self-consistent-field (SCF) calculation, we need to find out the relation between $\psi_\mathbf{k}(\mathbf{r})$ and $\psi_{V\mathbf{k}}(\mathbf{r})$. 
To start with, we first notice that
\begin{equation}
\{E|\mathbf{R}\}\{V|\mathbf{f}\}=\{V|\mathbf{f}\}\{E|V^{-1}\mathbf{R}\}
\label{eq:oper_change_order}
\end{equation}
which can be easily obtained from the following derivation,
\begin{equation}
\{E|\mathbf{R}\}\{V|\mathbf{f}\}\mathbf{r}
=V\mathbf{r}+\mathbf{f}+\mathbf{R}
=V(\mathbf{r}+V^{-1}\mathbf{R})+\mathbf{f}
=\{V|\mathbf{f}\}\{E|V^{-1}\mathbf{R}\}\mathbf{r} \, .
\end{equation}
Using Eq.~\eqref{eq:oper_change_order}, we have
\begin{equation}
\begin{aligned}
\{E|\mathbf{R}\}\{V|\mathbf{f}\}\psi_\mathbf{k}(\mathbf{r})&=\{V|\mathbf{f}\}\{E|V^{-1}\mathbf{R}\}\psi_\mathbf{k}(\mathbf{r})\\
&=\{V|\mathbf{f}\}e^{-i\mathbf{k}\cdot V^{-1}\mathbf{R}}\psi_\mathbf{k}(\mathbf{r})\\
&=e^{-iV\mathbf{k}\cdot\mathbf{R}}\{V|\mathbf{f}\}\psi_\mathbf{k}(\mathbf{r})\\
\end{aligned}
\end{equation}
which means $\{V|\mathbf{f}\}\psi_\mathbf{k}(\mathbf{r})$ is a Bloch function associated with the $\mathbf{k}$-point $V\mathbf{k}$.
Therefore, a generalized Bloch function at $\tilde{\mathbf{k}}=V\mathbf{k}$ will differ from the rotated Bloch function $\{V|\mathbf{f}\}\psi_\mathbf{k}(\mathbf{r})$
at most by a phase factor $\lambda=e^{i\theta}$,
\begin{equation}
\psi_{\tilde{\mathbf{k}}}(\mathbf{r})=\psi_{V\mathbf{k}}(\mathbf{r})=\lambda\{V|\mathbf{f}\}\psi_\mathbf{k}(\mathbf{r})\, .
\label{eq:rot-wfc}
\end{equation}
%
%
Combining Eqs.~\eqref{eq:rot-wfc}, and \eqref{eq:rot-bloch}, we obtain
\begin{equation}
    \psi_{\tilde{\mathbf{k}}}(\mathbf{r})=\lambda\{V|\mathbf{f}\}
    \sum_{\mathcal{U}\mu}c^\mathbf{k}_{\mathcal{U}\mu}\phi_{\mathcal{U}\mu}^{\mathbf{k}}   =\lambda\sum_{\mathcal{U}\mu}c^\mathbf{k}_{\mathcal{U}\mu}\sum_{\mathcal{U}'\mu'}\phi_{\mathcal{U}'\mu'}^{\tilde{
    \mathbf{k}}}(\mathbf{r})M_{\mathcal{U}'\mu',\mathcal{U}\mu}(\hat{P};\mathbf{k})\, .
\end{equation}
Further using  Eq.~\eqref{eq:wf_expansion} for $\tilde{\mathbf{k}}$,
\begin{equation}
    \psi_{\tilde{\mathbf{k}}}(\mathbf{r})=\sum_{\mathcal{U}'\mu'}c^\mathbf{\tilde{k}}_{\mathcal{U}'\mu'}\phi^{\tilde{\mathbf{k}}}_{\mathcal{U}'\mu'}(\mathbf{r}),
    \label{eq:KS_expansion_tildek}
\end{equation}
we obtain the following relation between the expansion coefficients at symmetry-linked $\mathbf{k}$ points,
\begin{equation}
\label{eq:coeff}
c^\mathbf{\tilde{k}}_{\mathcal{U'}\mu'}
=\lambda\sum_{\mathcal{U}\mu}M_{\mathcal{U'}\mu',\mathcal{U}\mu}(\hat{P};\mathbf{k}) c^\mathbf{k}_{\mathcal{U}\mu}
=\lambda[\mathbf{M}(\hat{P};\mathbf{k})\mathbf{c^k}]_{\mathcal{U'}\mu'} \, .
\end{equation}
%
Eq.~\eqref{eq:coeff} can be derived in an equivalent way by applying Eq.~\eqref{eq:rot-hamilt-bloch-matrix} to the KS equation.


From the charge density in AO basis, we can extract the expression of density matrix at a certain $\mathbf{k}$-point and its Fourier transformation to the real space \cite{soler_siesta_2002,li_large-scale_2016}: 
\begin{equation}
\begin{aligned}
\label{eq:charge-density}
\rho(\mathbf{r})&=\frac{1}{N_\mathbf{k}}\sum_{n\mathbf{k}}f_{n\mathbf{k}}\psi_{n\mathbf{k}}(\mathbf{r})\psi^*_{n\mathbf{k}}(\mathbf{r})
=\sum_{\mathcal{U}\mu}\sum_{\mathcal{V}\nu}\sum_\mathbf{R}
D_{\mathcal{U}\mu,\mathcal{V}\nu}(\mathbf{R})
\phi^\mathbf{0}_{\mathcal{U}\mu}(\mathbf{r})\phi^\mathbf{R}_{\mathcal{V}\nu}(\mathbf{r})\\
\end{aligned}
\end{equation}
where
\begin{equation}
D_{\mathcal{U}\mu,\mathcal{V}\nu}(\mathbf{R})
\equiv\frac{1}{N_\mathbf{k}}\sum_{n\mathbf{k}}f_{n\mathbf{k}}
c_{\mathcal{U}\mu, n\mathbf{k}}c^*_{\mathcal{V}\nu, n\mathbf{k}}e^{-i\mathbf{k}\cdot\mathbf{R}}
\equiv\frac{1}{N_\mathbf{k}}\sum_\mathbf{k}e^{-i\mathbf{k\cdot R}}D_{\mathcal{U}\mu,\mathcal{V}\nu}(\mathbf{k})\, .
\label{eq:dm-fouier}
\end{equation}
Using Eq.~\ref{eq:coeff} and the normalization condition $\lambda\lambda^{\ast}=1$, the relationship between
the density matrices at $\mathbf{k}$ and $\mathbf{\tilde{k}}$ can be derived,
\begin{equation}
\label{eq:dm-matrix}
\mathbf{D}(\mathbf{\tilde{k}})=\mathbf{c}^\mathbf{\tilde{k}} \left(\mathbf{c}^\mathbf{\tilde{k}}\right)^\dagger
=\mathbf{M}(\hat{P};\mathbf{k}) \mathbf{c}^\mathbf{k} \left(\mathbf{c}^\mathbf{k}\right)^\dagger\mathbf{M}^\dagger(\hat{P};\mathbf{k})
=\mathbf{M}(\hat{P};\mathbf{k})\mathbf{D}(\mathbf{k})\mathbf{M}^\dagger(\hat{P};\mathbf{k}) \, .
\end{equation}
%
Inserting Eq.~\eqref{eq:dm-matrix}, ~\eqref{eq:T2M} and Eq.~\eqref{eq:phase} to Eq.~\eqref{eq:dm-fouier} leads to the rules for rotating the density sub-matrices on the atom pair at a distance of lattice vector $\mathbf{R}$)
\begin{equation}
\begin{aligned}
D_{\mathcal{U}\mu,\mathcal{V}\nu}(\mathbf{R})
&=\frac{1}{N_\mathbf{k}}\sum_{\mathbf{k}\in\text{BZ}}e^{-i\mathbf{k\cdot R}}D_{\mathcal{U}\mu,\mathcal{V}\nu}(\mathbf{k})\\
&=\frac{1}{N_\mathbf{k}}\sum_{\mathbf{\tilde{k}}\in\text{BZ}}e^{-i\mathbf{\tilde{k}}\cdot (\mathbf{\tilde{R}}-\mathbf{O}^{\hat{P}}_\mathcal{V})}
\sum_{\mathcal{U}'\mu'}\sum_{\mathcal{V}'\nu'}
e^{i\mathbf{\tilde{k}}\cdot\mathbf{O}^{\hat{P}}_\mathcal{U}}
\delta_{\mathcal{U}'\tilde{\mathcal{U}}}\tilde{T}^*_{\mu'\mu}(V)
D_{\mathcal{U}'\mu'\mathcal{V}'\nu'}(\tilde{\mathbf{k}})\delta_{\mathcal{V}'\mathcal{\tilde{V}}}\tilde{T}_{\nu'\nu}(V)e^{-i\mathbf{\tilde{k}}\cdot\mathbf{O}^{\hat{P}}_\mathcal{V}}\\
&=\sum_{\mathcal{U}'\mu'}\sum_{\mathcal{V}'\nu'}
\delta_{\mathcal{U}'\tilde{\mathcal{U}}}\tilde{T}^*_{\mu'\mu}(V)
\frac{1}{N_\mathbf{k}}
\sum_{\mathbf{\tilde{k}\in\text{BZ}}}
e^{-i\mathbf{\tilde{k}}\cdot(V\mathbf{R}+\mathbf{O}^{\hat{P}}_\mathcal{V}-\mathbf{O}_\mathcal{U}^{\hat{P}})} D_{\mathcal{U}'\mu'\mathcal{V}'\nu'}(\tilde{\mathbf{k}})\delta_{\mathcal{V}'\mathcal{\tilde{V}}}\tilde{T}_{\nu'\nu}(V)\\
&=\sum_{\mu'}\sum_{\nu'}\tilde{T}^*_{\mu'\mu}(V)D_{\tilde{\mathcal{U}}\mu',\tilde{\mathcal{V}},\nu'}(V\mathbf{R}+\mathbf{O}^{\hat{P}}_\mathcal{V}-\mathbf{O}^{\hat{P}}_\mathcal{U})
\tilde{T}_{\nu'\nu}(V)
\end{aligned}
\end{equation}
or in matrix form,
\begin{equation}
\mathbf{D}_{\mathcal{UV}}(\mathbf{R})=\mathbf{\tilde{T}}^\dagger(V)\mathbf{D}_{\mathcal{\tilde{U}\tilde{V}}}(V\mathbf{R}+\mathbf{O}^{\hat{P}}_\mathcal{V}-\mathbf{O}^{\hat{P}}_\mathcal{U})\mathbf{\tilde{T}}(V) \, .
\label{eq:DM_rot_realspace}
\end{equation}
Note that the atom in lattice $\mathbf{R}$ is $\mathcal{V}$ according to Eq.~\eqref{eq:charge-density}, and hence
in the above derivation we have used $V\mathbf{R}=\mathbf{\tilde{R}}-\mathbf{O}^{\hat{P}}_\mathcal{V}$.
Consequently, the phase factor $e^{i\tilde{\mathbf{k}}\cdot\mathbf{O}^{\hat{P}}_\mathcal{V}}$ cancels with $e^{-i\tilde{\mathbf{k}}\cdot\mathbf{O}^{\hat{P}}_\mathcal{V}}$ in $M_{\mathcal{V}'\nu',\mathcal{V}\nu}$).

As a brief summary, in this section we derived the key equations behind the symmetry operations behind the Hamiltonian and density matrices, i.e., Eqs.~\eqref{eq:H-mat}, \eqref{eq:rot-hamilt-bloch-matrix}, \eqref{eq:dm-matrix} and \eqref{eq:DM_rot_realspace}, which show that the Hamiltonian and density matrices follow the same rules for the
transformation in both real and reciprocal spaces. 
In particular, Eq.~\eqref{eq:H-mat} and Eq.~\eqref{eq:dm-matrix} are used in our actual implementations that utilize
symmetries in HDF calculations.

\section{Algorithms and Implementation Details}
\label{sec:implement}

In this section, we discuss how to utilize the crystal space-group symmetry to speed up the calculation of the
EXX Hamiltonian. The algorithm for finding the irreducible sector in the BvK supercell is presented in Sec. \ref{imp:finding_sector}, which is the prerequisite for the application of symmetry in real space. Sec.~\ref{imp:perspective} compares the formulas of the EXX Hamiltonian in two perspectives under the LRI scheme, and finally in Sec.~\ref{imp:apply_selection} the detailed algorithm for calculating the irreducible-sector EXX Hamiltonian is described.

\subsection{Finding the Irreducible Sector}
\label{imp:finding_sector}
To exploit the symmetry for constructing the Hamiltonian matrix in real space, it is necessary to find the irreducible sector of
the BvK supercell at first, i.e. the smallest set of atomic pairs that can recover the full set of atomic pairs in the supercell via symmetry operations.
To this end, the symmetry of the BvK supercell needs to be first analyzed, which will be equal to the unit-cell symmetry when $\mathbf{k}$-points are equally sampled along all the three reciprocal lattice vectors, but lower otherwise. The set of symmetry operations of the BvK supercell $\{\hat{P}\}$  will be used to find the irreducible sector. 

Starting from the full sector, i.e. the set of all the distinct atom pairs $\{(\mathcal{U}\mathcal{V}\mathbf{R})\}$ within the BvK supercell (with $\mathcal{U}, \mathcal{V}$ labeling two atoms in the unit cell and $\mathbf{R}$ the lattice vector in the BvK supercell), we need to identify a minimal set of atom pairs $\{(\tilde{\mathcal{U}}\tilde{\mathcal{V}}\tilde{\mathbf{R}})\}$ such that acting all the elements in the spacegroup $\{\hat{P}\}$ on them can recover the full set $\{(\mathcal{U}\mathcal{V}\mathbf{R})\}$. The actual procedure to identify the irreducible sector is described in Algorithm~\ref{alg:alg1}\cite{86Dovesi}.
The essential step is to apply each of the space-group operations $\{\hat{P}\}$ on a selected atom pair $(\mathcal{U}\mathcal{V}\mathbf{R})$ to generate its ``star'' (the set of the atom pairs transformed from it), choose one representative element from the star (here we simply choose the generator $(\mathcal{U}\mathcal{V}\mathbf{R})$) and put it into the irreducible sector. The atom pairs covered by the existing stars will not be selected as the generator any more. 
The generation and selection 
process will stop once the existing stars cover the full set $\{(\mathcal{U}\mathcal{V}\mathbf{R})\}$.
\begin{algorithm}[t]
    \caption{Finding the Irreducible Sector}
    \label{alg:alg1}
    \KwIn{
        The full sector of atomic pairs: $S=\{(\mathcal{U}\mathcal{V}\mathbf{R})\}$, sorted in the order $|\mathbf{R}|$ from short to long;\newline
        The set of symmetry operations of the BvK supercell: $\{\hat{P}\}$.
    }
    \KwOut{
        The stars of pairs with the symmetry operation to their irreducible pair:  $\mathcal{S}=\{s_{(\tilde{\mathcal{U}}\tilde{\mathcal{V}}\mathbf{\tilde{R}})}=\{(\hat{P}, (\mathcal{U}\mathcal{V}\mathbf{R}))\}\}$;\newline
        Irreducible sector $\tilde{S}=\{(\tilde{\mathcal{U}}\tilde{\mathcal{V}}\mathbf{\tilde{R}})\}$.
    }
    \BlankLine
    \While{\textnormal{$S$ is not empty,}}{
        \ForEach{$\hat{P}$,}{
            Apply $\hat{P}^{-1}$ to the first element of $S$: $(\mathcal{U}\mathcal{V}\mathbf{R})=\hat{P}^{-1}(\tilde{\mathcal{U}}\tilde{\mathcal{V}}\mathbf{\tilde{R}})$.
            \BlankLine
            \If{$(\mathcal{U}\mathcal{V}\mathbf{R}) \in S$}{
                Save $(\hat{P}, (\mathcal{U}\mathcal{V}\mathbf{R}))$ into the star of $(\tilde{\mathcal{U}}\tilde{\mathcal{V}}\mathbf{\tilde{R}})$: $s_{(\tilde{\mathcal{U}}\tilde{\mathcal{V}}\mathbf{\tilde{R}})}$.

                Erase $(\mathcal{U}\mathcal{V}\mathbf{R})$ from $S$.
            }
        }
        Save the star $s_{(\tilde{\mathcal{U}}\tilde{\mathcal{V}}\mathbf{\tilde{R}})}$ into $\mathcal{S}$ and its first element $ (\tilde{\mathcal{U}}\tilde{\mathcal{V}}\mathbf{\tilde{R}})$ (with $\hat{P}=\{E|\mathbf{0}\}$) into the irreducible sector $\tilde{S}$.
    }
\end{algorithm}

\subsection{V-perspective calculation of the EXX Hamiltonian in the Irreducible Sector }
\label{imp:perspective}
\begin{figure}[h]
    \centering
    \includegraphics[width=\textwidth]{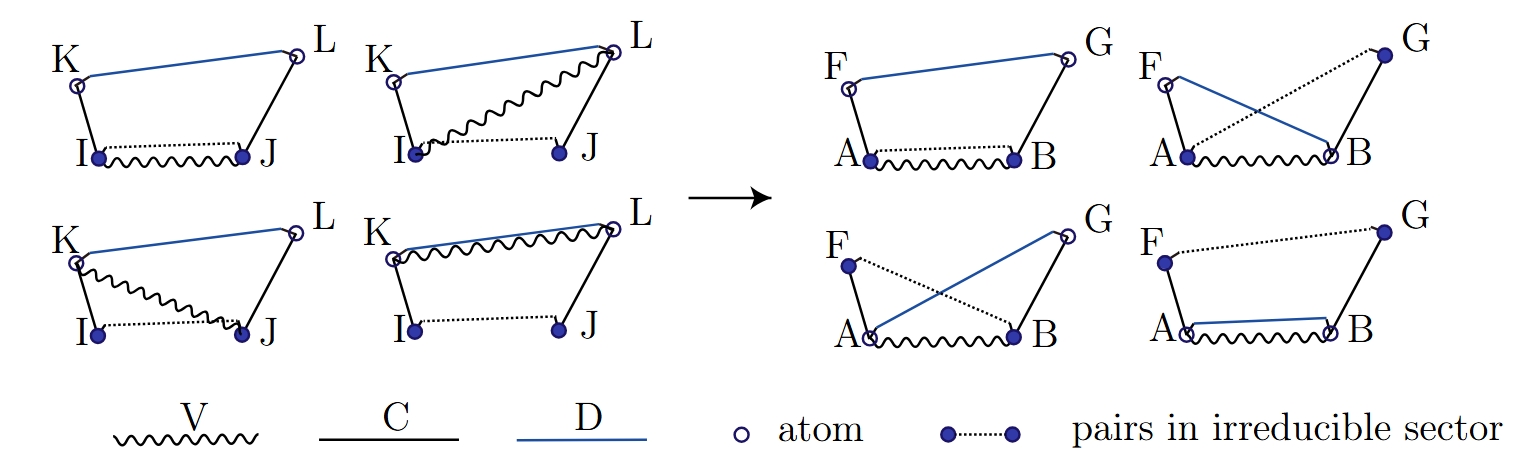}
    \caption{From $D$-perspective to $V$-perspective. The dot represent atoms; the wavy lines denote the Coulomb matrix $V$, the black solid lines the RI coefficient tensor $C$, and the blue solid lines the density matrix $D$. The dashed line in each graph signifies the atom pair to which the uncontracted external orbital indices belong, and the blue solid circles mark the atoms in the irreducible sector. }
    \label{fig:D-V-perspective}
\end{figure}

In this subsection, we shall discuss how to compute the EXX part of the Hamiltonian under the LRI scheme. 
The key is that we can restrict the calculation only to the irreducible sector of the BvK supercell, and rotate
the irreducible blocks to other regions when needed.
To simplify the notation, here we omit the lattice vector index $\mathbf{R}$ by defining the atom index $I$ in the full BvK supercell as the combination of the its index in the unit cell $\mathcal{U}$ and the lattice vector $\mathbf{R}$, i.e. $\phi_{Ii}\equiv\phi^\mathbf{R}_{\mathcal{U}i}$, and use the tilded ``$\tilde{I}$'' to denote that the atom $I\equiv\{\mathcal{U},\mathbf{R}\}$ lies in the irreducible sector. 
In this nomenclature, the irreducible-sector element of the EXX Hamiltonian is given by
\begin{equation}
    H^\text{EXX}_{\tilde{I}i,\tilde{J}j}
    =\sum_{KL}\sum_{kl}D_{Kk, Ll}(\phi_{\tilde{I}i}\phi_{Kk}|\phi_{\tilde{J}j}\phi_{Ll})\, ,
    \label{eq:H_exx_realspace}
\end{equation}
where $(\phi_{\tilde{I}i}\phi_{Kk}|\phi_{\tilde{J}j}\phi_{Ll})$ are the 2-electron 4-orbital Coulomb integrals, defined as
\begin{equation}
(\phi_{\tilde{I}i}\phi_{Kk}|\phi_{\tilde{J}j}\phi_{Ll})
=\int{d\mathbf{r}}\int{d\mathbf{r'}}
\phi_{\tilde{I}i}(\mathbf{r})\phi_{Kk}(\mathbf{r})v(|\mathbf{r}-\mathbf{r'}|)
\phi_{\tilde{J}j}(\mathbf{r'})\phi_{Ll}(\mathbf{r'}) \, .
\label{eq:4-center_integral}
\end{equation}
Note that the summation over the elements $Kk,Ll$  of the density matrix in Eq.~\eqref{eq:H_exx_realspace} 
is not restricted to the irreducible sector.
Following previous works \cite{levchenko_hybrid_2015,lin_accuracy_2020,lin_efficient_2021}, we adopt the
localized resolution of identity (LRI) technique \cite{15Ihrig-LRI} to compute the 4-orbital integrals,
which means the orbital products are expanded in terms of a set of auxiliary basis functions (ABFs) located on 
the two atoms of the products, 
\begin{equation}
    \phi_{Ii}(\mathbf{r})\phi_{Jj}(\mathbf{r})
=\sum_{\alpha\in I}C_{Ii,Jj}^{I\alpha}P_{I\alpha}(\mathbf{r})+\sum_{\alpha\in J}C_{Jj,Ii}^{J\alpha}P_{J\alpha}(\mathbf{r})\, ,
\label{eq:LRI}
\end{equation}
where $P_{I\alpha}(\mathbf{r})$'s are the atom-centered ABFs and  $C_{Ii,Jj}^{I\alpha}$ are the corresponding expansion
coefficient.
Combining Eqs.~\eqref{eq:H_exx_realspace} and \eqref{eq:LRI}, we arrive at
\begin{equation}
    \begin{aligned}
    H^{\text{EXX}-D}_{\tilde{I}i,\tilde{J}j}=\sum_{KL}\sum_{kl}D_{Kk, Ll}
    &\left[ \sum_{\alpha\in\tilde{I}}\sum_{\beta\in\tilde{J}}C_{\tilde{I}i, Kk}^{\tilde{I}\alpha}V_{\tilde{I}\alpha,\tilde{J}\beta}C_{\tilde{J}j, Ll}^{\tilde{J}\beta} \right.\\
    &+\sum_{\alpha\in\tilde{I}}\sum_{\beta\in L}C_{\tilde{I}i, Kk}^{\tilde{I}\alpha}V_{\tilde{I}\alpha,L\beta}C_{Ll,\tilde{J}j}^{L\beta}\\
    &+\sum_{\alpha\in K}\sum_{\beta\in\tilde{J}}C_{Kk,\tilde{I}i}^{K\alpha}V_{K\alpha,\tilde{J}\beta}C_{\tilde{J}j, Ll}^{\tilde{J}\beta}\\
    &\left. +\sum_{\alpha\in K}\sum_{\beta\in L}C_{Kk,\tilde{I}i}^{K\alpha}V_{K\alpha,L\beta}C_{Ll,\tilde{J}j}^{L\beta}
    \right]\, ,
    \end{aligned}
    \label{eq:D-perspective}
\end{equation}
where
\begin{equation}
V_{I\alpha,J\beta}=\int{d\mathbf{r}}\int{d\mathbf{r'}}
P_{I\alpha}(\mathbf{r})v(|\mathbf{r}-\mathbf{r'}|)P_{J\beta}(\mathbf{r'})
\end{equation}
is the Coulomb matrix in the representation of the auxiliary basis. In Eq.~\eqref{eq:D-perspective}, 
the superscript ``$D$'' in 
$H^{\text{EXX}-D}_{\tilde{I}i,\tilde{J}j}$ indicates that the 4 terms share the same density matrix $D_{Kk, Ll}$ distributed on the atoms $K$ and $L$. This
expression is called ``$D$-perspective'' in the following, and a pictorial representation of the four terms
in this perspective is shown in the left half of Fig.~\ref{fig:D-V-perspective}).
The splitting of the EXX Hamiltonian into four terms is the characteristic of the LRI approximation.

As a side remark, we note that the most memory-intensive quantity in the construction of the EXX Hamiltonian
(cf. Eq.~\eqref{eq:D-perspective}) is the LRI coefficients, i.e. the two-center 3-index $\mathbf{C}$-tensors. 
In principle, one can save memory by storing only the irreducible part of them and retrieving other needed parts
on the fly via rational operations. The formulation of such operations is derived in \Cref{ap:Cs}.
However, in the present work we have not put this feature in practical use.
In the practical implementation in the ABACUS code \cite{lin_efficient_2021}, instead of
adopting the ``$D$-perspective'', we followed an alternative, the so-called  ``$V$-perspective'' to
build the EXX Hamiltonian, as illustrated in the right panel of Fig.~\ref{fig:D-V-perspective}. The ``$V$-perspective'' can be obtained from the ``$D$-perspective'' by relabeling the atoms so that 
the Coulomb matrix always stays on the same pair of atoms. The choice of ``$V$-perspective'' 
facilitates the design of efficient
loop structure and parallelization scheme \cite{lin_efficient_2021,lin_force_2025}. 
Specifically, we relabel the atoms where the ABFs are located as $A, B$, the neighbor of $A$ as $F$ (connected by $C_{AF}^A$) and the neighbor of $B$ as $G$ (connected by $C_{BG}^B$). 
In this way, the $CVC$ part of all the four terms share the same label, i.e., 
$C_{AF}^AV_{AB}C_{BG}^B$. However, now there are four types of $D$'s, i.e. $D_{FG},  D_{FB}, D_{AG}, D_{AB}$, representing the four types of topological structure of $CVCD$ (see the right panel of Fig.~\ref{fig:D-V-perspective}), respectively. The abstract label $\{ABFG\}$ runs over all the 4-atom integrals $\{IJKL\}$ that need to be calculated after filtering out the tiny ones according to given thresholds \cite{haser_improvements_1989,lin_efficient_2021}. 
In this work, the symmetry constraints are applied in addition as another filter to screen out the 4-atom integrals that do not contribute to the irreducible sector.

As a specific example,  the second term of $H^{\text{EXX}-D}_{\tilde{I}i,\tilde{J}j}$ (with ABFs on $\tilde{I}$ and $L$) can be written as
\begin{equation}
\begin{aligned}
H_{\tilde{I}i,\tilde{J}j}^{\text{EXX}, D-V_{IL}}&=\sum_{\alpha\in \tilde{I}}\sum_{\beta\in L}\sum_{KL}\sum_{kl}C_{\tilde{I}i,Kk}^{\tilde{I}\alpha}V_{\tilde{I}\alpha,L\beta}C_{Ll,\tilde{J}j}^{L\beta}D_{Kk,Ll}\\
&=\left. \sum_{\alpha\in \tilde{A}}\sum_{\beta\in B}\sum_{FB}\sum_{kl}C_{\tilde{A}i,Fk}^{\tilde{A}\alpha}V_{\tilde{A}\alpha,B\beta}C_{Bl,\tilde{G}j}^{B\beta}D_{Fk,Bl}\right|_{A=I, B=L, F=K, G=J} \\
&\equiv \left.H_{\tilde{A}i,\tilde{G}j}^{\text{EXX}, V-D_{FB}}\right|_{A=I, B=L, F=K, G=J}
\end{aligned}
\end{equation}
After doing the same thing for all the 4 terms, we have the following correspondence between the
two perspectives (for brevity, omitting the "EXX" tag),
\begin{equation}
    \begin{aligned}
\mathbf{H}_{\tilde{I}, \tilde{J}}^{\text{D-}V_{IJ}}
&=\left.\mathbf{H}_{\tilde{A}, \tilde{B}}^{\text{V-}D_{FG}}\right|
_{A=I, B=J, F=K, G=L}\\
\mathbf{H}_{\tilde{I}, \tilde{J}}^{\text{D-}V_{IL}}
&=\left.\mathbf{H}_{\tilde{A}, \tilde{G}}^{\text{V-}D_{FB}}\right|
_{A=I, B=L, F=K, G=J}\\
\mathbf{H}_{\tilde{I}, \tilde{J}}^{\text{D-}V_{KJ}}
&=\left.\mathbf{H}_{\tilde{F}, \tilde{B}}^{\text{V-}D_{AG}}\right|
_{A=K, B=J, F=I, G=L}\\
\mathbf{H}_{\tilde{I}, \tilde{J}}^{\text{D-}V_{KL}}
&=\left.\mathbf{H}_{\tilde{F}, \tilde{G}}^{\text{V-}D_{AB}}\right|
_{A=K, B=L, F=I, G=J}\\
    \end{aligned}
\end{equation}
Summing over the relabeled 4 terms contributing to the irreducible pair $(\tilde{I}\tilde{J})$, 
we arrive at the $V$-perspective formula,
\begin{equation}
\begin{aligned}
&\quad H_{\tilde{I}i,\tilde{J}j}^{\text{EXX},V}\\
&=H^{\text{V-}D_{FG}}_{\tilde{A}i,\tilde{B}j}|_{ABFG=IJKL}
+H^{\text{V-}D_{FB}}_{\tilde{A}i,\tilde{G}j}|_{ABFG=ILKJ}\\
&\quad+H^{\text{V-}D_{AG}}_{\tilde{F}i,\tilde{B}j}|_{ABFG=KJIL}
+H^{\text{V-}D_{AB}}_{\tilde{F}i,\tilde{G}j}|_{ABFG=KLIJ}\\
&=\left.\sum_{FG}\sum_{kl}\sum_{\alpha\in\tilde{A}}\sum_{\beta\in\tilde{B}}C_{\tilde{A}i, Fk}^{\tilde{A}
\alpha}V_{\tilde{A}\alpha,\tilde{B}\beta}C_{\tilde{B}j, Gl}^{\tilde{B}\beta}D_{Fk, Gl}\right|_{ABFG=IJKL}\\
&+\left.\sum_{FB}\sum_{kj}\sum_{\alpha\in \tilde{A}}\sum_{\beta\in B}
C_{\tilde{A}i,Fk}^{\tilde{A}\alpha}V_{\tilde{A}\alpha,B\beta}C_{Bj,\tilde{G}l}^{B\beta}D_{Fk,Bj}\right|_{ABFG=ILKJ}\\
&+\left.\sum_{AG}\sum_{il}\sum_{\alpha\in A}\sum_{\beta\in \tilde{B}}
C_{Ai,\tilde{F}k}^{A\alpha}V_{A\alpha,\tilde{B}\beta}C_{\tilde{B}j,Gl}^{\tilde{B}\beta}D_{Ai,Gl}\right|_{ABFG=KJIL}\\
&+\left.\sum_{AB}\sum_{ij}\sum_{\alpha\in A}\sum_{\beta\in B} C_{Ai,\tilde{F}k}^{A\alpha}V_{A\alpha,B\beta}C_{Bj,\tilde{G}l}^{B\beta}D_{Ai,Bj}\right|_{ABFG=KLIJ}
\end{aligned}\label{eq:loop4}
\end{equation}
Now we observe a difference in the application of the irreducible-sector selection between the $D$-perspective Eq.~\eqref{eq:D-perspective} and the $V$-perspective Eq.~\eqref{eq:loop4} formulations illustrated by Fig.~\ref{fig:D-V-perspective}. Namely, in the former case, the four contributions are added to the same irreducible pair 
$(\tilde{I}\tilde{J})$ together, while the latter scheme places them into four different classes according to the location of the ABFs, and in each class the irreducible-sector pairs are labeled differently, namely $\tilde{A}\tilde{B}$, $\tilde{A}\tilde{G}$, $\tilde{F}\tilde{B}$ and $\tilde{F}\tilde{G}$. This prevents the different types of terms from being calculated together within a common 4-layer loop, as was done previously. However, as will be discussed in the next subsection, an improved algorithm allows the irreducible-sector selection to be applied once again.

\subsection{Applying the irreducible sector selection to the existing algorithm}
\label{imp:apply_selection}
\begin{figure}[htbp]
    \centering
    \includegraphics[width=0.5\textwidth]{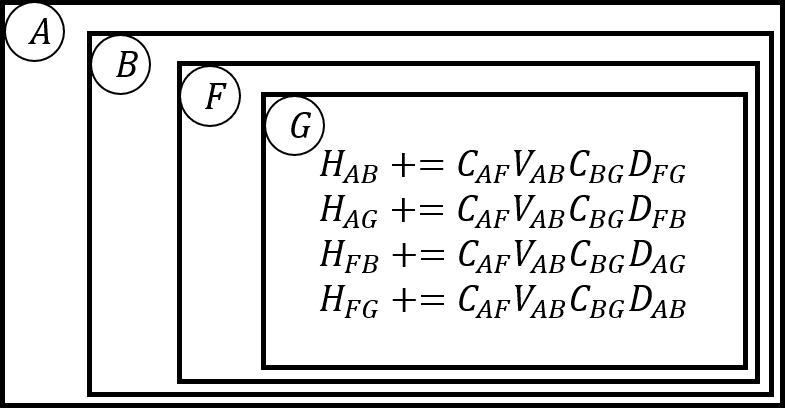}
    \caption{The ``loop4'' algorithm. Each box denotes a for-loop of the atom label in the upper-left corner. The 4 terms can be calculated together in the inner loop of the same structure, but it's hard to apply irreducible sector selection because the 2 loops of atoms in the irreducible sector are located at different layers for each term. }
    \label{fig:loop4}
\end{figure}
\begin{figure}[htbp]
    \centering
    \includegraphics[width=0.8\textwidth]{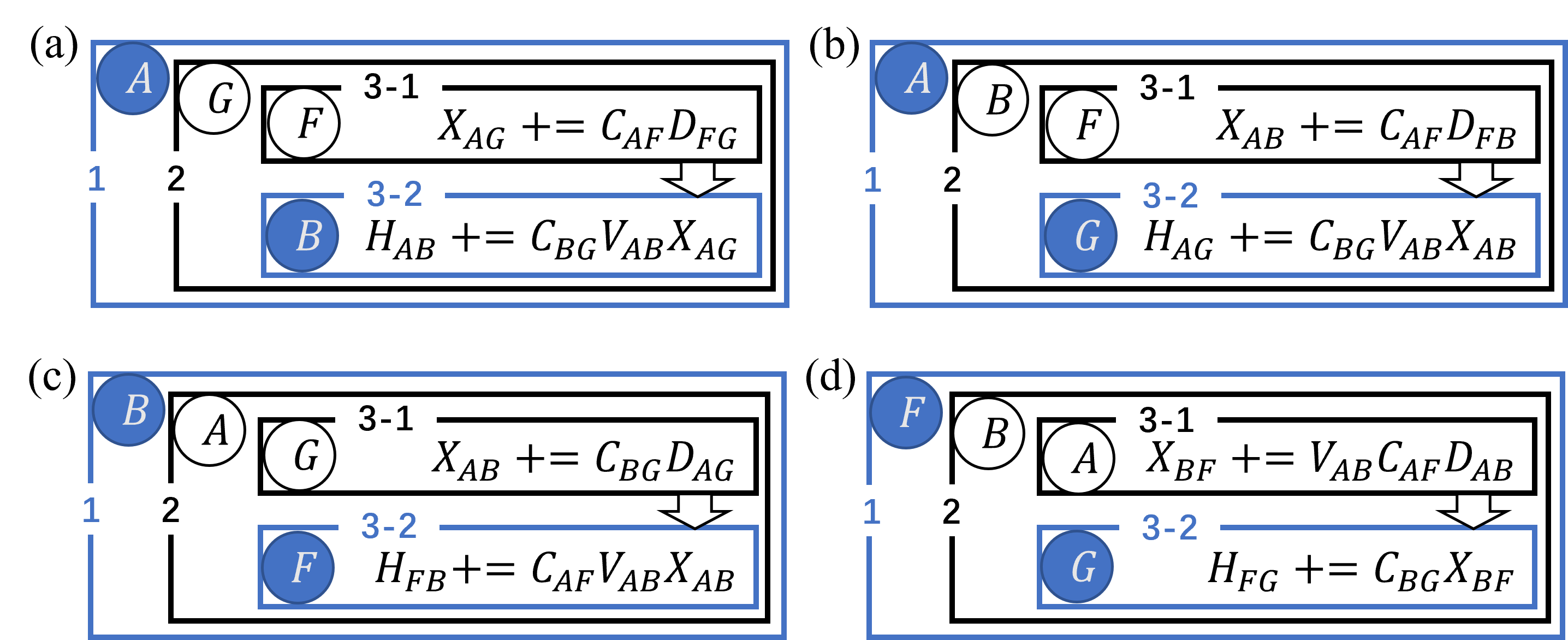}
    \caption{The ``loop3'' algorithm with irreducible sector selection: The blue boxes with solid circles mean that only the atoms in the irreducible sector are traversed in the loop over the enclosed atomic label, located at the same place for all the 4 terms. The label (1), (2), (3-1), (3-2) mark different locations of the 3-layer loop
    structure, to be referred to in the main text.
    The panel (a), (b), (c) and (d) correspond to the four terms listed in Eq.~\eqref{eq:loop3-term1} to ~\eqref{eq:loop3-term4}, respectively.}
    \label{fig:loop3}
\end{figure}
In the $V$-perspective algorithm developed originally \cite{lin_efficient_2021},  the 4-tensor products ``$\mathrm{CVCD}$'' are calculated inside a 4-layer loop, running over the 4 atoms $\{ABFG\}$, where the summation over the two atoms associated with the density matrix is carried out outside all the contractions over AOs and ABFs (see Fig.~\ref{fig:loop4}). 
The algorithm has recently been improved by summing up one of the two atoms before contracting the AOs and ABFs located on the other atoms (see Fig.~\ref{fig:loop3}). By so doing, the original 4-layer loop structure can be
reduced to a 3-layer one.
In the following, we shall refer to the previous algorithm as ``loop4'' and the improved one as ``loop3'', respectively. The ``loop3'' algorithm was proposed and implemented in LibRI by Lin \textit{et al.},  which will be published separately \cite{LibRI-paper}.

Here we briefly explain how the ``loop3'' algorithm facilitates imposing the restriction of the Hamiltonian matrix to the irreducible sector.
We take the first term of Eq.~\ref{eq:loop4} as an example, corresponding to panel (a) of Fig.~\ref{fig:loop3}. In the innermost loop of calculating this term, $C_{AF}^A$ and $D_{FG}$ can be first contracted by summing over $Fk$ to get $[CD]_{AG}$. In other words, $C_{BG}^B$ and $V_{AB}$ are factored out from the summation over $F$. 
Furthermore, the intermediate result $[CD]_{AG}$ does not depend on $B$, in contrast with
the original ``loop4'' structure where $C^A_{AF}\times D_{FG}$ is repeated $N_B$ times inside the loop over $B$. By contracting over atom $F$ before traversing atom $B$, the 4-layer loop can be flattened to a 3-layer one. 

Now we explain how the 3-layer loop structure is determined to evaluate the first term. Given the first-summed atom ($F$) in the first innermost loop (indicated by ``3-1'' in Fig.~\ref{fig:loop3}), and the atom $B$ independent of $F$ and $CD_{AG}$ should be traversed in the second innermost loop (indicated by ``3-2"). 
The order of the two outer loops over the atoms $A$ and $G$ still needs to be determined. 
Note that $A$ is one of the final external atomic indices while $G$ is to be contracted, so $G$ (loop 2) should be placed inside of $A$ (loop 1). The final order of the loops from outer to inner layers is thus $A\to G\to (F, B)$.

After a similar analysis of all other 3 terms, we can determine the loop structure separately for each of them (see panels of (b), (c), and (d) of Fig.~\ref{fig:loop3}) and express explicitly the corresponding formulas as follows (the order of summation is different from the ``loop4'' algorithm represented by Eq.~\eqref{eq:loop4} and Fig.~\ref{fig:loop4}): 
\begin{itemize}
 \item[] Term 1: $\tilde{A}\to G\to (F, \tilde{B})$
\begin{equation}
    H_{\tilde{A}i,\tilde{B}j}^\text{EXX, 1}
=\sum_{Gl}\sum_{\alpha\in \tilde{A},\beta\in \tilde{B}}C_{\tilde{B}j,Gl}^{\tilde{B}\beta}V_{\tilde{A}\alpha,\tilde{B}\beta}\sum_{Fk}C_{\tilde{A}i,Fk}^{\tilde{A}\alpha}D_{Fk,Gl}
\label{eq:loop3-term1}
\end{equation}
\item[] Term 2: $\tilde{A}\to B\to (F, \tilde{G})$
\begin{equation}
    H_{\tilde{A}i,\tilde{G}l}^\text{EXX, 2}
=\sum_{Bj}\sum_{\alpha\in \tilde{A}, \beta\in B}C_{Bj,\tilde{G}l}^{B\beta}V_{\tilde{A}\alpha,B\beta}\sum_{Fk}C_{\tilde{A}i,Fk}^{\tilde{A}\alpha}D_{Fk,Bj}
\end{equation}
\item[] Term 3: $\tilde{B}\to A\to (G, \tilde{F})$
\begin{equation}
    H_{\tilde{F}k,\tilde{B}j}^\text{EXX, 3}
=\sum_{Ai}\sum_{\alpha\in A, \beta\in \tilde{B}}C_{Ai,\tilde{F}k}^{A\alpha}V_{A\alpha,\tilde{B}\beta}
\sum_{Gl}C_{\tilde{B}j,Gl}^{\tilde{B}\beta}D_{Ai,Gl}
\end{equation}
\item[] Term 4: $\tilde{F}\to B\to (A, \tilde{G})$
\begin{equation}
    H_{\tilde{F}k,\tilde{G}l}^\text{EXX, 4}
=\sum_{Bj}\sum_{\beta\in B}C_{Bj,\tilde{G}l}^{B\beta}\sum_{Ai}\sum_{\alpha\in A}V_{A\alpha,B\beta}C_{Ai,\tilde{F}k}^{A\alpha}D_{Ai,Bj} \, .
\label{eq:loop3-term4}
\end{equation}
\end{itemize}

\begin{algorithm}[t]
    \caption{Constructing the EXX Hamiltonian with symmetry (irreducible sector selection) applied to the ``loop3'' algorithm, taking the first-type terms in $V$-perspective as an example. }
    \label{alg:alg2}    
    \BlankLine
    \KwIn{$\{C\}$, $\{V\}$, $\{D\}$, the irreducible sector $\tilde{S}=\{\tilde{I}\tilde{J}\}=\{\mathcal{\tilde{U}}\mathcal{\tilde{V}}\mathbf{\tilde{R}}\}$}
    \KwOut{the first type of contribution to $H^\text{EXX}$ in V-perspective}
    \ForEach{$A$}{
    \If{\textit{A is the first atom of any pair in $\tilde{S}$}}{
        \ForEach{$G$}{
            \ForEach{F}{$X_{AG}+=C_{AF}D_{FG}$}
            \ForEach{$B$}{
           \If{\textit{$(AB)\in\tilde{S}$ }}{$H_{AB}+=C_{BG}V_{AB}X_{AG}$}}
            }
            }
    }
    
\end{algorithm}

Algorithm~\ref{alg:alg2} illustrates the detailed implementation of Eq.~\eqref{eq:loop3-term1} as a typical example. 
The two ``if conditions'' applied before entering loop (1) and loop (3-2) are used to skip the pairs beyond the irreducible sector.
The key point is that, in the ``loop3'' algorithm (Fig.~\ref{fig:loop3} and Eqs.~\eqref{eq:loop3-term1}-\eqref{eq:loop3-term4}), the atomic pairs belonging to the irreducible sector (marked by ``$~ \tilde{~} ~$'') always appear at the same locations in the loop structure for computing the four terms, so that the irreducible-sector filtering can be applied at the same locations of the code structure: One is in the innermost second loop (3-2), 
the other is at the beginning of the outermost loop (1), as can be seen from the four panels in Fig.~\ref{fig:loop3}. 
For example, in term 1 (Fig.~\ref{fig:loop3} (a)), if atom $A$ is not the first atom of any irreducible-sector pair, the inner loops can be all be skipped because they do not contribute to the irreducible sector.


\section{Results and Discussion}
\label{sec:result}

We have implemented the algorithms described above in the ABACUS code \cite{chen_systematically_2010,li_large-scale_2016,lin_ab_2024}, which allows exploitation of
the space-group symmetry in its HDF calculations. In this section, we benchmark our implementation for several crystal systems with different symmetry groups, including high-symmetry systems like crystalline silicon ($O_h$), GaAs($T_d$) and systems with 3- and 6-fold axes like MoS$_2$ and graphene, as listed in Table~\ref{tab:group-cases}.
The ``4-Al'' configurations with group $D_{4h}$, $D_{2h}$, $C_{4v}$, and $C_{2h}$ are obtained by slightly changing the lattice constants and atomic positions from the original 4-atom face-centered-cubic (FCC) cell.
To keep the symmetry of the BvK supercell consistent with the primitive cell, uniform sampling of BZ is used in the above configurations. We also break the BvK-supercell symmetry by sampling $\mathbf{k}$-points differently along different reciprocal lattice vectors to test the acceleration effect at lower symmetries like $D_{2h}$ and $C_{2v}$. A typical 2D system (MoS$_2$) and a relatively large system containing heavy elements (PbTiO$_3$ with $2\times2\times2$ unit cells) are also included in the test cases.
\begin{table}[!ht]
    \centering
    \caption{The symmetry of tested cases: ``group'' refers to the point group (rotation part) of space group symmetry of the BvK supercell, and $N_\text{op}$ is the order (number of symmetry operations) of the space group.}
    \begin{tabular}{|c|c|c|c|c|c|c|c|c|}
    \hline
        \textbf{System} & \textbf{Si (diamond)} &  \textbf{Si (diamond)} & \textbf{GaAs} & \textbf{GaAs}  \\ \hline
        \textbf{$\mathbf{k}$-points} & $n\times n\times n$ & $n\times n\times 1$ & $n\times n\times n$ & $n\times n\times 1$ \\ \hline 
        \textbf{Group} & $O_h$ & $D_{2h}$ & $T_d$ & $C_{2v}$ \\ \hline 
        $N_\text{op}$ & 48 & 8 & 24 & 4 \\ \hline 
        \textbf{System} & \textbf{4[Al]} & \textbf{MoS$_2$} & \textbf{Graphene} & \textbf{8[PbTiO$_3$] } \\ \hline
        \textbf{Group} & $O_h$, $D_{4h}$, $D_{2h}$, $C_{4v}$, $C_{2h}$ & $D_{6h}$ & $D_{6h}$ & $O_h$, $D_{4h}$  \\ \hline
        $N_\text{op}$ & 48, 16, 8, 8, 4 & 24 & 24 & 48, 16  \\ \hline
    \end{tabular}
    \label{tab:group-cases}
\end{table}

All the following tests were performed on a server with 32 Intel(R) Xeon(R) Gold 6130 @ 2.10GHz CPU cores, using a single process and 32 threads.
The computational parameters are set to be the same except for whether to exploit the space-group symmetry ("symmetry-on") or not ("symmetry-off") for each configuration. Specifically, we use SG15 norm-conserving pseudo potentials \cite{hamann_norm-conserving_1979}\cite{SG15}, DZP NAO basis sets \cite{chen_systematically_2010,lin_strategy_2021} 
with cutoff radius of 10 Bohr. 
The standard Heyd-Scuseria-Ernzerhof (HSE) hybrid functional \cite{heyd_hybrid_2003} with mixing parameter $\alpha=0.25$ and screening parameter $\omega=0.11$ Bohr$^{-1}$ are used in all the calculations. 
Note that, in the ``symmetry-off'' case, neither the space-group symmetry nor the time-reversal symmetry is used.

\begin{figure}[htbp]
\begin{minipage}[t]{0.8\linewidth}
    \centering
    \includegraphics[width=\textwidth]{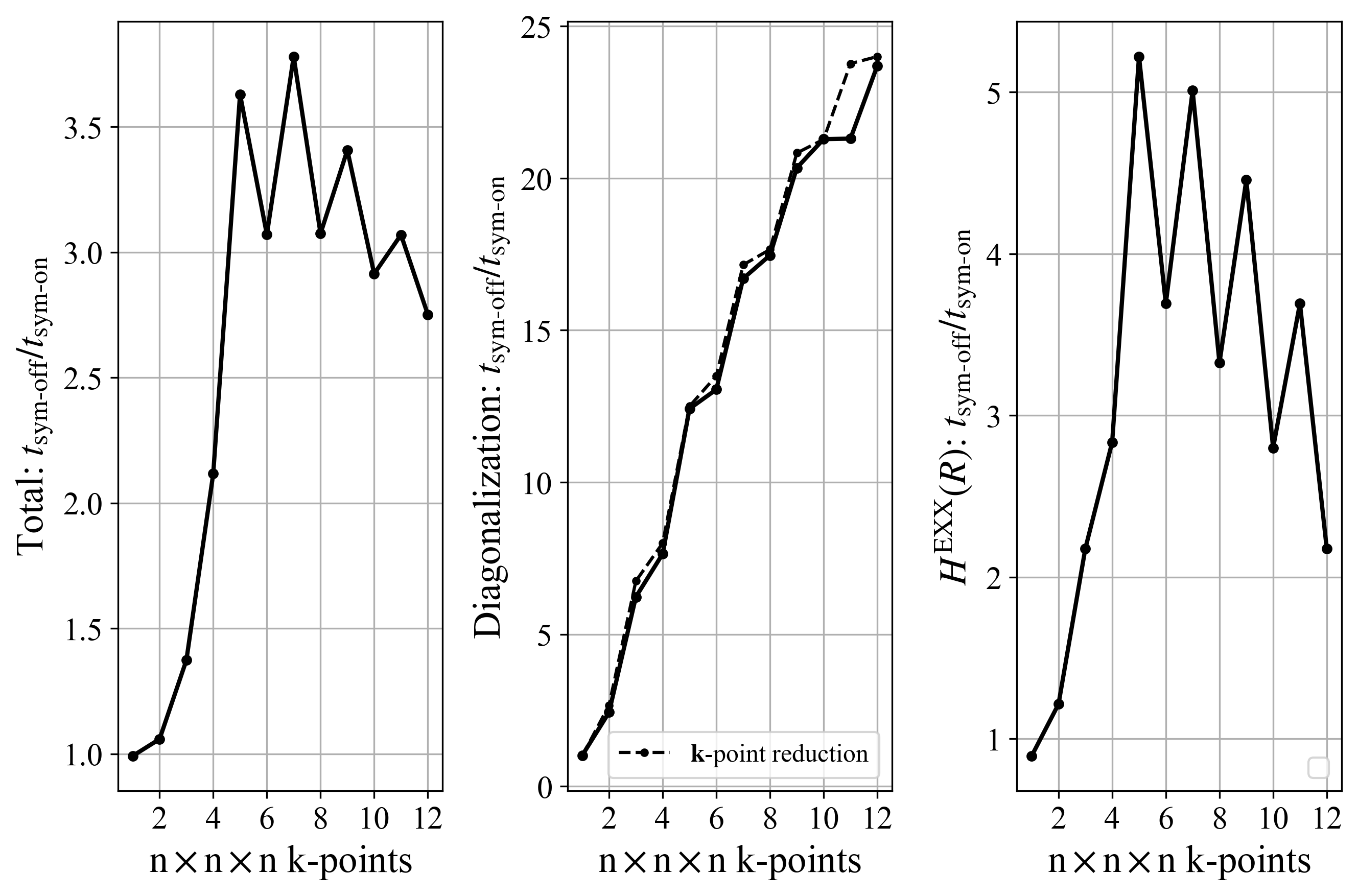}
    \caption{The speed-up ratio by exploiting symmetry ($t_\text{symmetry-off}/t_\text{symmetry-on}$) of total time, the time for diagonalization and the time for constructing EXX Hamiltonian per electronic step with respect to the number of $\mathbf{k}$-points along each direction of reciprocal space in crystalline silicon with 3D uniform $\mathbf{k}$-points.}
    \label{fig:Snnn}
\end{minipage}
\begin{minipage}[t]{0.8\linewidth}
    \centering
    \includegraphics[width=\textwidth]{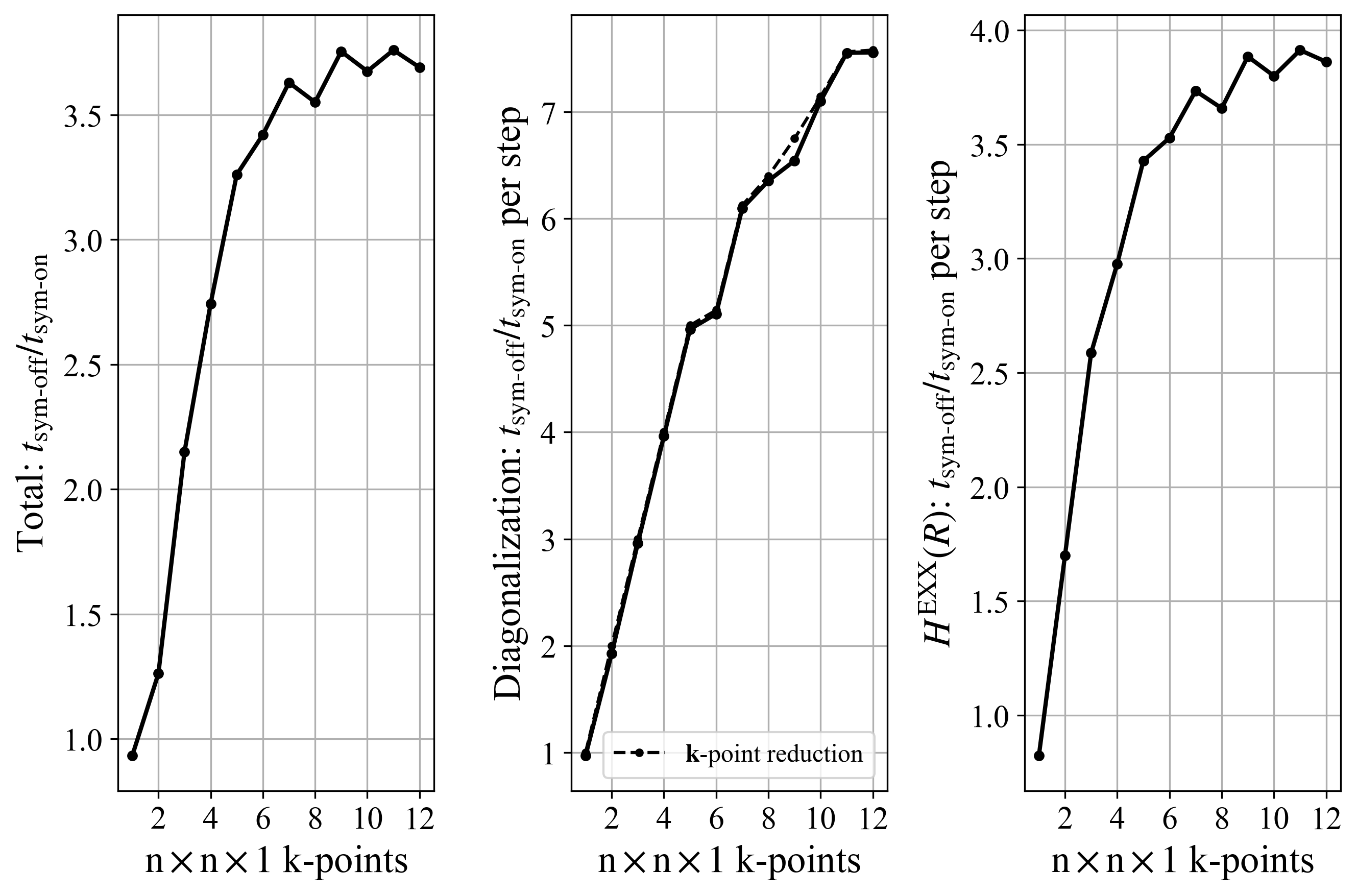}
    \caption{The speed-up ratio by exploiting symmetry ($t_\text{symmetry-off}/t_\text{symmetry-on}$) of total time, the time for diagonalization and the time for constructing EXX Hamiltonian per electronic step with respect to the number of $\mathbf{k}$-points along each direction of the reciprocal space in MoS$_2$ with 2D uniform $\mathbf{k}$-points.}
    \label{fig:MS2}
\end{minipage}
\end{figure}

In Fig.~\ref{fig:Snnn}, we present the results of a simple but representative system, crystalline silicon of diamond structure, while in Fig.~\ref{fig:MS2}, the results for a 2D system, MoS$_2$, are shown. In both figures, three sub-figures (panels) are used to demonstrate the speed-up by exploiting symmetry as a function of the number of $\mathbf{k}$ points along each direction in the full BZ.
As mentioned above, in HDF calculations, there are usually two major computational bottlenecks: One is to diagonalize the Hamiltonian matrix and the other is to construct the EXX part of the Hamiltonian matrix.  
The ratio of computational times with and without applying symmetries are given in the middle and right panels of Fig.~\ref{fig:Snnn}-~\ref{fig:MS2}, respectively, 
while the left panel presents the total time ratio of the calculations. Here the time ratio is counted for
one iteration of the self-consistent calculations.
Results for other systems show similar trend and are presented in Appendix \ref{ap:tests}.

\begin{figure}[h]
    \centering
    \includegraphics[width=0.6\textwidth]{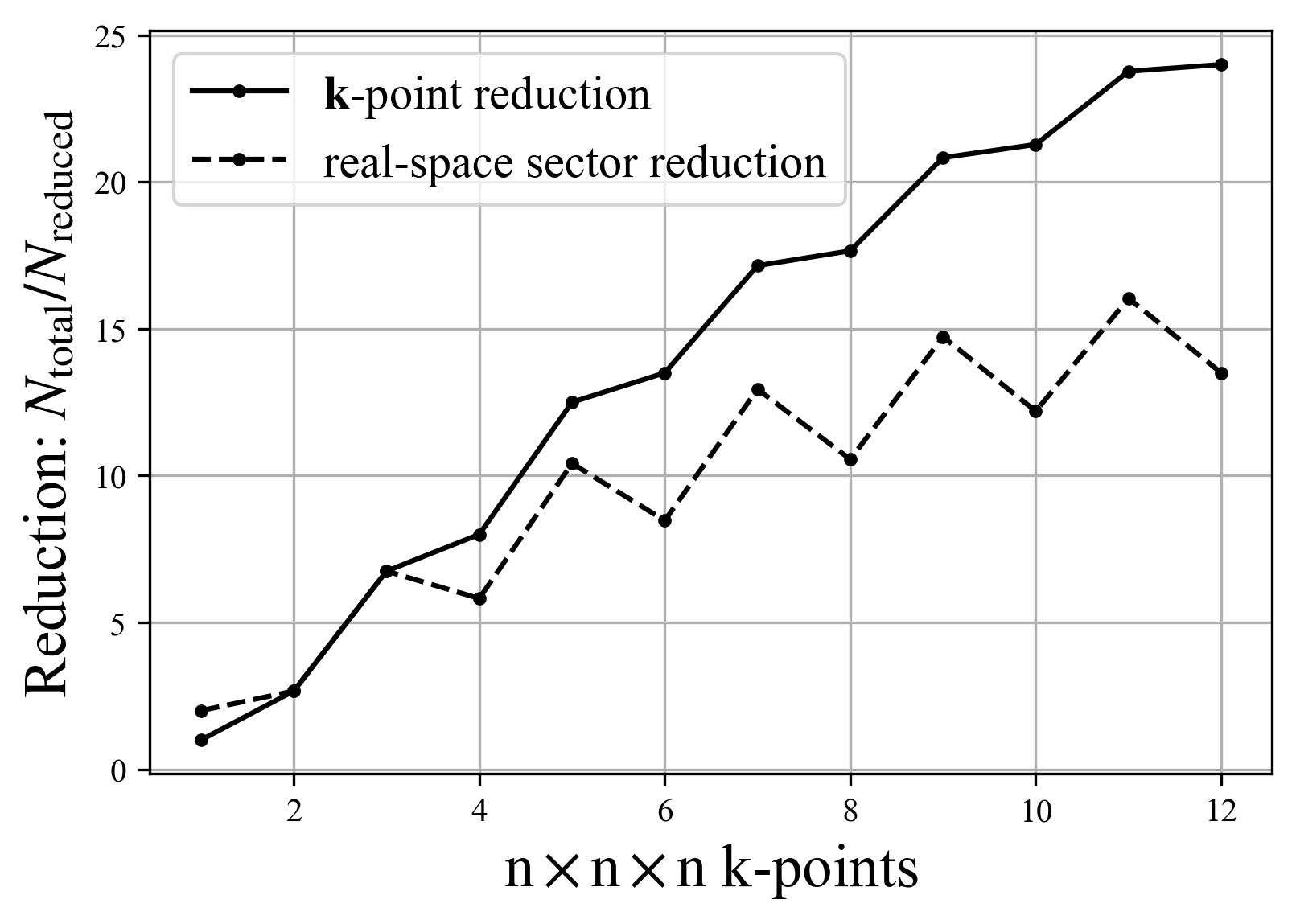}
    \caption{
    The $\mathbf{k}$-point reduction ($N_{\mathbf{k}}/N_{\mathbf{\tilde{k}}}$, the solid line) and 
    The sector reduction ($|\tilde{S}|/|S|$) 
    with respect to the number of $\mathbf{k}$-points along each direction of the reciprocal space in the crystalline silicon of diamond structure.}
    \label{fig:sector-size}
\end{figure}

The speed-up for solving the KS equations (diagonalization) is directly proportional to the ratio between the number of $\mathbf{k}$ points in the full BZ and the number of irreducible $\mathbf{k}$ points. 
This ratio is also plotted in the middle panel of Fig~\ref{fig:Snnn} as a reference.
For Si ($O_h$ symmetry), this ratio can reach more than 20 for dense $\mathbf{k}$ meshes, and correspondingly, the cost for diagonalization can also be accelerated by a factor of more than 20. 
The time reduction for constructing the $H^\text{EXX}(\mathbf{R})$ Hamiltonian is related to the ratio between the full real-space sector and the irreducible one, which is plotted in Fig.~\ref{fig:sector-size}. However, the relation is more involved.
For Si, one can achieve a speed-up by a factor of more than 4 for a $\mathbf{k}$ mesh from
$5\times 5 \times 5$ to $9\times 9 \times 9$, but the acceleration drops down for even dense $\mathbf{k}$ meshes 
(for reasons to be discussed below). 
Note that, for the present system, the construction of the EXX Hamiltonian is more expensive than the
matrix diagonalization, representing the bottleneck here.
Furthermore, there exist other time-consuming steps, mainly the construction of local Hamiltonian terms and the evaluation of charge density on the grid, which are not accelerated by this work. Therefore, the overall speed-up factor for the whole calculation is (slightly) less than the construction of the EXX Hamiltonian (see the left panel of Fig.~\ref{fig:Snnn}).

The speed-up of the 2D system MoS$_2$ ($D_{6h}$ symmetry) in Fig.~\ref{fig:MS2} shows similar behavior as Si when the $\mathbf{k}$ point mesh is coarser than $6\times 6\times 1$, while the reduction ratio of $\mathbf{k}$ points (and the diagonalization time) is much smaller due to the 2D nature. 
However, the speed-up ratio of the total time and the time for constructing the $H^\text{EXX}(\mathbf{R})$ keeps increasing rather than dropping down as in Si (Fig~\ref{fig:Snnn}).  
This results from the difference between 2D and 3D $\mathbf{k}$-point sampling. For comparison, Fig.~\ref{fig:Snn1} in Appendix \ref{ap:tests} displays the speed-up in Si with $n\times n\times 1$ $\mathbf{k}$ point sampling, 
showing no obvious decreasing trend until $12\times 12\times 1$ $\mathbf{k}$ points. 



While the behavior for the time reduction in the Hamiltonian diagonalization upon applying symmetries is trivial, 
it is not the case for the construction of the EXX Hamiltonian. 
In the latter case, the time savings come from the fact that we only need to loop over the atomic pairs in the irreducible sector. However, unlike the ``loop4'' algorithm illustrated in Fig.~\ref{fig:loop4}, the time saving
in the newly developed ``loop3'' algorithm (Fig.~\ref{fig:loop3}) is not directly proportional to the ratio
between the number of atomic pairs in the full sector and that in the irreducible sector. 
Inspection of the ``loop3'' structure in Fig.~\ref{fig:loop3} reveals that the speed-up due to symmetry occurs at 
the level of loop (1) and loop (3-2). The time reduction is related to the ratio of the atoms filtered out at
loop (1) and loop (3-2), and the relative time cost between loop (3-1) (which is not accelerated in this work) and loop (3-2). The actual speed-up effect also depends on the cutoff radius of AO basis functions and the filtering threshold in the low-scaling algorithm of constructing  $H^\text{EXX}(\mathbf{R})$ \cite{lin_efficient_2021,lin_force_2025}, and it is further complicated by the fact that the acceleration ratio is different for the four terms in Fig.~\ref{fig:loop3}. Overall, one can achieve a factor of 4-5 times speedup at best for high-symmetry systems (like Si) and it gets slightly worse for lower-symmetry systems.

Figure~\ref{fig:sector-size} shows how the sector reduction ($|\tilde{S}|/|S|$) increases with the number of unit cells within the BvK supercell (equal to the number of $\mathbf{k}$-points). For comparison, the behavior of the $\mathbf{k}$-point reduction ratios is also plotted.
Both reductions in real and reciprocal spaces are more effective at odd numbers than at even numbers of $\mathbf{k}$-points, which reflects the higher symmetry at odd numbers of $\mathbf{k}$-points and unit cells in BvK supercell.
However, both of them are less than the order (symmetry operations) of the point group of the space group, because different operations may map the atomic pair $(IJ)$ to the same irreducible pair $(\tilde{I}\tilde{J})$ (and similarly for $\mathbf{k}$-points).
Compared to the $\mathbf{k}$-point reduction, the real-space sector reduction grows slower with respect to the density of $\mathbf{k}$-point sampling, but it can be more than 1 even for the gamma-only case, when there are equivalent atom pairs in the unit cell. 
For example, in the unit cell of crystalline silicon with 2 atoms labeled 0 and 1, the atom pair 0-0 and 1-1 are equivalent, while 0-1 is equivalent to 1-0.
This feature can be useful in the calculation of large systems with space group symmetry.

However, because of the sparsity arising from the spatial locality of the orbitals (typically with a $7-10~\mathrm{a.u.}$ cutoff radius), the atomic pairs with a large $\mathbf{R}$ will not be calculated regardless of whether they are included in the irreducible sectors or not. Therefore, the reduction rate of the computation time cannot keep increasing with the density of $\mathbf{k}$-points. 
It is expected in principle that the speed-up ratio will saturate for dense $\mathbf{k}$ grids, but it decreases in practice for 3D systems, as the right panels of Fig.~\ref{fig:Snnn} and Fig.~\ref{fig:MS2}. An in-depth analysis,
as presented in Appendix~\ref{ap:why-down}, reveals that this behavior is caused by two factors. 
The first is the decrease in the time proportion of the accelerated loop (3-2) compared to the
unaccelerated loop (3-1), and the second is the increasing time cost of the filtering process within loop (3-2),
which is not reduced proportionally by the symmetry exploitation.
Our tests indicate that the speed-up in the construction of $H^\text{EXX}(\mathbf{R})$ is most effective for a large unit cell with a modest number of $\mathbf{k}$-points (up to $6\times 6 \times 6$ $\mathbf{k}$ mesh).

\begin{figure}[htbp]
    \centering
    \includegraphics[width=\textwidth]{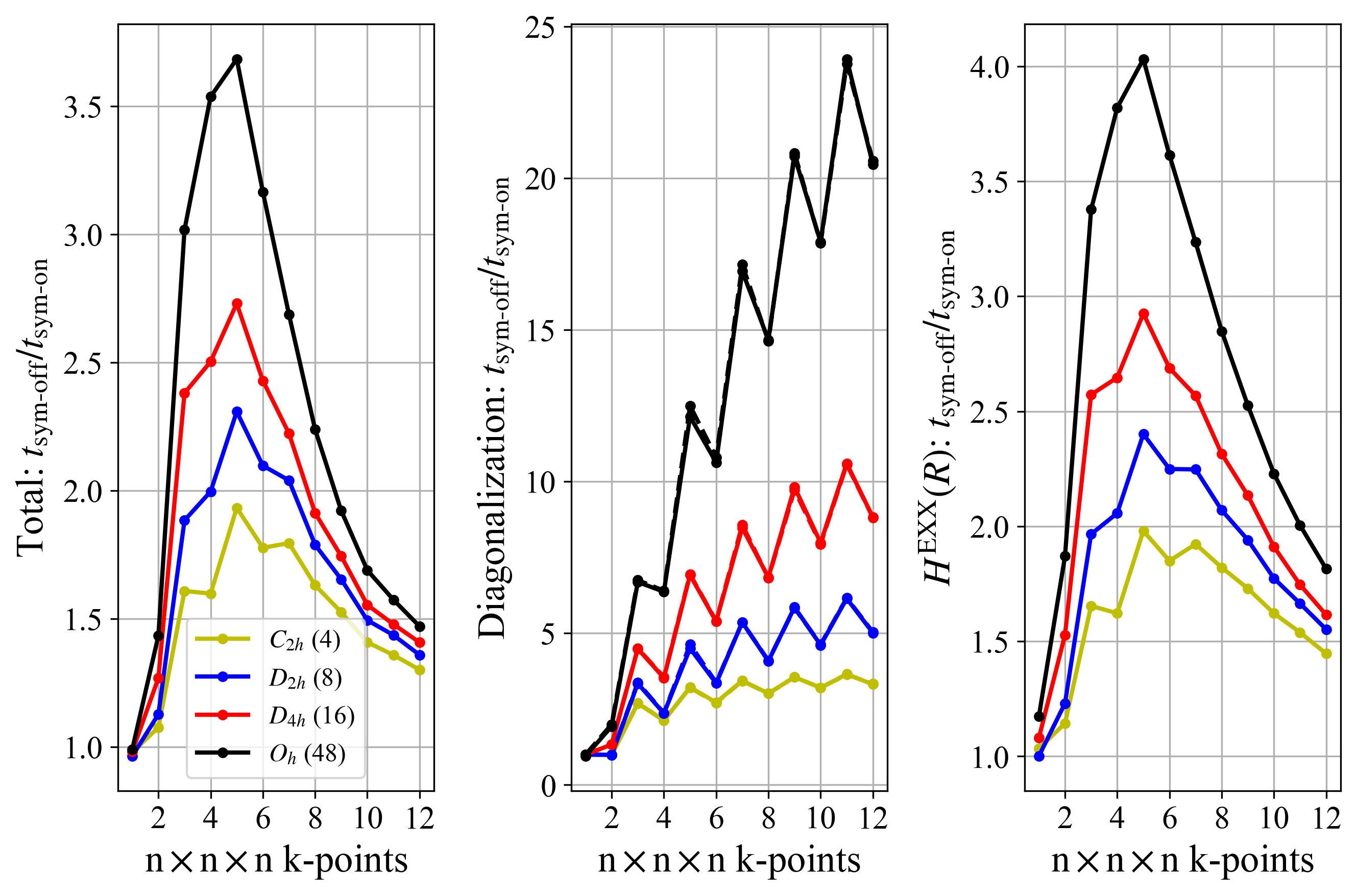}
    \caption{
    The speed-up ratios by exploiting symmetry ($t_\text{symmetry-off}/t_\text{symmetry-on}$) of total time, the time for diagonalization and the time for constructing EXX Hamiltonian with respect to the number of $\mathbf{k}$-points along each direction of the reciprocal space. The test examples here are 4-atom aluminum systems of 4 different symmetries ($O_h$, $D_{4h}$, {$D_{2h}$} $C_{2h}$).}
    \label{fig:Al4}
\end{figure}

Figure~\ref{fig:Al4} further shows the speed-up ratio  ($t_\text{symmetry-off}/t_\text{symmetry-on}$) for the Al systems by applying symmetries. As listed in Table~\ref{tab:group-cases}, all these systems have a unit cell containing 4 Al atoms, but with different space group symmetries. 
As expected, the higher the symmetry, the greater the speed-up ratio.
All the 4 types of structures achieve the best speed-up at $5\times 5\times 5$ $\mathbf{k}$-points. For the construction of $H^\text{EXX}(\mathbf{R})$ Hamiltonian, the structure of the highest $O_h$ symmetry (with 48 operations) displays 4 times speed-up, while the lowest $C_{2h}$ symmetry structure (with 4 operations) can still gain about a factor of 2 speed-up. Consistent with previous examples, for even denser $\mathbf{k}$ grid, the $H^\text{EXX}(\mathbf{R})$ speed-up of all 4 structures goes down, for reasons that have been discussed above. For this set of systems, the speed-up behavior of the total time (the left panel) closely follows that of the construction of $H^\text{EXX}(\mathbf{R})$, which dominates the calculations.

\CY
{
\begin{figure}[htbp]
    \centering
    \includegraphics[width=\textwidth]{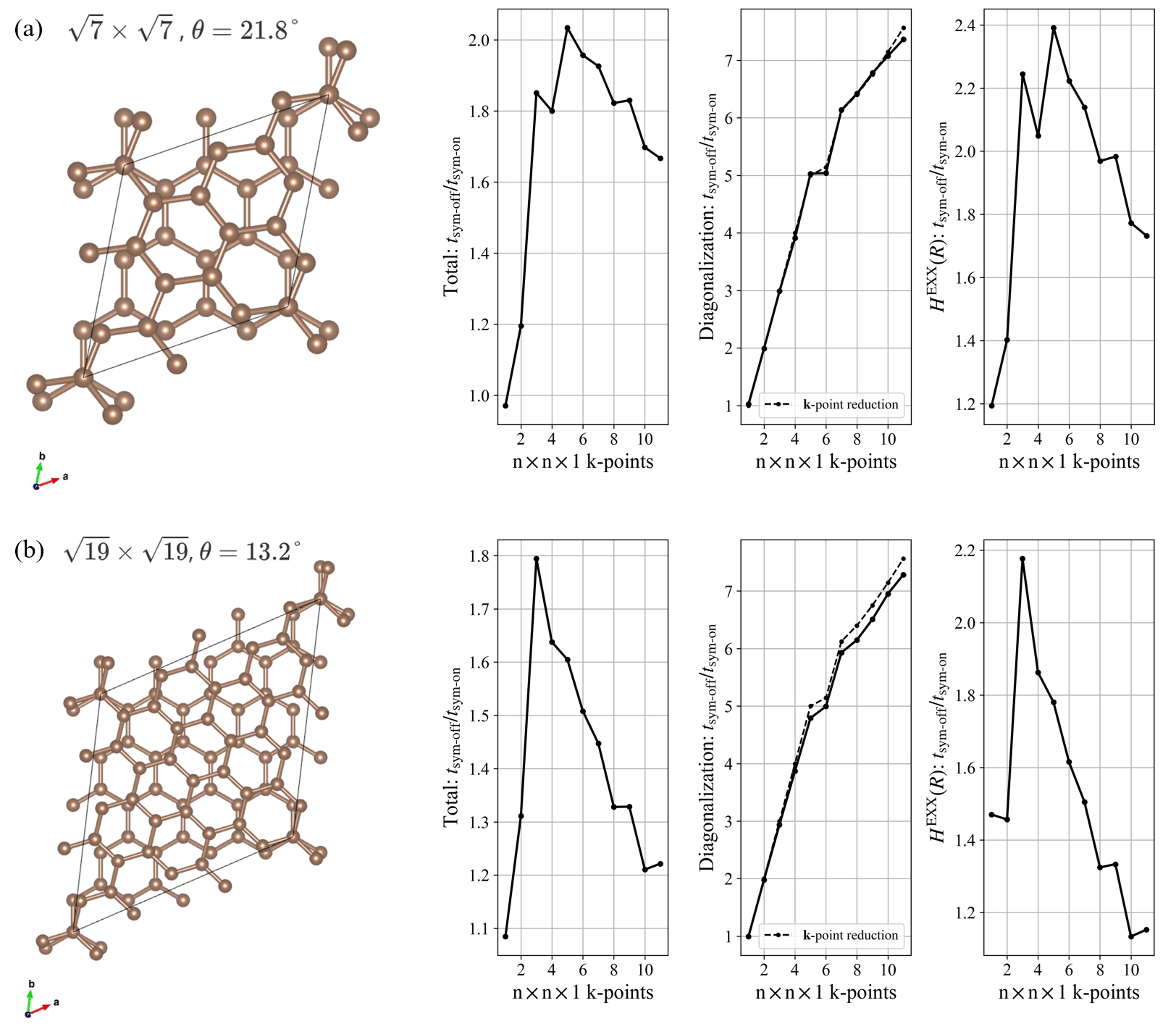}
    \caption{
    The speed-up ratios by exploiting symmetry ($t_\text{symmetry-off}/t_\text{symmetry-on}$) of total time, the time for diagonalization and the time for constructing EXX Hamiltonian with respect to the number of $\mathbf{k}$-points along each direction of 2D reciprocal space in two types of TBGs.}
    \label{fig:TBG}
\end{figure}
As a practical example, we test the performance of our implementation in two types of twisted bilayer graphene (TBG) of $D_3$ symmetry (6 symmetry operations) with 2D $\mathbf{k}$-point sampling. 
For the 28-atom $\sqrt{7}\times\sqrt{7}$ TBG with Moiré superlattice vectors $\mathbf{A}_1=2\mathbf{a}_1+\mathbf{a}_2$, $\mathbf{A}_2=\mathbf{a}_1+2\mathbf{a}_2$ (where $\mathbf{a}_1$ and $\mathbf{a}_2$ are the primitive graphene lattice vectors) and twist angle $\theta=21.8^\circ$  (panel (a) of Fig.~\ref{fig:TBG}), the best overall acceleration of 2 times and 2.4 times acceleration for $H^\text{EXX}(\mathbf{R})$ are achieved at $5\times5\times1$ $\mathbf{k}$ mesh. 
For the 76-atom $\sqrt{19}\times\sqrt{19}$ TBG with $\mathbf{A}_1=3\mathbf{a}_1+2\mathbf{a}_2$, $\mathbf{A}_2=2\mathbf{a}_1+3\mathbf{a}_2$ and $\theta=13.2^\circ$ (panel (b) of Fig.~\ref{fig:TBG}), the best overall acceleration of 1.8 times overall and 2.2 times acceleration for $H^\text{EXX}(\mathbf{R})$ are achieved at $3\times3\times1$ $\mathbf{k}$ mesh. 
In the latter case, the acceleration starts to drop earlier with respect to the density of $\mathbf{k}$ meshes and gets even lower than the gamma-only case at $8 \times 8\times 1$ $\mathbf{k}$ mesh, which can be attributed to the larger cell compared to the radius cutoff of the atomic orbitals and the increasing filter costs.
The diagonalization speeds up 7 times at $10\times 10\times 10$ $\mathbf{k}$ points for both types of TBGs, since the reductions of $\mathbf{k}$ points are the same for both systems.
}

Further test results for other systems listed in Table~\ref{tab:group-cases} are presented in \Cref{ap:tests}. These include GaAs with different $\mathbf{k}$-point sampling, graphene, Si with non-uniform $n\times n \times 1$ $\mathbf{k}$-point sampling, and a larger perovskite system (PbTiO$_3$ supercell with $2\times 2 \times 2$ unit cells). Again, the actual speed-up rate depends on the size and symmetry type of the systems, but the benefits of applying symmetries in HDF calculations can be seen clearly from these various test examples. 


\section{Conclusion}
\label{sec:conclusion}

In this work, we have derived the space-group symmetry transformation relations of the density matrix and the Hamiltonian matrix under the LCAO basis set framework. These relations are then utilized to accelerate HDF calculations. By exploiting symmetry, we can restrict the diagonalization of the HDF Hamiltonian to the IBZ on the one hand and limit the construction of the EXX part of the Hamiltonian to the irreducible sector of the BvK supercell on the other. These two steps represent the primary computational bottlenecks in typical HDF calculations, and our implementation significantly speeds up the overall process. For BZ reduction, rotating the density matrix under symmetry operations is a necessary step for HDF calculations, in contrast to KS-DFT calculations with local or semi-local exchange-correlation (XC) functionals. For real-space sector reduction, we have leveraged the existing code structure in ABACUS to calculate the EXX Hamiltonian using the LRI technique and the so-called ``loop3'' algorithm.

We tested the efficacy of our algorithm and implementation on a series of systems with varying symmetries. Depending on the symmetry order and the size of the BvK supercell, the speed-up from $\mathbf{k}$-point reduction ranges from a few times to over 20 times. The time savings for constructing the real-space EXX Hamiltonian are less significant, but a speed-up factor of 2 to 4 is still observed. Combining these two optimizations, an overall speed-up of HDF calculations by a factor of 2 to 7 can be achieved, depending on the system type and size. Our work not only significantly accelerates HDF calculations within the LCAO framework but also provides an infrastructure that facilitates symmetry analysis of such calculations. The algorithm can also be extended to more advanced methods, such as GW calculations, in a relatively straightforward manner.

\section*{Acknowledgments}
  We acknowledge the funding support by National Natural Science Foundation of China (Grant Nos 12134012, 12374067, 12188101 and 12204332) and Guangdong Basic and Applied Basic Research Foundation (Project Numbers 2021A1515110603).
  This work was also supported by the Strategic Priority Research Program of Chinese Academy of Sciences under Grant No. XDB0500201 and by the National Key Research and Development Program of China (Grant Nos. 2022YFA1403800 and 2023YFA1507004). We thank the electronic structure team (from AI for Science Institute, Beijing) for improving the ABACUS package from various aspects. M. C. gratefully acknowledges funding support from the National Natural Science Foundation of China (Grant Nos 12135002). The numerical calculations in this study were partly carried out on the ORISE Supercomputer.  
  
\appendix
\section{Symmetry operation on different representation}
\label{ap:opformulas}
To make the present paper self-contained, we start with a general discussion of the symmetry operations.
\subsection{Symmetry operation on position vectors}

Perform a symmetry operation, $\hat{P}=\{V|\mathbf{f}\}$, a rotation $V$ followed by a translation $f$ on the real space coordinate: 
\begin{equation}
\hat{P}\mathbf{r}=\{V|\mathbf{f}\}\mathbf{r}=V\mathbf{r}+\mathbf{f}
\end{equation}

The following relation between translation-rotation and rotation-translation operations is derived for future usage:
\begin{equation}
\label{eq:trans-rot}
\begin{aligned}
\{V|\mathbf{f}_1\}\{E|\mathbf{f}_2\}\mathbf{r}
&=\{V|\mathbf{f}_1\}(\mathbf{r+f_2})
=V(\mathbf{r+f_2})+f_1\\
&=V\mathbf{r}+V\mathbf{f_2}+\mathbf{f_1}\\
&=\{V|V\mathbf{f}_2+\mathbf{f}_1\} \mathbf{r}
\end{aligned}
\end{equation}

The inverse of $\hat{P}$ is given by $\{V|\mathbf{f}\}^{-1}=\{V^{-1}|-V^{-1}\mathbf{f}\}$: 
\begin{equation}
  \hat{P}^{-1}\mathbf{r}=V^{-1}(\mathbf{r-f})=V^{-1}\mathbf{r}-V^{-1}\mathbf{f}  
\end{equation}

The rotation operation upon the Cartesian coordinate system can be formulated with Euler angles $\alpha, \beta, \gamma$. When the coordinates are regarded as row-vectors, the expression of $V$ is:
\begin{equation}
    V=V(\alpha,\beta,\gamma)=
    \begin{pmatrix}
        \cos\alpha\cos\beta\cos\gamma-\sin\alpha\sin\gamma & \sin\alpha\cos\beta\cos\gamma+\cos\alpha\sin\gamma & -\sin\beta\cos\gamma\\
        -\cos\alpha\cos\beta\sin\gamma-\sin\alpha\cos\gamma & 
        -\sin\alpha\cos\beta\sin\gamma+\cos\alpha\cos\gamma & 
        \sin\beta\sin\gamma\\
        \cos\alpha\sin\beta & \sin\alpha\sin\beta & \cos\beta
    \end{pmatrix}
\end{equation}
\subsection{Symmetry operation on spherical harmonics}

If $\psi_j^m$ is an eigenvector of the angular momentum operator $\{\hat{J}, \hat{J}_z\}$, then the rotation operator $V(\alpha,\beta,\gamma)$ can be represented by
the Wigner $D$ matrix:\cite{57Rose}
\begin{equation}
    V\psi_j^m=\sum_{m'}\psi_j^{m'}D^j_{m'm}(V)
\end{equation}
\begin{equation}
    D^j_{m'm}(V)=\braket{\psi_j^{m'}|e^{-i\alpha\hat{J}_z}e^{-i\beta\hat{J}_y}e^{-i\gamma\hat{J}_z}|\psi_j^m}
=e^{-im'\alpha}e^{-im\gamma}d^j_{m'm}(\beta)
\end{equation}
\begin{equation}
    \begin{aligned}
    d^j_{m'm}(\beta)&=\sqrt{(j+m)!(j-m)!(j+m')!(j-m')!}\\
    &\times \sum_i(-1)^i[i!(j-m'-i)!(j+m-i)!(i-m+m')!]^{-1}\\
    &\times (\cos\frac{\beta}{2})^{2j+m-m'-2i}(-\sin\frac{\beta}{2})^{m'-m+2i}
    \end{aligned}
\end{equation}
where $\max(0, m-m')\leq i\leq\min(j+m, j-m')$ so that the factorial is well-defined.
Here, $D$ is unitary: 
\begin{equation}
    D^j(V^{-1})=\left[D^{j}(V)\right]^{-1}=\left[D^{j}(V)\right]^{\dagger}
\end{equation}
Without spin-orbital coupling (SOC), the orbital part and the spin part of the rotation can be separated,  
and we can take the spherical harmonics as the eigenfunctions of the orbital part of rotation with integer $j=l$:
\begin{equation}
    VY_l^m=\sum_{m'}Y_l^{m'}D^l_{m'm}(V)
\end{equation}

For operations including inversion: $V=RI$, since $Y_l^m(-\mathbf{\hat{r}})=(-1)^lY_l^m(\mathbf{\hat{r}})$, the representation can be obtained by its
rotation part $R$: $D^l_{m'm}(V)=(-1)^lD^l_{m'm}(R)$.

\subsection{Symmetry operation on real spherical harmonics}
\label{sec:rot-sph}
The real spherical harmonics used in ABACUS\cite{li_large-scale_2016} are linear combinations of spherical harmonics\cite{97Blanco-RealSphereHarm}: 
\begin{equation}
    S_l^m=\sum_{m'}\braket{Y_l^{m'}|S_l^{m'}}Y_l^{m'}=\sum_{m'}c_{m'm}Y_l^{m'}\, .
\end{equation}
Note that the expansion coefficient matrix $\mathbf{c}$ is unitary, and hence $Y_l^m=\sum_{m'}c^*_{mm'}S_l^{m'}$.

Applying a symmetry operation on $S_l^m$, we get a linear combination of $S_l^{m'}$ with $T_{lm'm}$ as coefficient, 
i.e. the $\mathbf{T}$ matrix is the symmetry operation under the representation of real spherical harmonics: 
\begin{equation}
\begin{aligned}
VS_l^m
&=\sum_{m'}c_{m'm}VY_l^{m'}\\
&=\sum_{m'm''}c_{m'm}D^l_{m'',m'}Y_l^{m''}\\
&=\sum_{m'm''m'''}c_{m'm}D^l_{m'',m'}c^*_{m'',m'''}S_l^{m'''}\\
&=\sum_{m'}S_l^{m'}[\mathbf{c}^\dagger \mathbf{D}^l \mathbf{c}]_{m'm}\\\
&\equiv \sum_{m'}S_l^{m'}T^l_{m'm}
\end{aligned}
\end{equation}
\subsection{Symmetry operation on atomic orbitals}
\label{sec:rot-ao}
An atomic orbital can be expressed as the product of radial functions and real spherical harmonics:
\begin{equation}
\phi_{nlm}(\mathbf{r})=f_{nl}(r)S_l^m(\hat{\mathbf{r}})\, .
\end{equation}
When applying a symmetry operation on the atomic orbital as the basis function, the rotation part can be represented by the $\mathbf{T}$ matrix: 
\begin{equation}
    V\phi_{nlm}(\mathbf{r})=f_{nl}(r)VS_l^m(\hat{\mathbf{r}})=f_{nl}(r)\sum_{m'}S_l^{m'}(\hat{\mathbf{r}})T^l_{m'm}
=\sum_{m'}\phi_{nlm'}(\mathbf{r})T^l_{m'm}\, .
\end{equation}

An orbital on atom $\mathcal{U}$ in cell $\mathbf{R}$ with position $\mathbf{s}_\mathcal{U}$ in its cell is denoted as: 
\begin{equation}
    \phi^\mathbf{R}_{\mathcal{U}\mu}(\mathbf{r})\equiv\phi_\mu(\mathbf{r-R}-s_\mathcal{U})
=\{E|\mathbf{R}+\mathbf{s}_\mathcal{U}\}\phi_{nlm}(\mathbf{r})
\end{equation}
where $\mu=\{nlm\}$ is the index of atomic orbital on atom $\mathcal{U}$.
Apply a symmetry operation on it, which rotates the atom position from $\mathbf{R}+\mathbf{s}_\mathcal{U}$ to $\mathbf{\tilde{R}}+\mathbf{s}_\mathcal{\tilde{U}}$
(using Eq.~\ref{eq:trans-rot}):
\begin{equation}
    \begin{aligned}
    \hat{P}\phi^\mathbf{R}_{\mathcal{U}\mu}(\mathbf{r})
    &=\{V|\mathbf{f}\}\{E|\mathbf{R}+\mathbf{s}_\mathcal{U}\}\phi_{nlm}(\mathbf{r})\\
    &=\{E|V(\mathbf{R}+\mathbf{s}_\mathcal{U})+\mathbf{f}\}V\phi_{nlm}(\mathbf{r})\\
    &=\{E|(V\mathbf{R}+\mathbf{O}^{\hat{P}}_\mathcal{U})+(V\mathbf{s}_\mathcal{U}+\mathbf{f}-\mathbf{O}^{\hat{P}}_\mathcal{U})\}
    \sum_{m'}\phi_{nlm'}(\mathbf{r})T^l_{m'm}(V)\\
    &=\sum_{m'}\phi^\mathbf{\tilde{R}}_{\mathcal{\tilde{U}}nlm'}(\mathbf{r})T^l_{m'm}(V)\\
    &=\sum_{\mathcal{U}'\mu'}\phi_{\mathcal{U}'\mu'}^{\tilde{\mathbf{R}}}(\mathbf{r})\delta_{\mathcal{U}'\tilde{\mathcal{U}}}\tilde{T}_{\mathcal{U}'\mu',\mathcal{U}\mu}(V)
    \end{aligned}
\end{equation}
 where $\tilde{T}_{\mu'\mu}=\delta_{nn'}\delta_{ll'}T^l_{m'm}(V)$ is the rotation matrix with the dimension of number of orbitals on atom $\mathcal{U}$, and 
\begin{equation}
\tilde{\mathbf{R}}=V\mathbf{R}+\mathbf{O}^{\hat{P}}_\mathcal{U}
\end{equation}
\begin{equation}
\mathbf{s}_\mathcal{\tilde{U}}=V\mathbf{s}_\mathcal{U}+\mathbf{f}-\mathbf{O}^{\hat{P}}_\mathcal{U}
\end{equation}
are its new lattice cell and new position in the new cell, and $\mathbf{O}^{\hat{P}}_\mathcal{U}$ is the cell to which its $\mathbf{R}=0$ image is transformed (by $\hat{P}$). 

\subsection{Symmetry operation on Bloch orbitals}
\label{sec:rot-bloch}
The Bloch orbitals for periodic systems are defined by:
\begin{equation}
    \phi^\mathbf{k}_{\mathcal{U}\mu}(\mathbf{r})=\frac{1}{\sqrt{N_\mathbf{k}}}\sum_\mathbf{R}e^{i\mathbf{k\cdot R}}\phi^\mathbf{R}_{\mathcal{U}\mu}(\mathbf{r})
\end{equation}
Expressing the phase factor with $\mathbf{\tilde{k}},\mathbf{\tilde{R}}$ after transformation and using the invariance of the inner product (where $\mathbf{K}$ is a reciprocal lattice vector), one has
\begin{equation}
\mathbf{k\cdot R}=(V\mathbf{k})\cdot V\mathbf{R}
=(\mathbf{\tilde{k}+K})(\mathbf{\tilde{R}}-\mathbf{O}_\mathcal{U}^{\hat{P}})
=\mathbf{\tilde{k}}\cdot\mathbf{\tilde{R}}
-\mathbf{\tilde{k}}\cdot \mathbf{O}_\mathcal{U}^{\hat{P}}
\end{equation}
Then, apply the symmetry operation on it:
\begin{equation}
\begin{aligned}
\hat{P}\phi_{\mathcal{U}\mu}^\mathbf{k}(\mathbf{r})
&=\frac{1}{\sqrt{N_\mathbf{k}}}\sum_\mathbf{R}e^{i\mathbf{k\cdot R}}\sum_{\mathcal{U}'\mu'}\phi_{\mathcal{U}'\mu'}^{\tilde{\mathbf{R}}}(\mathbf{r})\delta_{\mathcal{U}'\tilde{\mathcal{U}}}\tilde{T}_{\mathcal{U}'\mu',\mathcal{U}\mu}(V)\\
&=\sum_{\mathcal{U}'\mu'} e^{-i\mathbf{\tilde{k}\cdot O^{\hat{P}}_\mathcal{U}}}
\frac{1}{\sqrt{N_\mathbf{k}}}\sum_\mathbf{R}e^{i\mathbf{\tilde{k}\cdot \tilde{R}}}\phi_{\mathcal{U}'\mu'}^{\tilde{\mathbf{R}}}(\mathbf{r})\delta_{\mathcal{U}'\tilde{\mathcal{U}}}\tilde{T}_{\mathcal{U}'\mu',\mathcal{U}\mu}(V)\\
&=\sum_{\mathcal{U}'\mu'}\phi_{\mathcal{U}'\mu'}^{\tilde{\mathbf{k}}}(\mathbf{r})e^{-i\mathbf{\tilde{k}}\cdot\mathbf{O}^{\hat{P}}_\mathcal{U}}\delta_{\mathcal{U}'\tilde{\mathcal{U}}}\tilde{T}_{\mathcal{U}'\mu',\mathcal{U}\mu}(V)\\
&\equiv\sum_{\mathcal{U}'\mu'}\phi_{\mathcal{U}'\mu'}^{\tilde{\mathbf{k}}}(\mathbf{r})M_{\mathcal{U}'\mu',\mathcal{U}\mu}(\hat{P};\mathbf{k})
\end{aligned}
\end{equation}
where $\mathbf{M}$ is defined as the rotation matrix in the Bloch orbital representation. 

\section{The rotation of local-RI coefficient tensor $\mathbf{C}$}
\label{ap:Cs}
Applying a symmetry operation to the orbital products is equal to applying it respectively to each orbital and multiplying them:
\begin{equation}
    \label{eq:product}
    \hat{P}(\phi_1(\mathbf{r})\phi_2(\mathbf{r}))=\phi_1(\hat{P}^{-1}\mathbf{r})\phi_2(\hat{P}^{-1}\mathbf{r})=\hat{P}\phi_1(\mathbf{r})\hat{P}\phi_2(\mathbf{r})
\end{equation}
Then the symmetry rotation relation between the equivalent overlap matrices of the orbital products $\phi^\mathbf{0}_{\mathcal{U}\mu}\phi^\mathbf{R}_{\mathcal{V}\nu}$ and the ABFs on atom $\mathcal{U}$ can be derived as
\begin{equation}
    \begin{aligned}
    I^{\mathcal{U}\alpha}_{\mathcal{U}\mu,\mathcal{V}\nu}(\mathbf{R})
    &=\braket{\phi^\mathbf{0}_{\mathcal{U}\mu}\phi^\mathbf{R}_{\mathcal{V}\nu}|\mathcal{P}^\mathbf{0}_{\mathcal{U}\alpha}}
    =\braket{\phi^\mathbf{0}_{\mathcal{U}\mu}\mathcal{P}^\mathbf{0}_{\mathcal{U}\alpha}|\phi^\mathbf{R}_{\mathcal{V}\nu}}
    =\braket{\hat{P}\phi^\mathbf{0}_{\mathcal{U}\mu}\hat{P}\mathcal{P}^\mathbf{0}_{\mathcal{U}\alpha}|\hat{P}\phi^\mathbf{R}_{\mathcal{V}\nu}}\\
    &=\sum_{\mu'\nu'\alpha'}\tilde{T}^*_{\mathcal{\tilde{U}}\mu',\mathcal{U}\mu}(V)\tilde{T}^*_{\mathcal{\tilde{U}}\alpha',\mathcal{U}\alpha}(V)\braket{\phi^\mathbf{\tilde{0}}_{\tilde{\mathcal{U}}\mu'}\mathcal{P}^\mathbf{\tilde{0}}_{\tilde{\mathcal{U}}\alpha'}|\phi_{\tilde{\mathcal{V}}\nu'}^\mathbf{\tilde{R}}}\tilde{T}_{\mathcal{\tilde{V}}\nu',\mathcal{V}\nu}(V)\\
    &=\sum_{\mu'\nu'\alpha'}\tilde{T}^*_{\mathcal{\tilde{U}}\mu',\mathcal{U}\mu}(V)\tilde{T}^*_{\mathcal{\tilde{U}}\alpha',\mathcal{U}\alpha}(V)
    I^{\mathcal{\tilde{U}}\alpha'}_{\tilde{\mathcal{U}}\mu,\tilde{\mathcal{V}}\nu'}(\mathbf{\tilde{R}})
    \tilde{T}_{\mathcal{\tilde{V}}\nu',\mathcal{V}\nu}(V)
    \end{aligned}
\end{equation}
Similarly for the overlaps that the ABFs are on atom $\mathcal{V}$:
\begin{equation}
    \begin{aligned}
    I^{\mathcal{V}\alpha}_{\mathcal{U}\mu,\mathcal{V}\nu}(\mathbf{R})
    &=\braket{\phi^\mathbf{0}_{\mathcal{U}\mu}\phi^\mathbf{R}_{\mathcal{V}\nu}|\mathcal{P}^\mathbf{R}_{\mathcal{V}\alpha}}
    =\braket{\phi^\mathbf{0}_{\mathcal{U}\mu}|\phi^\mathbf{R}_{\mathcal{V}\nu}\mathcal{P}^\mathbf{R}_{\mathcal{V}\alpha}}
    =\braket{\hat{P}\phi^\mathbf{0}_{\mathcal{U}\mu}|\hat{P}\phi^\mathbf{R}_{\mathcal{V}\nu}\hat{P}\mathcal{P}^\mathbf{R}_{\mathcal{V}\alpha}}\\
    &=\sum_{\mu'\nu'\alpha'}\tilde{T}^*_{\mathcal{\tilde{U}}\mu',\mathcal{U}\mu}(V)\braket{\phi^\mathbf{\tilde{0}}_{\tilde{\mathcal{U}}\mu'}|\phi_{\tilde{\mathcal{V}}\nu'}^\mathbf{\tilde{R}}\mathcal{P}^\mathbf{\tilde{R}}_{\tilde{\mathcal{V}}\alpha'}}\tilde{T}_{\mathcal{\tilde{V}}\nu',\mathcal{V}\nu}(V)\tilde{T}_{\mathcal{\tilde{V}}\alpha',\mathcal{V}\alpha}(V)\\
    &=\sum_{\mu'\nu'\alpha'}\tilde{T}^*_{\mathcal{\tilde{U}}\mu',\mathcal{U}\mu}(V)
    I^{\mathcal{\tilde{V}}\alpha'}_{\tilde{\mathcal{U}}\mu',\tilde{\mathcal{V}}\nu'}(\mathbf{\tilde{R}})
    \tilde{T}_{\mathcal{\tilde{V}}\nu',\mathcal{\tilde{V}}\nu}(V)\tilde{T}_{\mathcal{\tilde{V}}\alpha',\mathcal{V}\alpha}(V)
    \end{aligned}
\end{equation}
where $\mathbf{\tilde{R}}=V\mathbf{R}+\mathbf{O}_\mathcal{V}^{\hat{P}}-\mathbf{O}_\mathcal{U}^{\hat{P}}$. 
Expressing the above two formulas in matrix form, one has 
\begin{equation}
    \mathbf{I}^{\mathcal{U}}_{\mathcal{UV}}(\mathbf{R})
    =[\tilde{\mathbf{T}}_{\mathcal{\tilde{U}}\mathcal{U}}^\text{ABF}(V)\otimes\tilde{\mathbf{T}}_{\mathcal{\tilde{U}}\mathcal{U}}^\text{AO}(V)]^\dagger
    \times
    \mathbf{I}^{\mathcal{\tilde{U}}}_\mathcal{\tilde{U}\tilde{V}}(\mathbf{\tilde{R}})
    \times
    \tilde{\mathbf{T}}_{\mathcal{\tilde{V}}\mathcal{V}}^\text{AO}(V)
\end{equation}
\begin{equation}
    \begin{aligned}
    \mathbf{I}^{\mathcal{V}}_{\mathcal{UV}}(\mathbf{R})
    &=\tilde{\mathbf{T}}_{\mathcal{\tilde{U}}\mathcal{U}}^{\text{AO}\dagger}(V)
    \times
    \mathbf{I}^{\mathcal{\tilde{V}}}_\mathcal{\tilde{U}\tilde{V}}(\mathbf{\tilde{R}})
    \times
    [\tilde{\mathbf{T}}_{\mathcal{\tilde{V}}\mathcal{V}}^\text{ABF}(V)\otimes\tilde{\mathbf{T}}_{\mathcal{\tilde{V}}\mathcal{V}}^\text{AO}(V)]\\
    \mathbf{I}^{\mathcal{V}}_{\mathcal{VU}}(-\mathbf{R})
    =[\mathbf{I}^{\mathcal{V}}_{\mathcal{UV}}(\mathbf{R})]^\dagger
    &=[\tilde{\mathbf{T}}_{\mathcal{\tilde{V}}\mathcal{V}}^\text{ABF}(V)\otimes\tilde{\mathbf{T}}_{\mathcal{\tilde{V}}\mathcal{V}}^\text{AO}(V)]^\dagger
    \times
    \mathbf{I}^{\mathcal{\tilde{V}}}_\mathcal{\tilde{V}\tilde{U}}(-\mathbf{\tilde{R}})
    \times
    \tilde{\mathbf{T}}_{\mathcal{\tilde{U}}\mathcal{U}}^\text{AO}(V)
    \end{aligned}
\end{equation}

The local RI coefficient tensors $C(\mathbf{R})$ are obtained by solving the following equation:
\begin{equation}
    \sum_{\alpha\in I}C_{Ii,Kk}^{I\alpha}(P_{I\alpha}|P_{F\zeta})+\sum_{\alpha\in K}C_{Kk,Ii}^{K\alpha}(P_{K\alpha}|P_{F\zeta})
=(\phi_{Ii}\phi_{Kk}|P_{F\zeta}),\quad F\in\{I,K\}
\end{equation}
The matrix form for $F=I$ and $F=K$ are respectively: 
\begin{equation}
    \label{eq:Cs-1}
    \mathbf{C}^\mathcal{U}_\mathcal{UV}(\mathbf{R})\mathbf{V}_{\mathcal{UU}}(\mathbf{0})
+ \mathbf{C}^\mathcal{V}_\mathcal{VU}(-\mathbf{R})\mathbf{V}_\mathcal{VU}(-\mathbf{R})=\mathbf{I}^\mathcal{U}_{\mathcal{UV}}(\mathbf{R})
\end{equation}
\begin{equation}
    \mathbf{C}^\mathcal{U}_\mathcal{UV}(\mathbf{R})\mathbf{V}_{\mathcal{UV}}(\mathbf{R})
+ \mathbf{C}^\mathcal{V}_\mathcal{VU}(-\mathbf{R})\mathbf{V}_\mathcal{VV}(\mathbf{0})=\mathbf{I}^\mathcal{V}_{\mathcal{VU}}(-\mathbf{R})\, .
\end{equation}
The above two equations apply to any atom pairs, so they also hold for atom pairs in the irreducible sector. 
Write Eq.~\ref{eq:Cs-1} on the irreducible pair $(\mathcal{\tilde{U}\tilde{V}}\mathbf{\tilde{R}})$ and rotate it out by symmetry operation $\hat{P}=\{V|\mathbf{t}\}$:
\begin{equation}
    \begin{aligned}
    \mathbf{C}^\mathcal{\tilde{U}}_\mathcal{\tilde{U}\tilde{V}}(\mathbf{\tilde{R}})\mathbf{V}_{\mathcal{\tilde{U}\tilde{U}}}(\mathbf{\tilde{0}})
    + \mathbf{C}^\mathcal{\tilde{V}}_\mathcal{\tilde{V}\tilde{U}}(-\mathbf{\tilde{R}})\mathbf{V}_\mathcal{\tilde{V}\tilde{U}}(-\mathbf{\tilde{R}})&=\mathbf{I}^\mathcal{\tilde{U}}_{\mathcal{\tilde{U}\tilde{V}}}(\mathbf{\tilde{R}})\\
    \tilde{\tilde{\mathbf{T}}}^\dagger\mathbf{C}^\mathcal{\tilde{U}}_\mathcal{\tilde{U}\tilde{V}}(\mathbf{\tilde{R}})\tilde{\mathbf{T}}
    \tilde{\mathbf{T}}^\dagger\mathbf{V}_{\mathcal{\tilde{U}\tilde{U}}}(\mathbf{\tilde{0}})\tilde{\mathbf{T}}
    +\tilde{\tilde{\mathbf{T}}}^\dagger \mathbf{C}^\mathcal{\tilde{V}}_\mathcal{\tilde{V}\tilde{U}}(-\mathbf{\tilde{R}})\tilde{\mathbf{T}}
    \tilde{\mathbf{T}}^\dagger\mathbf{V}_\mathcal{\tilde{V}\tilde{U}}(-\mathbf{\tilde{R}})\tilde{\mathbf{T}}
    &=\tilde{\tilde{\mathbf{T}}}^\dagger
    \mathbf{I}^{\mathcal{\tilde{U}}}_\mathcal{\tilde{U}\tilde{V}}(\mathbf{\tilde{R}})
    \tilde{\mathbf{T}}\\
    [\tilde{\tilde{\mathbf{T}}}^\dagger\mathbf{C}^\mathcal{\tilde{U}}_\mathcal{\tilde{U}\tilde{V}}(\mathbf{\tilde{R}})\tilde{\mathbf{T}}]
    \mathbf{V}_\mathcal{UU}(\mathbf{0})
     +[\tilde{\tilde{\mathbf{T}}}^\dagger \mathbf{C}^\mathcal{\tilde{V}}_\mathcal{\tilde{V}\tilde{U}}(-\mathbf{\tilde{R}})\tilde{\mathbf{T}}]
    \mathbf{V}_\mathcal{VU}(-\mathbf{R})
    &=\mathbf{I}^{\mathcal{U}}_{\mathcal{UV}}(\mathbf{R})
\end{aligned}
\end{equation}
That is, $[\tilde{\mathbf{T}}^\dagger_{(\alpha)}\mathbf{C}^\mathcal{\tilde{U}}_\mathcal{\tilde{U}\tilde{V}}(\mathbf{\tilde{R}})\tilde{\mathbf{T}};~\tilde{\mathbf{T}}^\dagger_{(\alpha)} \mathbf{C}^\mathcal{\tilde{V}}_\mathcal{\tilde{V}\tilde{U}}(-\mathbf{\tilde{R}})\tilde{\mathbf{T}}]$
and $[\mathbf{C}^\mathcal{U}_\mathcal{UV}(\mathbf{R}); \mathbf{C}^\mathcal{V}_\mathcal{VU}(-\mathbf{R})]$
satisfy the same equation (Eq.~\eqref{eq:Cs-1}).
Therefore, we can rotate the local RI coefficient tensor $C(\mathbf{R})$ in the same way as 
the (2,1)-orbital two-center integral $I(\mathbf{R})$.

\section{Discussion about irreducible quads}
\label{ap:IRQ}
Following the idea of Charles et al \cite{10Charles-IRQuads}, we consider a symmetry operation $\hat{Q}=\{Q|\mathbf{t}\}$ that does not change the irreducible pair $(IJ)$, namely $(IJ)$-invariant, 
transforming 
$(\tilde{I}\tilde{J},KL)$
to the irreducible quad 
$(\tilde{I}\tilde{J},\tilde{K}\tilde{L})$.
The ERI in the irreducible sector writes: 
\begin{equation}
    \begin{aligned}
    &\quad \braket{\phi_{\tilde{I}i}, \phi_{Kk}|\hat{V}|
    \phi_{\tilde{J}j},\phi_{Ll}}\\
    &=\braket{\phi_{\tilde{I}i}, \phi_{Kk}|\hat{Q}^{-1}\hat{Q}\hat{V}\hat{Q}^{-1}\hat{Q}|
    \phi_{\tilde{J}j},\phi_{Ll}}\\
    &=\braket{\hat{Q}\phi_{\tilde{I}i}, \hat{Q}\phi_{Kk}|\hat{V}|
    \hat{Q}\phi_{\tilde{I}j},\hat{Q}\phi_{Ll}}\\
    &=\sum_{i'j'}\sum_{k'l'}\tilde{T}^*_{\tilde{I}i',\tilde{I}i}(Q)\tilde{T}^*_{\tilde{K}k',Kk}(Q)\braket{\phi_{\tilde{I}i}, \phi_{\tilde{K}k'}|\hat{V}|
    \phi_{\tilde{J}j},\phi_{\tilde{L}l'}}\tilde{T}_{\tilde{J}j',\tilde{J}j}(Q)\tilde{T}_{\tilde{L}l',Ll}(Q).\\
    \end{aligned}
\end{equation}
%
Multiply the ERI with the density matrix on $KL$ and sum for $kl$: 
\begin{equation}
\begin{aligned}
&\quad \sum_{kl}D_{Kk,Ll}\braket{\phi_{\tilde{I}i}, \phi_{Kk}|\hat{V}|
\phi_{\tilde{J}j},\phi_{Ll}}\\
&=\sum_{kl}\left(\sum_{k'l'}\tilde{T}_{\tilde{K}k',Kk}(Q)D_{\tilde{K}k',\tilde{L}l'}\tilde{T}^*_{\tilde{L}l',Ll}(Q)\right)\times \\
&\qquad \qquad \left(\sum_{i'j'}\tilde{T}^*_{\tilde{I}i',\tilde{I}i}(Q)\tilde{T}_{\tilde{J}j',\tilde{J}j}(Q)
\sum_{k''l''}\tilde{T}^*_{\tilde{K}k'',Kk}(Q)
\braket{\phi_{\tilde{I}i'}, \phi_{\tilde{K}k''}|\hat{V}|
\phi_{\tilde{J}j'},\phi_{\tilde{L}l''}}\tilde{T}_{\tilde{L}l'',Ll}(Q)\right) \\
&=\sum_{i'j'}\tilde{T}^*_{\tilde{I}i',\tilde{I}i}(Q)\tilde{T}_{\tilde{J}j',\tilde{J}j}(Q)
\sum_{k'k''}\left(\sum_k\tilde{T}_{\tilde{K}k',Kk}(Q)\tilde{T}^*_{\tilde{K}k'',Kk}(Q)\right)\times \\
&\qquad \qquad \sum_{l'l''}\left(\sum_l\tilde{T}^*_{\tilde{L}l',Ll}(Q)\tilde{T}_{\tilde{L}l'',Ll}(Q)\right)
D_{\tilde{K}k',\tilde{L}l'}
\braket{\phi_{\tilde{I}i'}, \phi_{\tilde{K}k''}|\hat{V}|
\phi_{\tilde{J}j'},\phi_{\tilde{L}l''}}\\
&=\sum_{i'j'}\tilde{T}^*_{\tilde{I}i',\tilde{I}i}(Q)\tilde{T}_{\tilde{J}j',\tilde{J}j}(Q)\sum_{k'k''}\delta_{k'k''}\sum_{l'l''}\delta_{l'l''}
D_{\tilde{K}k',\tilde{L}l'}
\braket{\phi_{\tilde{I}i'}, \phi_{\tilde{K}k''}|\hat{V}|
\phi_{\tilde{J}j'},\phi_{\tilde{L}l''}}\\
&=\sum_{i'j'}\tilde{T}^*_{\tilde{I}i',\tilde{I}i}(Q)\tilde{T}_{\tilde{J}j',\tilde{J}j}(Q)\sum_{kl}
D_{\tilde{K}k,\tilde{L}l}
\braket{\phi_{\tilde{I}i'}, \phi_{\tilde{K}k}|\hat{V}|
\phi_{\tilde{J}j'},\phi_{\tilde{L}l}}\\
\end{aligned}
\end{equation}
where the unitary of $T$ is used: (consistent with Elder's derivation \cite{elder_use_1973})
\begin{equation}
    [\tilde{T}\tilde{T}^\dagger]_{ij}=[\tilde{T}^*\tilde{T}^T]_{ij}=\delta_{ij}
\end{equation}
Then the irreducible part of EXX Hamiltonian can be expressed as:
\begin{equation}
\begin{aligned}
H_{\tilde{I}i,\tilde{I}j}^\text{exx}
&= \sum_{KL}\sum_{kl}
D_{Kk,Ll}
\braket{\phi_{\tilde{I}i}, \phi_{Kk}|\hat{V}|
\phi_{\tilde{J}j},\phi_{Ll}}\\
&=\sum_{\hat{Q}\in S_{{\tilde{I}\tilde{J}}}}
\sum_{\tilde{K}\tilde{L}}
\left(\sum_{i'j'}\tilde{T}^*_{\tilde{I}i',\tilde{I}i}(Q)\tilde{T}_{\tilde{J}j',\tilde{J}j}(Q)\right)
\sum_{kl}
D_{\tilde{K}k,\tilde{L}l}
\braket{\phi_{\tilde{I}i'}, \phi_{\tilde{K}k}|\hat{V}|
\phi_{\tilde{J}j'},\phi_{\tilde{L}l}}\\
\end{aligned}
\end{equation}
where $\hat{Q}\in S_{{\tilde{I}\tilde{J}}}$
are all the $({\tilde{I}\tilde{J}})$-invariant symmetry operations.


Like the symmetrization of the charge density calculated by only irreducible k-point density matrices, we can symmetrize the $\tilde{H}_{\tilde{I}\tilde{J}}=\sum_{\tilde{K}\tilde{L}}\sum_{kl}D_{\tilde{K}k,\tilde{L}l}\braket{\phi_{\tilde{I}i}, \phi_{\tilde{K}k}|\hat{V}|\phi_{\tilde{J}j},\phi_{\tilde{L}l}}$
summed over the irreducible quads $(\tilde{K}\tilde{L})$ by all the $(\tilde{I}\tilde{J})$-invariant symmetry operations to obtain the symmetry-invariant ${H}_{\tilde{I}\tilde{J}}$.
However, in the case of charge density symmetrization, the correct weight of each irreducible $\mathbf{k}$-point should be multiplied onto the corresponding $\mathbf{k}$-space density matrix. Similarly, the correct weight of each irreducible quad $(\tilde{I}\tilde{J},\tilde{K}\tilde{L})$ should be multiplied onto $D_{\tilde{K}\tilde{L}}$ when summing over $(\tilde{K}\tilde{L})$. 
Unfortunately, in the ``loop3'' algorithm described in Sec.~\ref{sec:implement}, there is always one of $\tilde{I}$ and $\tilde{J}$ unknown when summing over $(\tilde{K}\tilde{L})$ in loop (3-1) (see Fig.~\ref{fig:loop3}), so we cannot find out the weight of $(\tilde{I}\tilde{J},\tilde{K}\tilde{L})$ in the full set of $(\tilde{I}\tilde{J})$-invariant quads, $(\tilde{I}\tilde{J}, \{KL\})$. In other words, it is the constraint condition of 
$(\tilde{I}\tilde{J})$-invariance that makes the reduction of $(KL)$ depend on a given irreducible sector pair $(\tilde{I}\tilde{J})$, which cannot be implemented based on the "loop3" algorithm.
If the 4-atom tensors of irreducible quads 
$\sum_{kl}
D_{\tilde{K}k,\tilde{L}l}
\braket{\phi_{\tilde{I}i}, \phi_{\tilde{K}k}|\hat{V}|
\phi_{\tilde{J}j},\phi_{\tilde{L}l}}$ 
are directly calculated and stored,
the above equation can be used to reduce quads to achieve further acceleration.

It should be pointed out, however, in the ``loop3'' algorithm, the way to accelerate loop (3-1) is to reduce the 3-atom tuple (triads). For example, for $H_{AB}$ terms shown in panel (a) of Fig.~\ref{fig:loop3}, we can calculate $X_{AG}$ using irreducible $\{AGF\}$ triads with correct weights, and then symmetrize it. It is a possible feature to be implemented in future works.


\section{Why the $H^\text{EXX}(\mathbf{R})$ speed-up ratio goes down with denser $\mathbf{k}$-points}
\label{ap:why-down}

As shown in Figs.~\ref{fig:Snnn} and \ref{fig:Al4}, the speed-up ratio of $H^\text{EXX}(\mathbf{R})$ goes down with denser $\mathbf{k}$ meshes. This can be attributed to a combined effect of a decrease of the proportion of the accelerated part, loop (3-2) in the whole ``loop3'' algorithm, and a reduction of the speed-up ratio of loop (3-2) itself.

\begin{figure}[htbp]
\begin{minipage}[t]{0.8\linewidth}
    \centering
    \includegraphics[width=\textwidth]{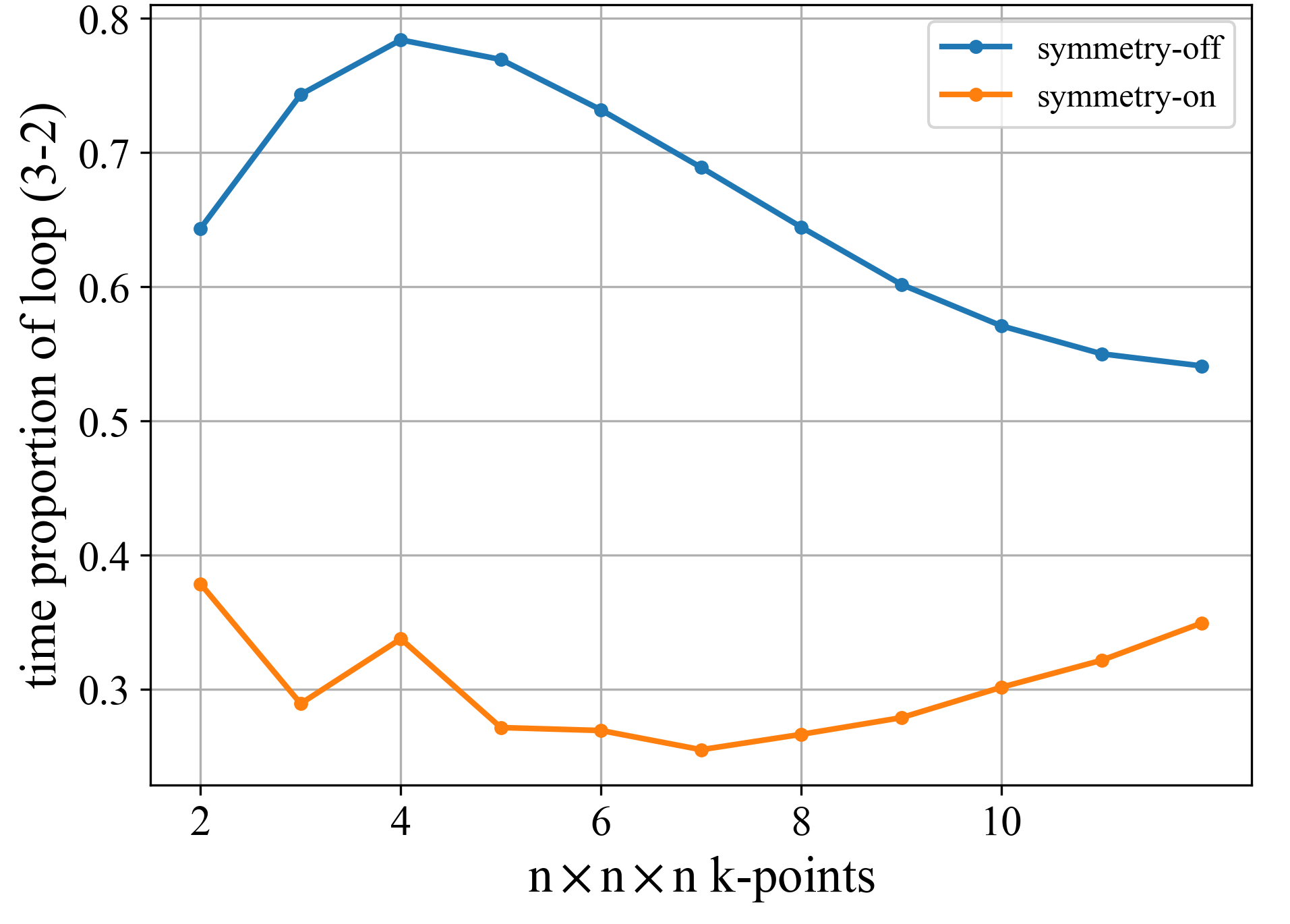}
    \caption{The time proportion of loop (3-2) in the whole $H^\text{EXX}(\mathbf{R})$ calculation with ``loop3'' algorithm in 4-atom FCC aluminum with respect to the density of $\mathbf{k}$ point sampling. }
    \label{fig:proportion-loop32}
\end{minipage}
\end{figure}
For the former, as shown in Fig.~\ref{fig:proportion-loop32}, without applying symmetry, the time proportion of loop (3-2) in the whole $H^\text{EXX}(\mathbf{R})$ calculation decreases from 0.78 to 0.54 for $\mathbf{k}$ meshes denser than $4\times4\times4$. Turning on symmetry reduces the time proportion of loop (3-2) to below 0.4. However, the other part, loop (3-1), is not accelerated by exploiting symmetry in this work, which is the first cause of the decrease of overall $H^\text{EXX}(\mathbf{R})$ speed-up ratio for denser $\mathbf{k}$ points. (The method and difficulty of accelerating loop (3-1) by exploiting symmetry have already been discussed in Appendix ~\ref{ap:IRQ}). 

The increase of time proportion of loop (3-2) with symmetry turned on for dense $\mathbf{k}$ meshes can be attributed to the decreasing speed-up ratio of loop (3-2) itself, which is another cause of the decline of speed-up ratio of $H^\text{EXX}(\mathbf{R})$ calculation. 
\begin{figure}[htbp]
\begin{minipage}[t]{0.8\linewidth}
    \centering
    \includegraphics[width=\textwidth]{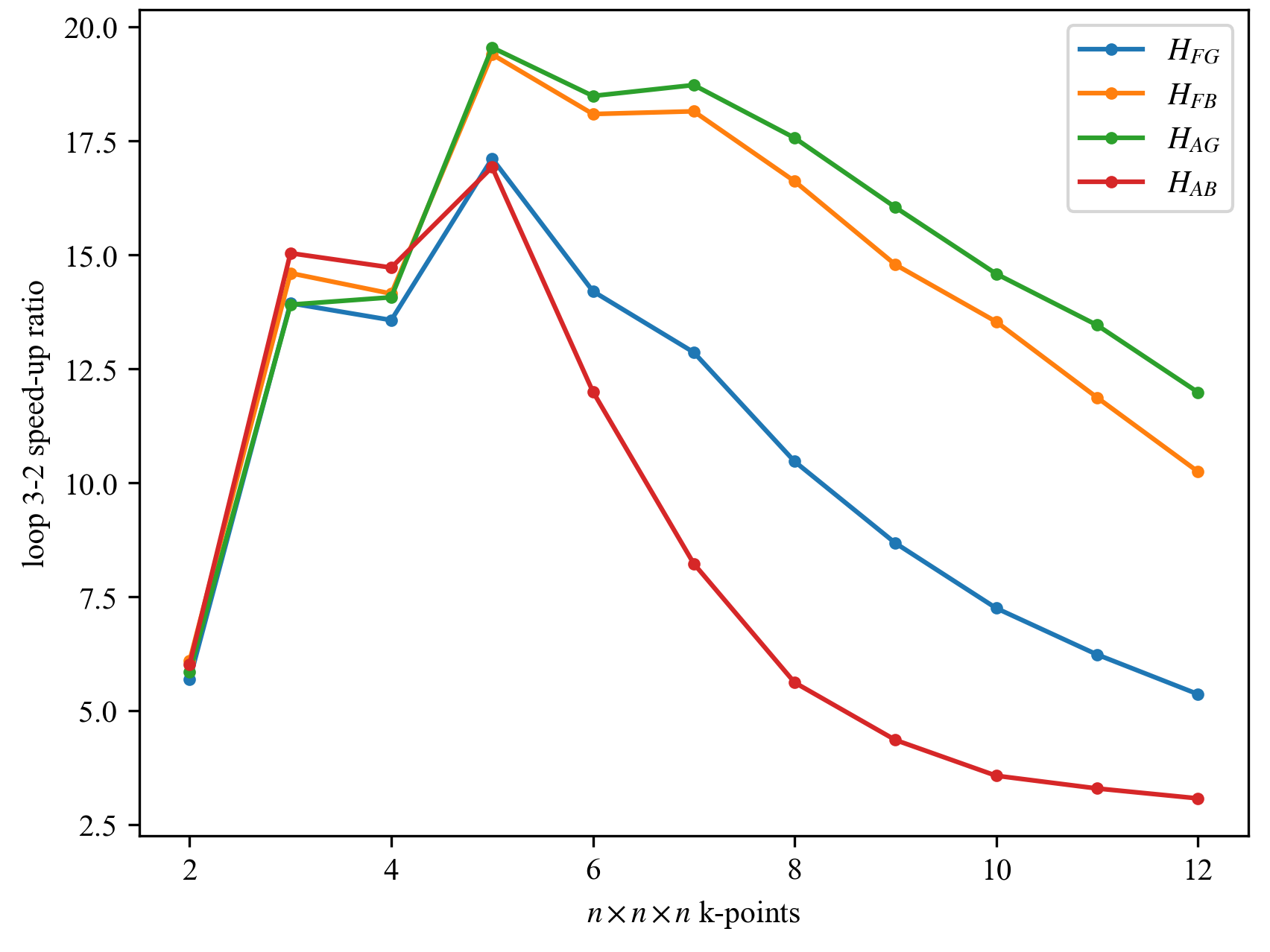}
    \caption{The speed-up ratio of four types of terms under V-perspective in loop (3-2).}
    \label{fig:loop32-4terms}
\end{minipage}
\end{figure}
In order to exclude the impact of possible multi-thread load imbalance on the time cost, we use a single thread to explore the cause of the decreasing speed-up ratio of loop (3-2) itself. As shown in Fig. ~\ref{fig:loop32-4terms}, even with a single thread, the speed-up ratio of all four types of terms decreases with $\mathbf{k}$ point mesh denser than $5\times5\times5$. This means that the dropping of the speed-up ratio is not due to load
imbalance.

\begin{figure}[htbp]
\begin{minipage}[t]{0.8\linewidth}
    \centering
    \includegraphics[width=\textwidth]{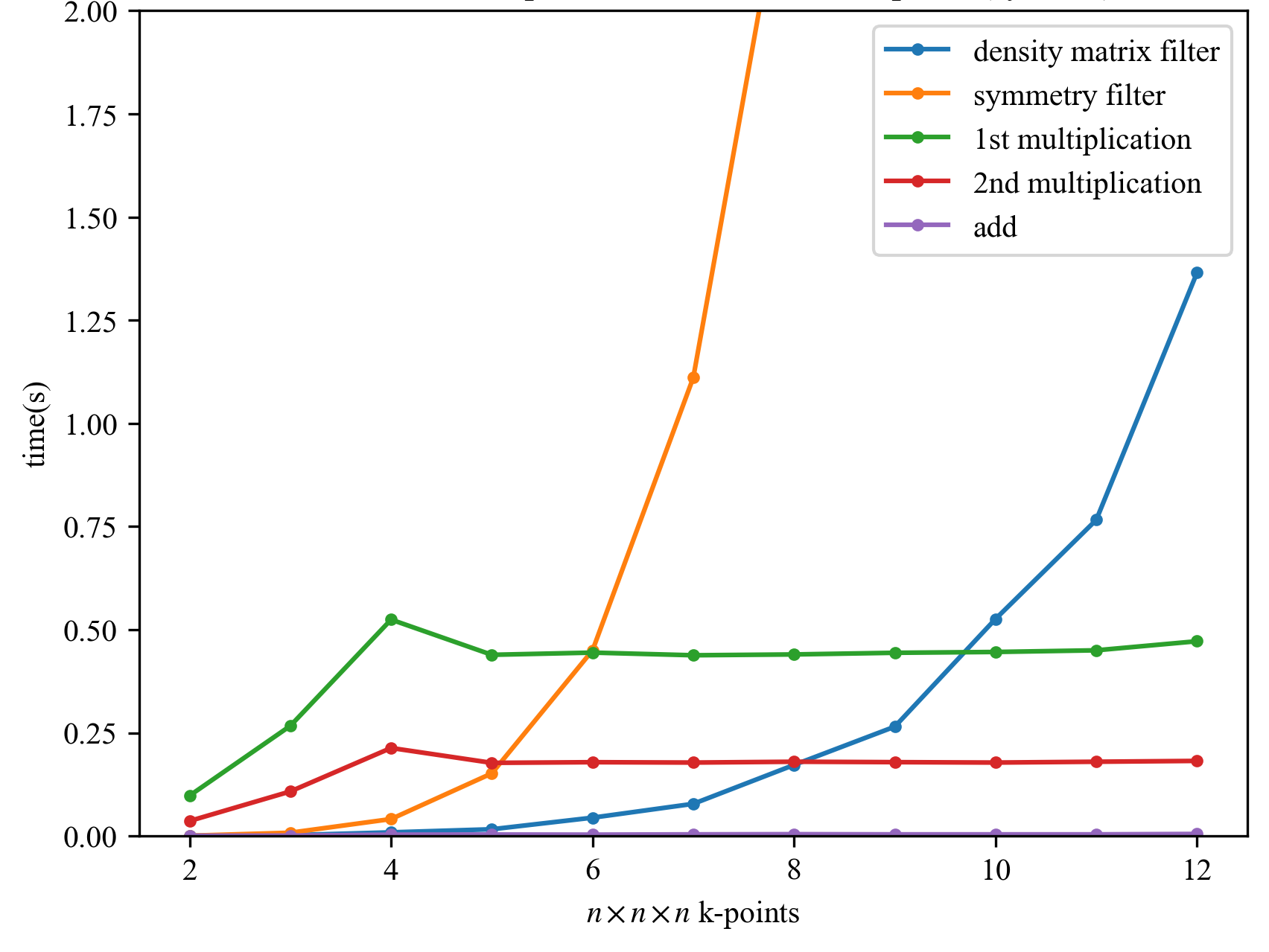}
    \caption{The time cost of each step in loop (3-2) for $H_{AB}$ type of terms with symmetry turned on.}
    \label{fig:a2b2-sym-on}
\end{minipage}
\end{figure}
\begin{figure}[htbp]
\begin{minipage}[t]{0.8\linewidth}
    \centering
    \includegraphics[width=\textwidth]{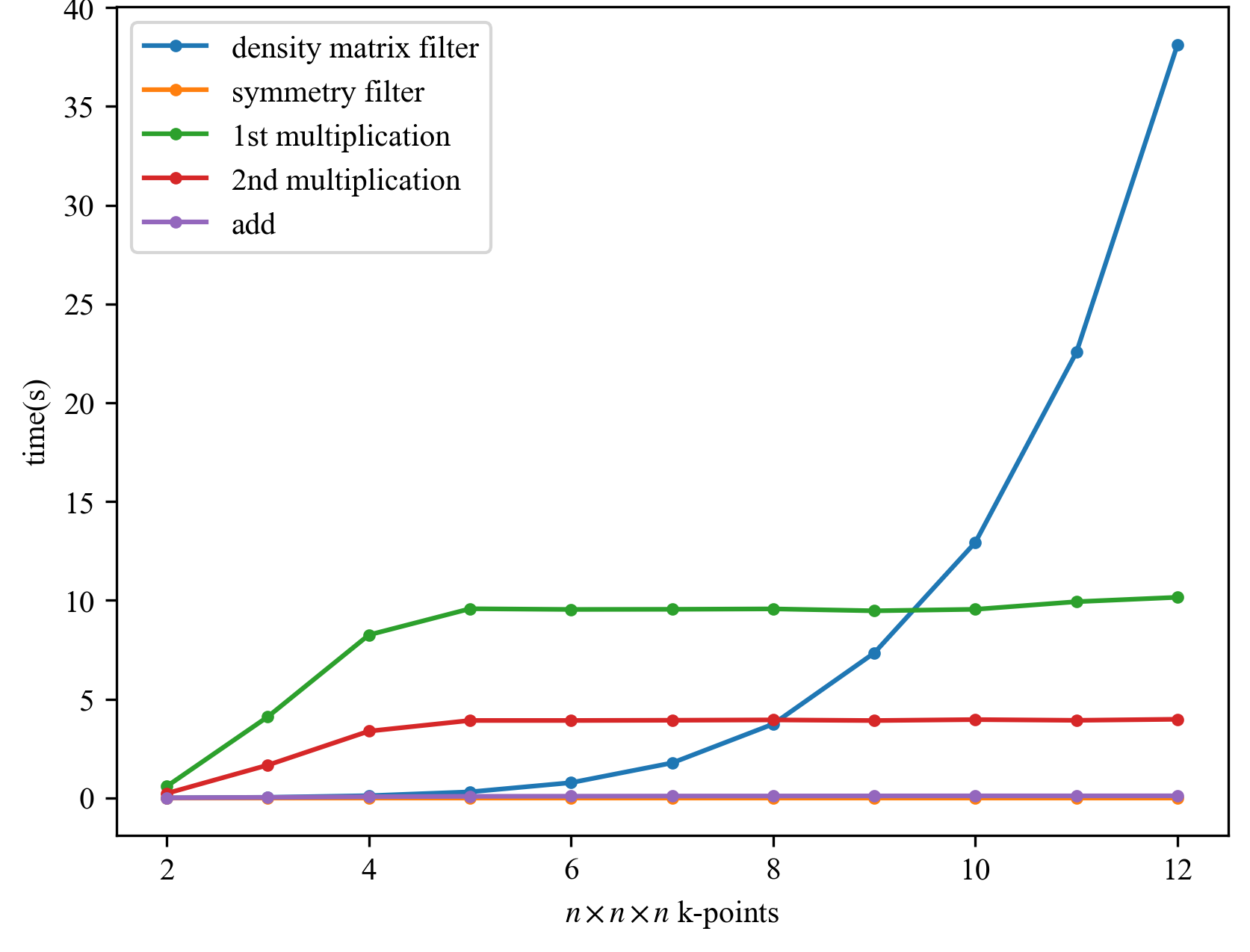}
    \caption{The time cost of each step in loop (3-2) for $H_{AB}$ type of terms with symmetry turned off.}
    \label{fig:a2b2-sym-off}
\end{minipage}
\end{figure}
\begin{figure}[htbp]
\begin{minipage}[t]{0.8\linewidth}
    \centering
    \includegraphics[width=\textwidth]{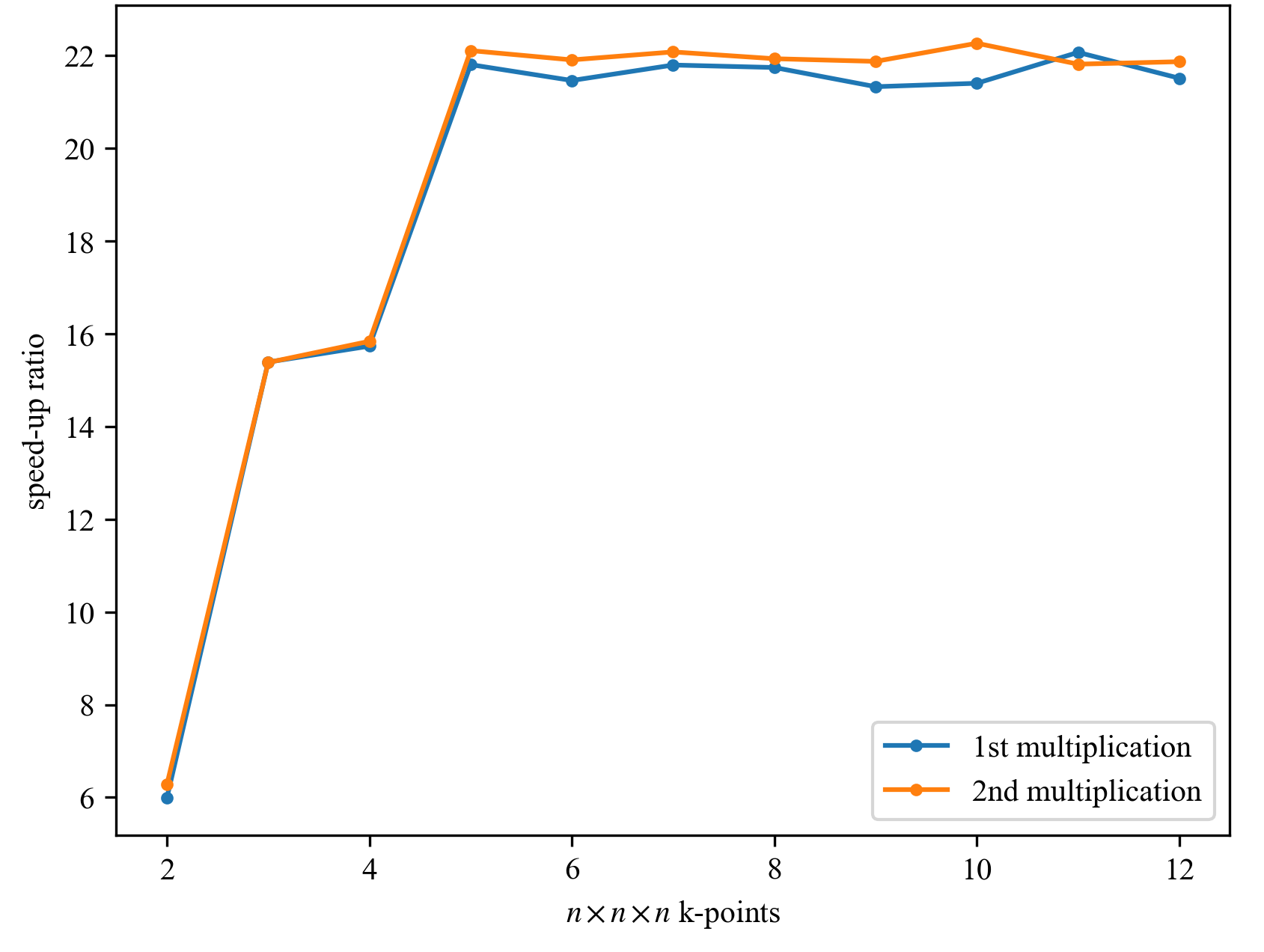}
    \caption{The speed-up factor of the two multiplications in loop (3-2) for $H_{AB}$ type of terms by exploiting symmetry.}
    \label{fig:a2b2-mul}
\end{minipage}
\end{figure}
We take the $H_{AB}$ type of terms [panel (a) of Fig. ~\ref{fig:loop3}] that exhibits the biggest drop as an example to unravel why the speed-up ratio goes down. 
The loop (3-2) procedure for $H_{AB}$ can be divided into 5 steps: symmetry filter, density matrix filter, the first multiplication, the second multiplication, and the final addition to the total $H^\text{EXX}(\mathbf{R})$, where the density matrix filter is to judge whether the density matrix block of a given atom pair ($D_{FB}(\mathbf{R})$) needs to be included in the calculation of $H^\text{EXX}(\mathbf{R})$ at the current screening threshold.
In Fig.~\ref{fig:a2b2-sym-on} and Fig.~\ref{fig:a2b2-sym-off}, the timings of each step are plotted with respect to the density of $\mathbf{k}$-point mesh for the symmetry being switched on and off, respectively.
It can be found that the time for filtering the density matrix increases rapidly and becomes dominating at dense $\mathbf{k}$ points, while the time for the two multiplications increases much slower. 
When the symmetry is turned on, the irreducible sector filter costs more time than the density matrix filter, because the former is called first in our implementation. If the call order of the two filters is switched, the density matrix filter will become the more expensive one.
Such a behavior is consistent with our expectation that the number of density matrix blocks ($D_{FB}(\mathbf{R})$) to be filtered is proportional to the size of the BvK supercell ($O(N^3)$ for 3D $\mathbf{k}$ mesh and $O(N^2)$ for 2D $\mathbf{k}$ mesh), while the multiplications after the filtering almost stop growing when the BvK supercell is large enough so that $D_{FB}(\mathbf{R})$ matrix elements with $|\mathbf{R}|$ beyond a certain range are always filtered out. This is also supported by Fig.~\ref{fig:a2b2-mul} that the speed-up ratios of the two multiplications keep unchanged at $5\times5\times5$ or denser $\mathbf{k}$ points.
Furthermore, it can also explain why the decrease is not so evident in 2D $\mathbf{k}$-point sampling as in the 3D counterpart, since in the former case, the time cost for the filtering process is not yet dominating.

In summary, the reduced speed-up ratio is attributed to the decreasing proportion of the accelerated loop (3-2) and the growing time cost of the filtering process within loop (3-2) itself. This also suggests that there is still room for further improvement in our implementation. 

\section{Test results for other systems}
\label{ap:tests}

\begin{figure}[htbp]
\begin{minipage}[t]{0.8\linewidth}
    \centering
    \includegraphics[width=\textwidth]{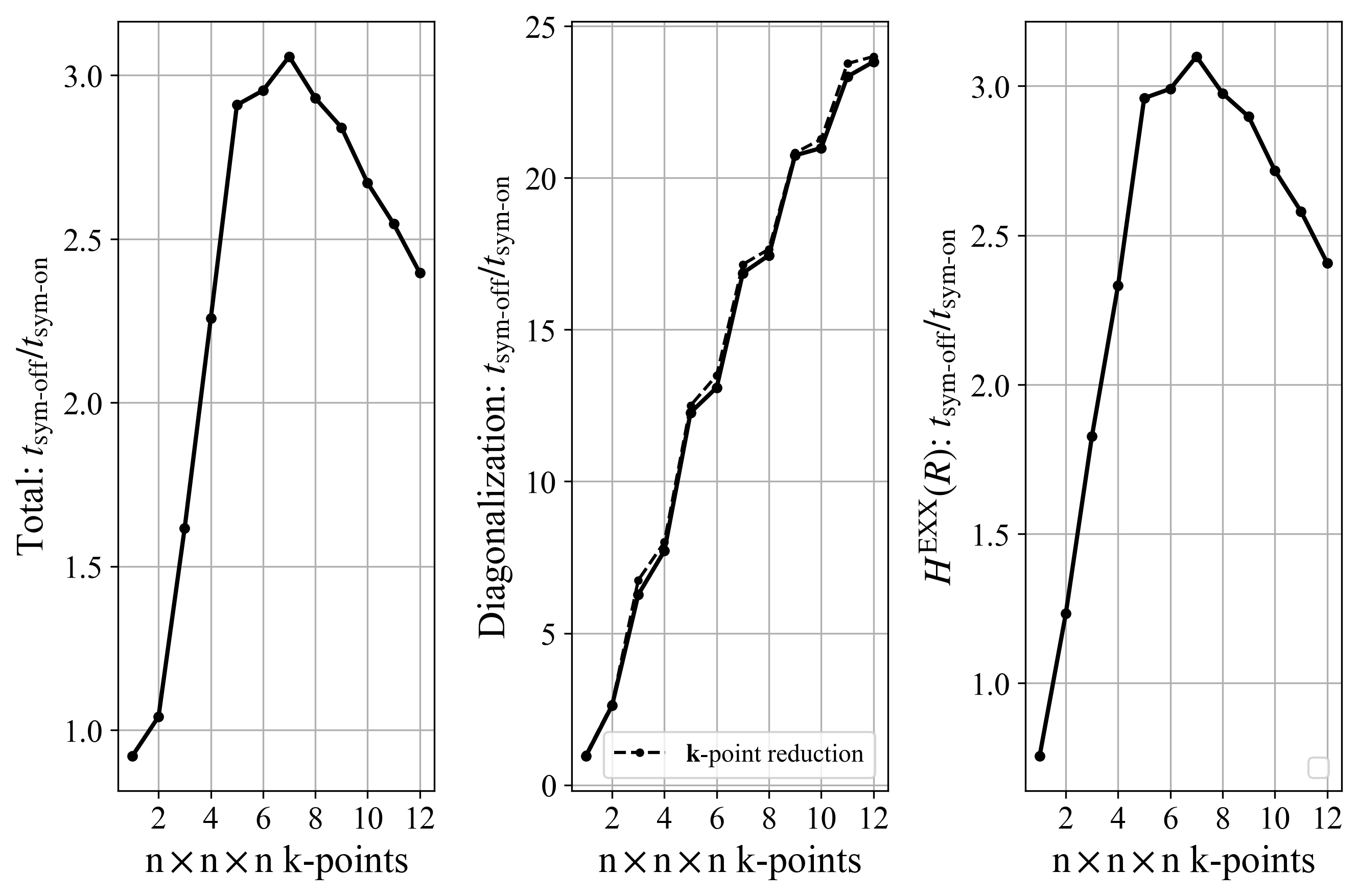}
    \caption{The speed-up ratios ($t_\text{symmetry-off}/t_\text{symmetry-on}$) for GaAs with 3D uniform $\mathbf{k}$-points. Presented are the ratios of total time, the time for diagonalization and the time for constructing EXX Hamiltonian per electronic step with respect to the number of $\mathbf{k}$-points in each direction of reciprocal space. }
    \label{fig:GAnnn}
\end{minipage}
\end{figure}
\begin{figure}[htbp]
\begin{minipage}[t]{0.8\linewidth}
    \centering
    \includegraphics[width=\textwidth]{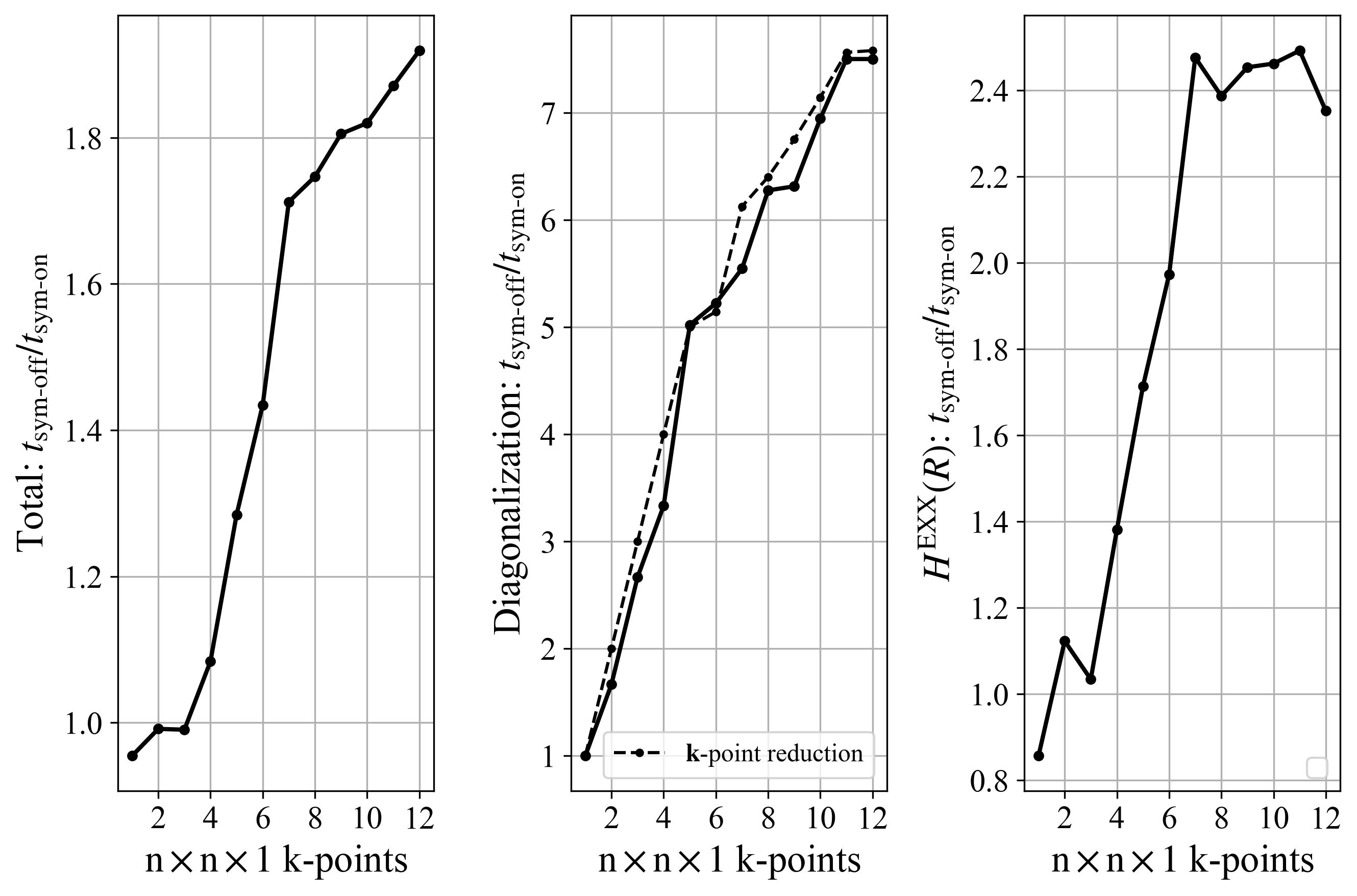}
    \caption{The speed-up ratios ($t_\text{symmetry-off}/t_\text{symmetry-on}$) for graphene with 2D uniform $n\times n\times 1$ $\mathbf{k}$-points by exploiting symmetry ($D_{6h}$). The legends are the same as Fig.~\ref{fig:GAnnn}.}
    \label{fig:GP}
\end{minipage}
\end{figure}
\begin{figure}[htbp]
\begin{minipage}[t]{0.8\linewidth}
    \centering
    \includegraphics[width=\textwidth]{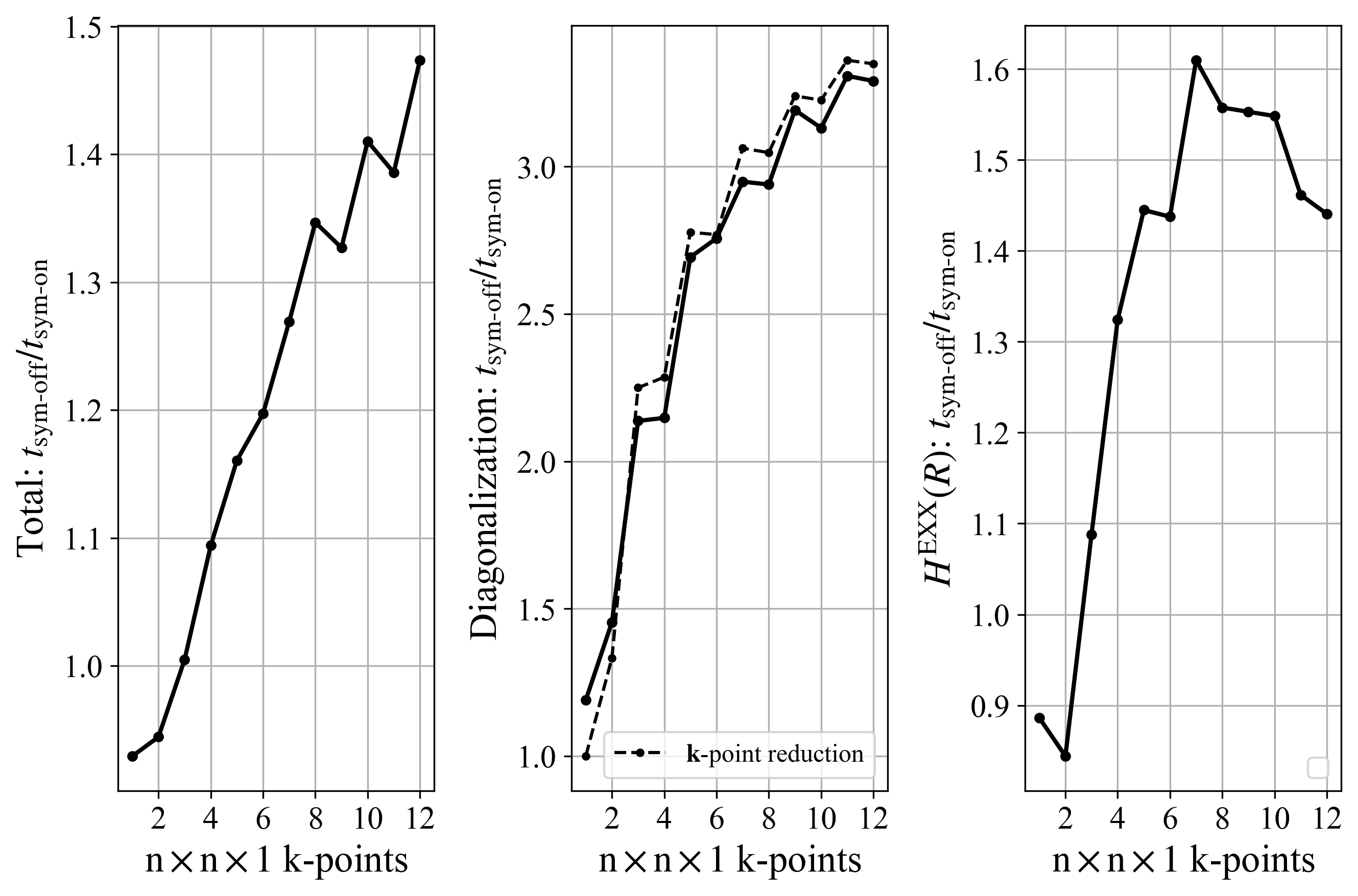}
    \caption{The speed-up ratios by exploiting symmetry ($t_\text{symmetry-off}/t_\text{symmetry-on}$) crystalline silicon with 2D uniform $n\times n\times 1$ $\mathbf{k}$-points. The legends are the same as Fig.~\ref{fig:GAnnn}.  }
    \label{fig:Snn1}
\end{minipage}
\end{figure}
In Figs.~\ref{fig:GAnnn}-\ref{fig:Snn1}, we present the speed-up ratios for GaAs with 3D uniform $\mathbf{k}$-point sampling, graphene with 2D $\mathbf{k}$-point sampling, and Si with $n\times n \times 1$ $\mathbf{k}$-point sampling, respectively. As a 3D system with $O_h$ symmetry, the behavior of the total and $H^\text{EXX}(\mathbf{R})$ speed-up 
ratios for GaAs with respect to the number of $\mathbf{k}$-points is similar to that of Si (Fig.~\ref{fig:Snnn}).
As a 2D system, the monotonic speed-up trend in graphene is similar to that in MoS$_2$. 

We note that, for the cases with $1\times1\times1$ and sometimes $2\times2\times2$ k-points in such small systems, the speed-up factor can be smaller than 1. This can be attributed to the extra steps like the charge density symmetrization and the rotation between blocks of $H^\text{EXX}(\mathbf{R})$, compared to the symmetry-off counterpart. The runtime environment also affects the timings. As the $\mathbf{k}$-point grid gets denser, the above factors become negligible.

For Si with 2D $n\times n\times 1$ $\mathbf{k}$-point sampling (Fig.~\ref{fig:Snn1}), the decreasing trend of the speedup effect for constructing $H^\text{EXX}(\mathbf{R})$ (for dense $\mathbf{k}$ grids) is not as obvious as its counterpart with 3D $\mathbf{k}$-point sampling  (Fig.~\ref{fig:Snnn}). Furthermore, unlike the 3D counterpart, the total speed-up ratio keeps increasing even at a $\mathbf{k}$ grid as dense as $12\times12\times1$.
\begin{table}[!ht]
    \caption{Test results including time for diagonalization ($t_\text{diag}$) and $H^\text{EXX}(\mathbf{R})$ construction ($t_{H^\text{EXX}(\mathbf{R})}$) in PbTiO$_3$ with $2\times2\times2$ unit cells of $D_{4h}$ and $O_h$ symmetry. The sector size is the number of atom pairs of $H^\text{EXX}(\mathbf{R})$ that need to be calculated.
    }
    \centering
    \begin{tabular}{|c|c|c|c|c|c|c|}
    \hline
         \textbf{Configuration} & \textbf{symmetry}  & \textbf{IBZ-kpoints} &  \textbf{$\mathbf{t_{H^\text{EXX}(\mathbf{R})}}$ (s)} &  \textbf{$\mathbf{t_\text{diag}}$(s)}  &  \textbf{Energy (eV)} & \textbf{Sector size} \\ \hline
        $D_{4h}$ & on & 18 & 3123.28 & 171.26 &  -35240.3  & 13625   \\ \hline
        $D_{4h}$ & off & 64 & 8500.52 & 611.91 &  -35240.3  & 102400  \\ \hline
        $O_h$ & on & 10 & 2259.38 & 97.51 & -35241.8  &  4873    \\ \hline
        $O_h$ & off & 64 & 8525.64 & 570.26 &  -35241.8  & 102400   \\ \hline
    \end{tabular}
    \label{tab:Pb}
\end{table}

In a larger case, PbTiO$_3$ of $2\times2\times2$ unit cells (see Table ~\ref{tab:Pb}), 
the sectors are reduced by $7.5$ and $21$ times respectively in the configuration of $D_{4h}$ and $O_h$ group, speeding up the calculation of $H^\text{EXX}(\mathbf{R})$ by $2.7$ and $3.7$ times.




\bibliography{main}

\providecommand{\latin}[1]{#1}
\makeatletter
\providecommand{\doi}
  {\begingroup\let\do\@makeother\dospecials
  \catcode`\{=1 \catcode`\}=2 \doi@aux}
\providecommand{\doi@aux}[1]{\endgroup\texttt{#1}}
\makeatother
\providecommand*\mcitethebibliography{\thebibliography}
\csname @ifundefined\endcsname{endmcitethebibliography}  {\let\endmcitethebibliography\endthebibliography}{}
\begin{mcitethebibliography}{102}
\providecommand*\natexlab[1]{#1}
\providecommand*\mciteSetBstSublistMode[1]{}
\providecommand*\mciteSetBstMaxWidthForm[2]{}
\providecommand*\mciteBstWouldAddEndPuncttrue
  {\def\EndOfBibitem{\unskip.}}
\providecommand*\mciteBstWouldAddEndPunctfalse
  {\let\EndOfBibitem\relax}
\providecommand*\mciteSetBstMidEndSepPunct[3]{}
\providecommand*\mciteSetBstSublistLabelBeginEnd[3]{}
\providecommand*\EndOfBibitem{}
\mciteSetBstSublistMode{f}
\mciteSetBstMaxWidthForm{subitem}{(\alph{mcitesubitemcount})}
\mciteSetBstSublistLabelBeginEnd
  {\mcitemaxwidthsubitemform\space}
  {\relax}
  {\relax}

\bibitem[Hohenberg and Kohn(1964)Hohenberg, and Kohn]{hohenberg_inhomogeneous_1964}
Hohenberg,~P.; Kohn,~W. Inhomogeneous {Electron} {Gas}. \emph{Physical Review} \textbf{1964}, \emph{136}, B864--B871, Publisher: American Physical Society\relax
\mciteBstWouldAddEndPuncttrue
\mciteSetBstMidEndSepPunct{\mcitedefaultmidpunct}
{\mcitedefaultendpunct}{\mcitedefaultseppunct}\relax
\EndOfBibitem
\bibitem[Kohn and Sham(1965)Kohn, and Sham]{kohn_self-consistent_1965}
Kohn,~W.; Sham,~L.~J. Self-{Consistent} {Equations} {Including} {Exchange} and {Correlation} {Effects}. \emph{Physical Review} \textbf{1965}, \emph{140}, A1133--A1138, Publisher: American Physical Society\relax
\mciteBstWouldAddEndPuncttrue
\mciteSetBstMidEndSepPunct{\mcitedefaultmidpunct}
{\mcitedefaultendpunct}{\mcitedefaultseppunct}\relax
\EndOfBibitem
\bibitem[Perdew and Schmidt(2001)Perdew, and Schmidt]{perdew_jacobs_2001}
Perdew,~J.~P.; Schmidt,~K. Jacob’s ladder of density functional approximations for the exchange-correlation energy. \emph{AIP Conference Proceedings} \textbf{2001}, \emph{577}, 1--20\relax
\mciteBstWouldAddEndPuncttrue
\mciteSetBstMidEndSepPunct{\mcitedefaultmidpunct}
{\mcitedefaultendpunct}{\mcitedefaultseppunct}\relax
\EndOfBibitem
\bibitem[Becke(1993)]{becke_densityfunctional_1993}
Becke,~A.~D. Density‐functional thermochemistry. {III}. {The} role of exact exchange. \emph{The Journal of Chemical Physics} \textbf{1993}, \emph{98}, 5648--5652\relax
\mciteBstWouldAddEndPuncttrue
\mciteSetBstMidEndSepPunct{\mcitedefaultmidpunct}
{\mcitedefaultendpunct}{\mcitedefaultseppunct}\relax
\EndOfBibitem
\bibitem[Zhang \latin{et~al.}(2020)Zhang, Cui, Wang, and Jiang]{zhang_hybrid_2020}
Zhang,~M.-Y.; Cui,~Z.-H.; Wang,~Y.-C.; Jiang,~H. Hybrid functionals with system-dependent parameters: {Conceptual} foundations and methodological developments. \emph{WIREs Computational Molecular Science} \textbf{2020}, \emph{10}, e1476, \_eprint: https://onlinelibrary.wiley.com/doi/pdf/10.1002/wcms.1476\relax
\mciteBstWouldAddEndPuncttrue
\mciteSetBstMidEndSepPunct{\mcitedefaultmidpunct}
{\mcitedefaultendpunct}{\mcitedefaultseppunct}\relax
\EndOfBibitem
\bibitem[Seidl \latin{et~al.}(1996)Seidl, Görling, Vogl, Majewski, and Levy]{seidl_generalized_1996}
Seidl,~A.; Görling,~A.; Vogl,~P.; Majewski,~J.~A.; Levy,~M. Generalized {Kohn}-{Sham} schemes and the band-gap problem. \emph{Physical Review B} \textbf{1996}, \emph{53}, 3764--3774, Publisher: American Physical Society\relax
\mciteBstWouldAddEndPuncttrue
\mciteSetBstMidEndSepPunct{\mcitedefaultmidpunct}
{\mcitedefaultendpunct}{\mcitedefaultseppunct}\relax
\EndOfBibitem
\bibitem[Lee \latin{et~al.}(1988)Lee, Yang, and Parr]{lee_development_1988}
Lee,~C.; Yang,~W.; Parr,~R.~G. Development of the {Colle}-{Salvetti} correlation-energy formula into a functional of the electron density. \emph{Physical Review B} \textbf{1988}, \emph{37}, 785--789, Publisher: American Physical Society\relax
\mciteBstWouldAddEndPuncttrue
\mciteSetBstMidEndSepPunct{\mcitedefaultmidpunct}
{\mcitedefaultendpunct}{\mcitedefaultseppunct}\relax
\EndOfBibitem
\bibitem[Perdew \latin{et~al.}(1996)Perdew, Burke, and Ernzerhof]{perdew_generalized_1996}
Perdew,~J.~P.; Burke,~K.; Ernzerhof,~M. Generalized {Gradient} {Approximation} {Made} {Simple}. \emph{Physical Review Letters} \textbf{1996}, \emph{77}, 3865--3868, Publisher: American Physical Society\relax
\mciteBstWouldAddEndPuncttrue
\mciteSetBstMidEndSepPunct{\mcitedefaultmidpunct}
{\mcitedefaultendpunct}{\mcitedefaultseppunct}\relax
\EndOfBibitem
\bibitem[Sun \latin{et~al.}(2015)Sun, Ruzsinszky, and Perdew]{sun_strongly_2015}
Sun,~J.; Ruzsinszky,~A.; Perdew,~J. Strongly {Constrained} and {Appropriately} {Normed} {Semilocal} {Density} {Functional}. \emph{Physical Review Letters} \textbf{2015}, \emph{115}, 036402, Publisher: American Physical Society\relax
\mciteBstWouldAddEndPuncttrue
\mciteSetBstMidEndSepPunct{\mcitedefaultmidpunct}
{\mcitedefaultendpunct}{\mcitedefaultseppunct}\relax
\EndOfBibitem
\bibitem[Perdew and Zunger(1981)Perdew, and Zunger]{perdew_self-interaction_1981}
Perdew,~J.~P.; Zunger,~A. Self-interaction correction to density-functional approximations for many-electron systems. \emph{Physical Review B} \textbf{1981}, \emph{23}, 5048--5079, Publisher: American Physical Society\relax
\mciteBstWouldAddEndPuncttrue
\mciteSetBstMidEndSepPunct{\mcitedefaultmidpunct}
{\mcitedefaultendpunct}{\mcitedefaultseppunct}\relax
\EndOfBibitem
\bibitem[Mori-Sánchez \latin{et~al.}(2008)Mori-Sánchez, Cohen, and Yang]{mori-sanchez_localization_2008}
Mori-Sánchez,~P.; Cohen,~A.~J.; Yang,~W. Localization and {Delocalization} {Errors} in {Density} {Functional} {Theory} and {Implications} for {Band}-{Gap} {Prediction}. \emph{Physical Review Letters} \textbf{2008}, \emph{100}, 146401, Publisher: American Physical Society\relax
\mciteBstWouldAddEndPuncttrue
\mciteSetBstMidEndSepPunct{\mcitedefaultmidpunct}
{\mcitedefaultendpunct}{\mcitedefaultseppunct}\relax
\EndOfBibitem
\bibitem[Almlöf \latin{et~al.}(1982)Almlöf, Faegri~Jr., and Korsell]{almlof_principles_1982}
Almlöf,~J.; Faegri~Jr.,~K.; Korsell,~K. Principles for a direct {SCF} approach to {LICAO}–{MOab}-initio calculations. \emph{Journal of Computational Chemistry} \textbf{1982}, \emph{3}, 385--399, \_eprint: https://onlinelibrary.wiley.com/doi/pdf/10.1002/jcc.540030314\relax
\mciteBstWouldAddEndPuncttrue
\mciteSetBstMidEndSepPunct{\mcitedefaultmidpunct}
{\mcitedefaultendpunct}{\mcitedefaultseppunct}\relax
\EndOfBibitem
\bibitem[Häser and Ahlrichs(1989)Häser, and Ahlrichs]{haser_improvements_1989}
Häser,~M.; Ahlrichs,~R. Improvements on the direct {SCF} method. \emph{Journal of Computational Chemistry} \textbf{1989}, \emph{10}, 104--111, \_eprint: https://onlinelibrary.wiley.com/doi/pdf/10.1002/jcc.540100111\relax
\mciteBstWouldAddEndPuncttrue
\mciteSetBstMidEndSepPunct{\mcitedefaultmidpunct}
{\mcitedefaultendpunct}{\mcitedefaultseppunct}\relax
\EndOfBibitem
\bibitem[Burant \latin{et~al.}(1996)Burant, Scuseria, and Frisch]{burant_linear_1996}
Burant,~J.~C.; Scuseria,~G.~E.; Frisch,~M.~J. A linear scaling method for {Hartree}–{Fock} exchange calculations of large molecules. \emph{The Journal of Chemical Physics} \textbf{1996}, \emph{105}, 8969--8972\relax
\mciteBstWouldAddEndPuncttrue
\mciteSetBstMidEndSepPunct{\mcitedefaultmidpunct}
{\mcitedefaultendpunct}{\mcitedefaultseppunct}\relax
\EndOfBibitem
\bibitem[Schwegler and Challacombe(1996)Schwegler, and Challacombe]{schwegler_linear_1996}
Schwegler,~E.; Challacombe,~M. Linear scaling computation of the {Hartree}–{Fock} exchange matrix. \emph{The Journal of Chemical Physics} \textbf{1996}, \emph{105}, 2726--2734\relax
\mciteBstWouldAddEndPuncttrue
\mciteSetBstMidEndSepPunct{\mcitedefaultmidpunct}
{\mcitedefaultendpunct}{\mcitedefaultseppunct}\relax
\EndOfBibitem
\bibitem[Schwegler \latin{et~al.}(1997)Schwegler, Challacombe, and Head-Gordon]{schwegler_linear_1997}
Schwegler,~E.; Challacombe,~M.; Head-Gordon,~M. Linear scaling computation of the {Fock} matrix. {II}. {Rigorous} bounds on exchange integrals and incremental {Fock} build. \emph{The Journal of Chemical Physics} \textbf{1997}, \emph{106}, 9708--9717\relax
\mciteBstWouldAddEndPuncttrue
\mciteSetBstMidEndSepPunct{\mcitedefaultmidpunct}
{\mcitedefaultendpunct}{\mcitedefaultseppunct}\relax
\EndOfBibitem
\bibitem[Ochsenfeld \latin{et~al.}(1998)Ochsenfeld, White, and Head-Gordon]{ochsenfeld_linear_1998}
Ochsenfeld,~C.; White,~C.~A.; Head-Gordon,~M. Linear and sublinear scaling formation of {Hartree}–{Fock}-type exchange matrices. \emph{The Journal of Chemical Physics} \textbf{1998}, \emph{109}, 1663--1669\relax
\mciteBstWouldAddEndPuncttrue
\mciteSetBstMidEndSepPunct{\mcitedefaultmidpunct}
{\mcitedefaultendpunct}{\mcitedefaultseppunct}\relax
\EndOfBibitem
\bibitem[Rudberg \latin{et~al.}(2011)Rudberg, Rubensson, and Sałek]{rudberg_kohnsham_2011}
Rudberg,~E.; Rubensson,~E.~H.; Sałek,~P. Kohn-Sham Density Functional Theory Electronic Structure Calculations with Linearly Scaling Computational Time and Memory Usage. \emph{Journal of Chemical Theory and Computation} \textbf{2011}, \emph{7}, 340--350, PMID: 26596156\relax
\mciteBstWouldAddEndPuncttrue
\mciteSetBstMidEndSepPunct{\mcitedefaultmidpunct}
{\mcitedefaultendpunct}{\mcitedefaultseppunct}\relax
\EndOfBibitem
\bibitem[Pisani and Dovesi(1980)Pisani, and Dovesi]{pisani_exact-exchange_1980}
Pisani,~C.; Dovesi,~R. Exact-exchange {Hartree}–{Fock} calculations for periodic systems. {I}. {Illustration} of the method. \emph{International Journal of Quantum Chemistry} \textbf{1980}, \emph{17}, 501--516, \_eprint: https://onlinelibrary.wiley.com/doi/pdf/10.1002/qua.560170311\relax
\mciteBstWouldAddEndPuncttrue
\mciteSetBstMidEndSepPunct{\mcitedefaultmidpunct}
{\mcitedefaultendpunct}{\mcitedefaultseppunct}\relax
\EndOfBibitem
\bibitem[Pisani \latin{et~al.}(1988)Pisani, Dovesi, and Roetti]{pisani_hartree-fock_1988}
Pisani,~C.; Dovesi,~R.; Roetti,~C. In \emph{Hartree-{Fock} {Ab} {Initio} {Treatment} of {Crystalline} {Systems}}; Berthier,~G., Dewar,~M. J.~S., Fischer,~H., Fukui,~K., Hall,~G.~G., Hinze,~J., Jaffé,~H.~H., Jortner,~J., Kutzelnigg,~W., Ruedenberg,~K., Tomasi,~J., Eds.; Lecture {Notes} in {Chemistry}; Springer Berlin Heidelberg: Berlin, Heidelberg, 1988; Vol.~48\relax
\mciteBstWouldAddEndPuncttrue
\mciteSetBstMidEndSepPunct{\mcitedefaultmidpunct}
{\mcitedefaultendpunct}{\mcitedefaultseppunct}\relax
\EndOfBibitem
\bibitem[Dovesi \latin{et~al.}(2014)Dovesi, Orlando, Erba, Zicovich-Wilson, Civalleri, Casassa, Maschio, Ferrabone, De~La~Pierre, D'Arco, Noël, Causà, Rérat, and Kirtman]{dovesi_c_2014}
Dovesi,~R.; Orlando,~R.; Erba,~A.; Zicovich-Wilson,~C.~M.; Civalleri,~B.; Casassa,~S.; Maschio,~L.; Ferrabone,~M.; De~La~Pierre,~M.; D'Arco,~P.; Noël,~Y.; Causà,~M.; Rérat,~M.; Kirtman,~B. CRYSTAL14: A program for the ab initio investigation of crystalline solids. \emph{International Journal of Quantum Chemistry} \textbf{2014}, \emph{114}, 1287--1317\relax
\mciteBstWouldAddEndPuncttrue
\mciteSetBstMidEndSepPunct{\mcitedefaultmidpunct}
{\mcitedefaultendpunct}{\mcitedefaultseppunct}\relax
\EndOfBibitem
\bibitem[Guidon \latin{et~al.}(2008)Guidon, Schiffmann, Hutter, and VandeVondele]{guidon_ab_2008}
Guidon,~M.; Schiffmann,~F.; Hutter,~J.; VandeVondele,~J. Ab initio molecular dynamics using hybrid density functionals. \emph{The Journal of Chemical Physics} \textbf{2008}, \emph{128}, 214104\relax
\mciteBstWouldAddEndPuncttrue
\mciteSetBstMidEndSepPunct{\mcitedefaultmidpunct}
{\mcitedefaultendpunct}{\mcitedefaultseppunct}\relax
\EndOfBibitem
\bibitem[Guidon \latin{et~al.}(2009)Guidon, Hutter, and VandeVondele]{guidon_robust_2009}
Guidon,~M.; Hutter,~J.; VandeVondele,~J. Robust Periodic Hartree-Fock Exchange for Large-Scale Simulations Using Gaussian Basis Sets. \emph{Journal of Chemical Theory and Computation} \textbf{2009}, \emph{5}, 3010--3021, PMID: 26609981\relax
\mciteBstWouldAddEndPuncttrue
\mciteSetBstMidEndSepPunct{\mcitedefaultmidpunct}
{\mcitedefaultendpunct}{\mcitedefaultseppunct}\relax
\EndOfBibitem
\bibitem[Paier \latin{et~al.}(2009)Paier, Diaconu, Scuseria, Guidon, VandeVondele, and Hutter]{paier_accurate_2009}
Paier,~J.; Diaconu,~C.~V.; Scuseria,~G.~E.; Guidon,~M.; VandeVondele,~J.; Hutter,~J. Accurate {Hartree}-{Fock} energy of extended systems using large {Gaussian} basis sets. \emph{Physical Review B} \textbf{2009}, \emph{80}, 174114, Publisher: American Physical Society\relax
\mciteBstWouldAddEndPuncttrue
\mciteSetBstMidEndSepPunct{\mcitedefaultmidpunct}
{\mcitedefaultendpunct}{\mcitedefaultseppunct}\relax
\EndOfBibitem
\bibitem[Paier \latin{et~al.}(2006)Paier, Marsman, Hummer, Kresse, Gerber, and Ángyán]{paier_screened_2006}
Paier,~J.; Marsman,~M.; Hummer,~K.; Kresse,~G.; Gerber,~I.~C.; Ángyán,~J.~G. Screened hybrid density functionals applied to solids. \emph{The Journal of Chemical Physics} \textbf{2006}, \emph{124}, 154709\relax
\mciteBstWouldAddEndPuncttrue
\mciteSetBstMidEndSepPunct{\mcitedefaultmidpunct}
{\mcitedefaultendpunct}{\mcitedefaultseppunct}\relax
\EndOfBibitem
\bibitem[Lin(2016)]{lin_adaptively_2016}
Lin,~L. Adaptively {Compressed} {Exchange} {Operator}. \emph{Journal of Chemical Theory and Computation} \textbf{2016}, \emph{12}, 2242--2249, Publisher: American Chemical Society\relax
\mciteBstWouldAddEndPuncttrue
\mciteSetBstMidEndSepPunct{\mcitedefaultmidpunct}
{\mcitedefaultendpunct}{\mcitedefaultseppunct}\relax
\EndOfBibitem
\bibitem[Betzinger \latin{et~al.}(2010)Betzinger, Friedrich, and Blügel]{betzinger_hybrid_2010}
Betzinger,~M.; Friedrich,~C.; Blügel,~S. Hybrid functionals within the all-electron {FLAPW} method: {Implementation} and applications of {PBE0}. \emph{Physical Review B} \textbf{2010}, \emph{81}, 195117, Publisher: American Physical Society\relax
\mciteBstWouldAddEndPuncttrue
\mciteSetBstMidEndSepPunct{\mcitedefaultmidpunct}
{\mcitedefaultendpunct}{\mcitedefaultseppunct}\relax
\EndOfBibitem
\bibitem[Wu \latin{et~al.}(2009)Wu, Selloni, and Car]{wu_order-n_2009}
Wu,~X.; Selloni,~A.; Car,~R. Order-\${N}\$ implementation of exact exchange in extended insulating systems. \emph{Physical Review B} \textbf{2009}, \emph{79}, 085102, Publisher: American Physical Society\relax
\mciteBstWouldAddEndPuncttrue
\mciteSetBstMidEndSepPunct{\mcitedefaultmidpunct}
{\mcitedefaultendpunct}{\mcitedefaultseppunct}\relax
\EndOfBibitem
\bibitem[Hu \latin{et~al.}(2017)Hu, Lin, and Yang]{hu_interpolative_2017}
Hu,~W.; Lin,~L.; Yang,~C. Interpolative {Separable} {Density} {Fitting} {Decomposition} for {Accelerating} {Hybrid} {Density} {Functional} {Calculations} with {Applications} to {Defects} in {Silicon}. \emph{Journal of Chemical Theory and Computation} \textbf{2017}, \emph{13}, 5420--5431, Publisher: American Chemical Society\relax
\mciteBstWouldAddEndPuncttrue
\mciteSetBstMidEndSepPunct{\mcitedefaultmidpunct}
{\mcitedefaultendpunct}{\mcitedefaultseppunct}\relax
\EndOfBibitem
\bibitem[Wu \latin{et~al.}(2022)Wu, Qin, Hu, and Yang]{Wu/Qin/etal:2022}
Wu,~K.; Qin,~X.; Hu,~W.; Yang,~J. Low-Rank Approximations Accelerated Plane-Wave Hybrid Functional Calculations with k-Point Sampling. \emph{Journal of Chemical Theory and Computation} \textbf{2022}, \emph{18}, 206--218, PMID: 34918919\relax
\mciteBstWouldAddEndPuncttrue
\mciteSetBstMidEndSepPunct{\mcitedefaultmidpunct}
{\mcitedefaultendpunct}{\mcitedefaultseppunct}\relax
\EndOfBibitem
\bibitem[Shang \latin{et~al.}(2010)Shang, Li, and Yang]{shang_implementation_2010}
Shang,~H.; Li,~Z.; Yang,~J. Implementation of {Exact} {Exchange} with {Numerical} {Atomic} {Orbitals}. \emph{The Journal of Physical Chemistry A} \textbf{2010}, \emph{114}, 1039--1043, Publisher: American Chemical Society\relax
\mciteBstWouldAddEndPuncttrue
\mciteSetBstMidEndSepPunct{\mcitedefaultmidpunct}
{\mcitedefaultendpunct}{\mcitedefaultseppunct}\relax
\EndOfBibitem
\bibitem[Qin \latin{et~al.}(2020)Qin, Liu, Hu, and Yang]{qin_interpolative_2020}
Qin,~X.; Liu,~J.; Hu,~W.; Yang,~J. Interpolative {Separable} {Density} {Fitting} {Decomposition} for {Accelerating} {Hartree}–{Fock} {Exchange} {Calculations} within {Numerical} {Atomic} {Orbitals}. \emph{The Journal of Physical Chemistry A} \textbf{2020}, \emph{124}, 5664--5674, Publisher: American Chemical Society\relax
\mciteBstWouldAddEndPuncttrue
\mciteSetBstMidEndSepPunct{\mcitedefaultmidpunct}
{\mcitedefaultendpunct}{\mcitedefaultseppunct}\relax
\EndOfBibitem
\bibitem[Feyereisen \latin{et~al.}(1993)Feyereisen, Fitzgerald, and Komornicki]{feyereisen_use_1993}
Feyereisen,~M.; Fitzgerald,~G.; Komornicki,~A. Use of approximate integrals in ab initio theory. {An} application in {MP2} energy calculations. \emph{Chemical Physics Letters} \textbf{1993}, \emph{208}, 359--363\relax
\mciteBstWouldAddEndPuncttrue
\mciteSetBstMidEndSepPunct{\mcitedefaultmidpunct}
{\mcitedefaultendpunct}{\mcitedefaultseppunct}\relax
\EndOfBibitem
\bibitem[Vahtras \latin{et~al.}(1993)Vahtras, Almlöf, and Feyereisen]{vahtras_integral_1993}
Vahtras,~O.; Almlöf,~J.; Feyereisen,~M.~W. Integral approximations for {LCAO}-{SCF} calculations. \emph{Chemical Physics Letters} \textbf{1993}, \emph{213}, 514--518\relax
\mciteBstWouldAddEndPuncttrue
\mciteSetBstMidEndSepPunct{\mcitedefaultmidpunct}
{\mcitedefaultendpunct}{\mcitedefaultseppunct}\relax
\EndOfBibitem
\bibitem[Weigend \latin{et~al.}(1998)Weigend, Häser, Patzelt, and Ahlrichs]{weigend_ri-mp2_1998}
Weigend,~F.; Häser,~M.; Patzelt,~H.; Ahlrichs,~R. {RI}-{MP2}: optimized auxiliary basis sets and demonstration of efficiency. \emph{Chemical Physics Letters} \textbf{1998}, \emph{294}, 143--152\relax
\mciteBstWouldAddEndPuncttrue
\mciteSetBstMidEndSepPunct{\mcitedefaultmidpunct}
{\mcitedefaultendpunct}{\mcitedefaultseppunct}\relax
\EndOfBibitem
\bibitem[Whitten(1973)]{whitten_coulombic_1973}
Whitten,~J.~L. Coulombic potential energy integrals and approximations. \emph{The Journal of Chemical Physics} \textbf{1973}, \emph{58}, 4496--4501\relax
\mciteBstWouldAddEndPuncttrue
\mciteSetBstMidEndSepPunct{\mcitedefaultmidpunct}
{\mcitedefaultendpunct}{\mcitedefaultseppunct}\relax
\EndOfBibitem
\bibitem[Dunlap \latin{et~al.}(1979)Dunlap, Connolly, and Sabin]{dunlap_approximations_1979}
Dunlap,~B.~I.; Connolly,~J. W.~D.; Sabin,~J.~R. On some approximations in applications of {X$\alpha$} theory. \emph{The Journal of Chemical Physics} \textbf{1979}, \emph{71}, 3396--3402\relax
\mciteBstWouldAddEndPuncttrue
\mciteSetBstMidEndSepPunct{\mcitedefaultmidpunct}
{\mcitedefaultendpunct}{\mcitedefaultseppunct}\relax
\EndOfBibitem
\bibitem[Dunlap \latin{et~al.}(2010)Dunlap, Rösch, and Trickey]{dunlap_variational_2010}
Dunlap,~B.~I.; Rösch,~N.; Trickey,~S.~B. Variational fitting methods for electronic structure calculations. \emph{Molecular Physics} \textbf{2010}, Publisher: Taylor \& Francis Group\relax
\mciteBstWouldAddEndPuncttrue
\mciteSetBstMidEndSepPunct{\mcitedefaultmidpunct}
{\mcitedefaultendpunct}{\mcitedefaultseppunct}\relax
\EndOfBibitem
\bibitem[Lee \latin{et~al.}(2020)Lee, Lin, and Head-Gordon]{lee_systematically_2020}
Lee,~J.; Lin,~L.; Head-Gordon,~M. Systematically {Improvable} {Tensor} {Hypercontraction}: {Interpolative} {Separable} {Density}-{Fitting} for {Molecules} {Applied} to {Exact} {Exchange}, {Second}- and {Third}-{Order} {Møller}–{Plesset} {Perturbation} {Theory}. \emph{Journal of Chemical Theory and Computation} \textbf{2020}, \emph{16}, 243--263, Publisher: American Chemical Society\relax
\mciteBstWouldAddEndPuncttrue
\mciteSetBstMidEndSepPunct{\mcitedefaultmidpunct}
{\mcitedefaultendpunct}{\mcitedefaultseppunct}\relax
\EndOfBibitem
\bibitem[Henneke \latin{et~al.}(2020)Henneke, Lin, Vorwerk, Draxl, Klein, and Yang]{henneke_fast_2020}
Henneke,~F.; Lin,~L.; Vorwerk,~C.; Draxl,~C.; Klein,~R.; Yang,~C. Fast optical absorption spectra calculations for periodic solid state systems. \emph{Communications in Applied Mathematics and Computational Science} \textbf{2020}, \emph{15}, 89--113, Publisher: Mathematical Sciences Publishers\relax
\mciteBstWouldAddEndPuncttrue
\mciteSetBstMidEndSepPunct{\mcitedefaultmidpunct}
{\mcitedefaultendpunct}{\mcitedefaultseppunct}\relax
\EndOfBibitem
\bibitem[Sodt \latin{et~al.}(2006)Sodt, Subotnik, and Head-Gordon]{sodt_linear_2006}
Sodt,~A.; Subotnik,~J.~E.; Head-Gordon,~M. Linear scaling density fitting. \emph{The Journal of Chemical Physics} \textbf{2006}, \emph{125}, 194109\relax
\mciteBstWouldAddEndPuncttrue
\mciteSetBstMidEndSepPunct{\mcitedefaultmidpunct}
{\mcitedefaultendpunct}{\mcitedefaultseppunct}\relax
\EndOfBibitem
\bibitem[Sodt and Head-Gordon(2008)Sodt, and Head-Gordon]{sodt_hartree-fock_2008}
Sodt,~A.; Head-Gordon,~M. Hartree-{Fock} exchange computed using the atomic resolution of the identity approximation. \emph{The Journal of Chemical Physics} \textbf{2008}, \emph{128}, 104106\relax
\mciteBstWouldAddEndPuncttrue
\mciteSetBstMidEndSepPunct{\mcitedefaultmidpunct}
{\mcitedefaultendpunct}{\mcitedefaultseppunct}\relax
\EndOfBibitem
\bibitem[Pisani \latin{et~al.}(2005)Pisani, Busso, Capecchi, Casassa, Dovesi, Maschio, Zicovich-Wilson, and Schütz]{pisani_local-mp2_2005}
Pisani,~C.; Busso,~M.; Capecchi,~G.; Casassa,~S.; Dovesi,~R.; Maschio,~L.; Zicovich-Wilson,~C.; Schütz,~M. Local-{MP2} electron correlation method for nonconducting crystals. \emph{The Journal of Chemical Physics} \textbf{2005}, \emph{122}, 094113\relax
\mciteBstWouldAddEndPuncttrue
\mciteSetBstMidEndSepPunct{\mcitedefaultmidpunct}
{\mcitedefaultendpunct}{\mcitedefaultseppunct}\relax
\EndOfBibitem
\bibitem[Pisani \latin{et~al.}(2008)Pisani, Maschio, Casassa, Halo, Schütz, and Usvyat]{pisani_periodic_2008}
Pisani,~C.; Maschio,~L.; Casassa,~S.; Halo,~M.; Schütz,~M.; Usvyat,~D. Periodic local {MP2} method for the study of electronic correlation in crystals: {Theory} and preliminary applications. \emph{Journal of Computational Chemistry} \textbf{2008}, \emph{29}, 2113--2124, \_eprint: https://onlinelibrary.wiley.com/doi/pdf/10.1002/jcc.20975\relax
\mciteBstWouldAddEndPuncttrue
\mciteSetBstMidEndSepPunct{\mcitedefaultmidpunct}
{\mcitedefaultendpunct}{\mcitedefaultseppunct}\relax
\EndOfBibitem
\bibitem[Reine \latin{et~al.}(2008)Reine, Tellgren, Krapp, Kjærgaard, Helgaker, Jansik, Høst, and Salek]{reine_variational_2008}
Reine,~S.; Tellgren,~E.; Krapp,~A.; Kjærgaard,~T.; Helgaker,~T.; Jansik,~B.; Høst,~S.; Salek,~P. Variational and robust density fitting of four-center two-electron integrals in local metrics. \emph{The Journal of Chemical Physics} \textbf{2008}, \emph{129}, 104101\relax
\mciteBstWouldAddEndPuncttrue
\mciteSetBstMidEndSepPunct{\mcitedefaultmidpunct}
{\mcitedefaultendpunct}{\mcitedefaultseppunct}\relax
\EndOfBibitem
\bibitem[Ren \latin{et~al.}(2012)Ren, Rinke, Blum, Wieferink, Tkatchenko, Sanfilippo, Reuter, and Scheffler]{ren_resolution--identity_2012}
Ren,~X.; Rinke,~P.; Blum,~V.; Wieferink,~J.; Tkatchenko,~A.; Sanfilippo,~A.; Reuter,~K.; Scheffler,~M. Resolution-of-identity approach to {Hartree}–{Fock}, hybrid density functionals, {RPA}, {MP2} and {GW} with numeric atom-centered orbital basis functions. \emph{New Journal of Physics} \textbf{2012}, \emph{14}, 053020, Publisher: IOP Publishing\relax
\mciteBstWouldAddEndPuncttrue
\mciteSetBstMidEndSepPunct{\mcitedefaultmidpunct}
{\mcitedefaultendpunct}{\mcitedefaultseppunct}\relax
\EndOfBibitem
\bibitem[Merlot \latin{et~al.}(2013)Merlot, Kjærgaard, Helgaker, Lindh, Aquilante, Reine, and Pedersen]{merlot_attractive_2013}
Merlot,~P.; Kjærgaard,~T.; Helgaker,~T.; Lindh,~R.; Aquilante,~F.; Reine,~S.; Pedersen,~T.~B. Attractive electron–electron interactions within robust local fitting approximations. \emph{Journal of Computational Chemistry} \textbf{2013}, \emph{34}, 1486--1496, \_eprint: https://onlinelibrary.wiley.com/doi/pdf/10.1002/jcc.23284\relax
\mciteBstWouldAddEndPuncttrue
\mciteSetBstMidEndSepPunct{\mcitedefaultmidpunct}
{\mcitedefaultendpunct}{\mcitedefaultseppunct}\relax
\EndOfBibitem
\bibitem[Ihrig \latin{et~al.}(2015)Ihrig, Wieferink, Zhang, Ropo, Ren, Rinke, Scheffler, and Blum]{ihrig_accurate_2015}
Ihrig,~A.~C.; Wieferink,~J.; Zhang,~I.~Y.; Ropo,~M.; Ren,~X.; Rinke,~P.; Scheffler,~M.; Blum,~V. Accurate localized resolution of identity approach for linear-scaling hybrid density functionals and for many-body perturbation theory. \emph{New Journal of Physics} \textbf{2015}, \emph{17}, 093020, Publisher: IOP Publishing\relax
\mciteBstWouldAddEndPuncttrue
\mciteSetBstMidEndSepPunct{\mcitedefaultmidpunct}
{\mcitedefaultendpunct}{\mcitedefaultseppunct}\relax
\EndOfBibitem
\bibitem[Levchenko \latin{et~al.}(2015)Levchenko, Ren, Wieferink, Johanni, Rinke, Blum, and Scheffler]{levchenko_hybrid_2015}
Levchenko,~S.~V.; Ren,~X.; Wieferink,~J.; Johanni,~R.; Rinke,~P.; Blum,~V.; Scheffler,~M. Hybrid functionals for large periodic systems in an all-electron, numeric atom-centered basis framework. \emph{Computer Physics Communications} \textbf{2015}, \emph{192}, 60--69\relax
\mciteBstWouldAddEndPuncttrue
\mciteSetBstMidEndSepPunct{\mcitedefaultmidpunct}
{\mcitedefaultendpunct}{\mcitedefaultseppunct}\relax
\EndOfBibitem
\bibitem[Wirz \latin{et~al.}(2017)Wirz, Reine, and Pedersen]{wirz_resolution---identity_2017}
Wirz,~L.~N.; Reine,~S.~S.; Pedersen,~T.~B. On {Resolution}-of-the-{Identity} {Electron} {Repulsion} {Integral} {Approximations} and {Variational} {Stability}. \emph{Journal of Chemical Theory and Computation} \textbf{2017}, \emph{13}, 4897--4906, Publisher: American Chemical Society\relax
\mciteBstWouldAddEndPuncttrue
\mciteSetBstMidEndSepPunct{\mcitedefaultmidpunct}
{\mcitedefaultendpunct}{\mcitedefaultseppunct}\relax
\EndOfBibitem
\bibitem[Kokott \latin{et~al.}(2024)Kokott, Merz, Yao, Carbogno, Rossi, Havu, Rampp, Scheffler, and Blum]{Kokott/etal:2024}
Kokott,~S.; Merz,~F.; Yao,~Y.; Carbogno,~C.; Rossi,~M.; Havu,~V.; Rampp,~M.; Scheffler,~M.; Blum,~V. Efficient all-electron hybrid density functionals for atomistic simulations beyond 10000 atoms. \emph{The Journal of Chemical Physics} \textbf{2024}, \emph{161}, 024112\relax
\mciteBstWouldAddEndPuncttrue
\mciteSetBstMidEndSepPunct{\mcitedefaultmidpunct}
{\mcitedefaultendpunct}{\mcitedefaultseppunct}\relax
\EndOfBibitem
\bibitem[Weigend(2002)]{weigend_fully_2002}
Weigend,~F. A fully direct {RI}-{HF} algorithm: {Implementation}, optimised auxiliary basis sets, demonstration of accuracy and efficiency. \emph{Physical Chemistry Chemical Physics} \textbf{2002}, \emph{4}, 4285--4291, Publisher: The Royal Society of Chemistry\relax
\mciteBstWouldAddEndPuncttrue
\mciteSetBstMidEndSepPunct{\mcitedefaultmidpunct}
{\mcitedefaultendpunct}{\mcitedefaultseppunct}\relax
\EndOfBibitem
\bibitem[Eshuis \latin{et~al.}(2010)Eshuis, Yarkony, and Furche]{eshuis_fast_2010}
Eshuis,~H.; Yarkony,~J.; Furche,~F. Fast computation of molecular random phase approximation correlation energies using resolution of the identity and imaginary frequency integration. \emph{The Journal of Chemical Physics} \textbf{2010}, \emph{132}, 234114\relax
\mciteBstWouldAddEndPuncttrue
\mciteSetBstMidEndSepPunct{\mcitedefaultmidpunct}
{\mcitedefaultendpunct}{\mcitedefaultseppunct}\relax
\EndOfBibitem
\bibitem[Del~Ben \latin{et~al.}(2013)Del~Ben, Hutter, and VandeVondele]{del_ben_electron_2013}
Del~Ben,~M.; Hutter,~J.; VandeVondele,~J. Electron {Correlation} in the {Condensed} {Phase} from a {Resolution} of {Identity} {Approach} {Based} on the {Gaussian} and {Plane} {Waves} {Scheme}. \emph{Journal of Chemical Theory and Computation} \textbf{2013}, \emph{9}, 2654--2671, Publisher: American Chemical Society\relax
\mciteBstWouldAddEndPuncttrue
\mciteSetBstMidEndSepPunct{\mcitedefaultmidpunct}
{\mcitedefaultendpunct}{\mcitedefaultseppunct}\relax
\EndOfBibitem
\bibitem[Bussy and Hutter(2024)Bussy, and Hutter]{Bussy/Hutter:2024}
Bussy,~A.; Hutter,~J. Efficient periodic resolution-of-the-identity Hartree–Fock exchange method with k-point sampling and Gaussian basis sets. \emph{The Journal of Chemical Physics} \textbf{2024}, \emph{160}, 064116\relax
\mciteBstWouldAddEndPuncttrue
\mciteSetBstMidEndSepPunct{\mcitedefaultmidpunct}
{\mcitedefaultendpunct}{\mcitedefaultseppunct}\relax
\EndOfBibitem
\bibitem[Sun \latin{et~al.}(2017)Sun, Berkelbach, McClain, and Chan]{sun_gaussian_2017}
Sun,~Q.; Berkelbach,~T.~C.; McClain,~J.~D.; Chan,~G. K.-L. Gaussian and plane-wave mixed density fitting for periodic systems. \emph{The Journal of Chemical Physics} \textbf{2017}, \emph{147}, 164119\relax
\mciteBstWouldAddEndPuncttrue
\mciteSetBstMidEndSepPunct{\mcitedefaultmidpunct}
{\mcitedefaultendpunct}{\mcitedefaultseppunct}\relax
\EndOfBibitem
\bibitem[Beebe and Linderberg(1977)Beebe, and Linderberg]{beebe_simplifications_1977}
Beebe,~N. H.~F.; Linderberg,~J. Simplifications in the generation and transformation of two-electron integrals in molecular calculations. \emph{International Journal of Quantum Chemistry} \textbf{1977}, \emph{12}, 683--705, \_eprint: https://onlinelibrary.wiley.com/doi/pdf/10.1002/qua.560120408\relax
\mciteBstWouldAddEndPuncttrue
\mciteSetBstMidEndSepPunct{\mcitedefaultmidpunct}
{\mcitedefaultendpunct}{\mcitedefaultseppunct}\relax
\EndOfBibitem
\bibitem[Koch \latin{et~al.}(2003)Koch, Sánchez~de Merás, and Pedersen]{koch_reduced_2003}
Koch,~H.; Sánchez~de Merás,~A.; Pedersen,~T.~B. Reduced scaling in electronic structure calculations using {Cholesky} decompositions. \emph{The Journal of Chemical Physics} \textbf{2003}, \emph{118}, 9481--9484\relax
\mciteBstWouldAddEndPuncttrue
\mciteSetBstMidEndSepPunct{\mcitedefaultmidpunct}
{\mcitedefaultendpunct}{\mcitedefaultseppunct}\relax
\EndOfBibitem
\bibitem[Hohenstein \latin{et~al.}(2012)Hohenstein, Parrish, and Martínez]{hohenstein_tensor_2012}
Hohenstein,~E.~G.; Parrish,~R.~M.; Martínez,~T.~J. Tensor hypercontraction density fitting. {I}. {Quartic} scaling second- and third-order {Møller}-{Plesset} perturbation theory. \emph{The Journal of Chemical Physics} \textbf{2012}, \emph{137}, 044103\relax
\mciteBstWouldAddEndPuncttrue
\mciteSetBstMidEndSepPunct{\mcitedefaultmidpunct}
{\mcitedefaultendpunct}{\mcitedefaultseppunct}\relax
\EndOfBibitem
\bibitem[Parrish \latin{et~al.}(2012)Parrish, Hohenstein, Martínez, and Sherrill]{parrish_tensor_2012}
Parrish,~R.~M.; Hohenstein,~E.~G.; Martínez,~T.~J.; Sherrill,~C.~D. Tensor hypercontraction. {II}. {Least}-squares renormalization. \emph{The Journal of Chemical Physics} \textbf{2012}, \emph{137}, 224106\relax
\mciteBstWouldAddEndPuncttrue
\mciteSetBstMidEndSepPunct{\mcitedefaultmidpunct}
{\mcitedefaultendpunct}{\mcitedefaultseppunct}\relax
\EndOfBibitem
\bibitem[Lu and Ying(2015)Lu, and Ying]{lu_compression_2015}
Lu,~J.; Ying,~L. Compression of the electron repulsion integral tensor in tensor hypercontraction format with cubic scaling cost. \emph{Journal of Computational Physics} \textbf{2015}, \emph{302}, 329--335\relax
\mciteBstWouldAddEndPuncttrue
\mciteSetBstMidEndSepPunct{\mcitedefaultmidpunct}
{\mcitedefaultendpunct}{\mcitedefaultseppunct}\relax
\EndOfBibitem
\bibitem[Lin \latin{et~al.}(2021)Lin, Ren, and He]{lin_efficient_2021}
Lin,~P.; Ren,~X.; He,~L. Efficient {Hybrid} {Density} {Functional} {Calculations} for {Large} {Periodic} {Systems} {Using} {Numerical} {Atomic} {Orbitals}. \emph{Journal of Chemical Theory and Computation} \textbf{2021}, \emph{17}, 222--239\relax
\mciteBstWouldAddEndPuncttrue
\mciteSetBstMidEndSepPunct{\mcitedefaultmidpunct}
{\mcitedefaultendpunct}{\mcitedefaultseppunct}\relax
\EndOfBibitem
\bibitem[Lin \latin{et~al.}(2020)Lin, Ren, and He]{lin_accuracy_2020}
Lin,~P.; Ren,~X.; He,~L. Accuracy of {Localized} {Resolution} of the {Identity} in {Periodic} {Hybrid} {Functional} {Calculations} with {Numerical} {Atomic} {Orbitals}. \emph{The Journal of Physical Chemistry Letters} \textbf{2020}, \emph{11}, 3082--3088, Publisher: American Chemical Society\relax
\mciteBstWouldAddEndPuncttrue
\mciteSetBstMidEndSepPunct{\mcitedefaultmidpunct}
{\mcitedefaultendpunct}{\mcitedefaultseppunct}\relax
\EndOfBibitem
\bibitem[Li \latin{et~al.}(2016)Li, Liu, Chen, Lin, Ren, Lin, Yang, and He]{li_large-scale_2016}
Li,~P.; Liu,~X.; Chen,~M.; Lin,~P.; Ren,~X.; Lin,~L.; Yang,~C.; He,~L. Large-scale ab initio simulations based on systematically improvable atomic basis. \emph{Computational Materials Science} \textbf{2016}, \emph{112}, 503--517\relax
\mciteBstWouldAddEndPuncttrue
\mciteSetBstMidEndSepPunct{\mcitedefaultmidpunct}
{\mcitedefaultendpunct}{\mcitedefaultseppunct}\relax
\EndOfBibitem
\bibitem[Lin \latin{et~al.}(2024)Lin, Ren, Liu, and He]{lin_ab_2024}
Lin,~P.; Ren,~X.; Liu,~X.; He,~L. Ab initio electronic structure calculations based on numerical atomic orbitals: {Basic} fomalisms and recent progresses. \emph{WIREs Computational Molecular Science} \textbf{2024}, \emph{14}, e1687\relax
\mciteBstWouldAddEndPuncttrue
\mciteSetBstMidEndSepPunct{\mcitedefaultmidpunct}
{\mcitedefaultendpunct}{\mcitedefaultseppunct}\relax
\EndOfBibitem
\bibitem[Chen \latin{et~al.}(2010)Chen, Guo, and He]{chen_systematically_2010}
Chen,~M.; Guo,~G.-C.; He,~L. Systematically improvable optimized atomic basis sets for \textit{ab initio} calculations. \emph{Journal of Physics: Condensed Matter} \textbf{2010}, \emph{22}, 445501\relax
\mciteBstWouldAddEndPuncttrue
\mciteSetBstMidEndSepPunct{\mcitedefaultmidpunct}
{\mcitedefaultendpunct}{\mcitedefaultseppunct}\relax
\EndOfBibitem
\bibitem[Chen \latin{et~al.}(2011)Chen, Guo, and He]{chen_electronic_2011}
Chen,~M.; Guo,~G.-C.; He,~L. Electronic structure interpolation via atomic orbitals. \emph{Journal of Physics: Condensed Matter} \textbf{2011}, \emph{23}, 325501\relax
\mciteBstWouldAddEndPuncttrue
\mciteSetBstMidEndSepPunct{\mcitedefaultmidpunct}
{\mcitedefaultendpunct}{\mcitedefaultseppunct}\relax
\EndOfBibitem
\bibitem[Lin \latin{et~al.}(2021)Lin, Ren, and He]{lin_strategy_2021}
Lin,~P.; Ren,~X.; He,~L. Strategy for constructing compact numerical atomic orbital basis sets by incorporating the gradients of reference wavefunctions. \emph{Physical Review B} \textbf{2021}, \emph{103}, 235131\relax
\mciteBstWouldAddEndPuncttrue
\mciteSetBstMidEndSepPunct{\mcitedefaultmidpunct}
{\mcitedefaultendpunct}{\mcitedefaultseppunct}\relax
\EndOfBibitem
\bibitem[Dresselhaus \latin{et~al.}(2007)Dresselhaus, Dresselhaus, and Jorio]{dresselhaus_group_2007}
Dresselhaus,~M.~S.; Dresselhaus,~G.; Jorio,~A. \emph{Group {Theory}: {Application} to the {Physics} of {Condensed} {Matter}}; Springer Science \& Business Media, 2007; Google-Books-ID: sKaH8vrfmnQC\relax
\mciteBstWouldAddEndPuncttrue
\mciteSetBstMidEndSepPunct{\mcitedefaultmidpunct}
{\mcitedefaultendpunct}{\mcitedefaultseppunct}\relax
\EndOfBibitem
\bibitem[Dacre(1970)]{dacre_use_1970}
Dacre,~P.~D. On the use of symmetry in {SCF} calculations. \emph{Chemical Physics Letters} \textbf{1970}, \emph{7}, 47--48\relax
\mciteBstWouldAddEndPuncttrue
\mciteSetBstMidEndSepPunct{\mcitedefaultmidpunct}
{\mcitedefaultendpunct}{\mcitedefaultseppunct}\relax
\EndOfBibitem
\bibitem[Dupuis and King(1977)Dupuis, and King]{dupuis_molecular_1977}
Dupuis,~M.; King,~H.~F. Molecular symmetry and closed‐shell SCF calculations. {I}. \emph{International Journal of Quantum Chemistry} \textbf{1977}, \emph{11}, 613--625\relax
\mciteBstWouldAddEndPuncttrue
\mciteSetBstMidEndSepPunct{\mcitedefaultmidpunct}
{\mcitedefaultendpunct}{\mcitedefaultseppunct}\relax
\EndOfBibitem
\bibitem[Ferrero \latin{et~al.}(2008)Ferrero, Rérat, Orlando, and Dovesi]{ferrero_coupled_2008}
Ferrero,~M.; Rérat,~M.; Orlando,~R.; Dovesi,~R. Coupled perturbed {Hartree}-{Fock} for periodic systems: {The} role of symmetry and related computational aspects. \emph{The Journal of Chemical Physics} \textbf{2008}, \emph{128}, 014110\relax
\mciteBstWouldAddEndPuncttrue
\mciteSetBstMidEndSepPunct{\mcitedefaultmidpunct}
{\mcitedefaultendpunct}{\mcitedefaultseppunct}\relax
\EndOfBibitem
\bibitem[Orlando \latin{et~al.}(2014)Orlando, De~La~Pierre, Zicovich-Wilson, Erba, and Dovesi]{orlando_full_2014}
Orlando,~R.; De~La~Pierre,~M.; Zicovich-Wilson,~C.~M.; Erba,~A.; Dovesi,~R. On the full exploitation of symmetry in periodic (as well as molecular) self-consistent-field \textit{ab initio} calculations. \emph{The Journal of Chemical Physics} \textbf{2014}, \emph{141}, 104108\relax
\mciteBstWouldAddEndPuncttrue
\mciteSetBstMidEndSepPunct{\mcitedefaultmidpunct}
{\mcitedefaultendpunct}{\mcitedefaultseppunct}\relax
\EndOfBibitem
\bibitem[Dovesi \latin{et~al.}(2018)Dovesi, Erba, Orlando, Zicovich-Wilson, Civalleri, Maschio, Rérat, Casassa, Baima, Salustro, and Kirtman]{dovesi_quantum-mechanical_2018}
Dovesi,~R.; Erba,~A.; Orlando,~R.; Zicovich-Wilson,~C.~M.; Civalleri,~B.; Maschio,~L.; Rérat,~M.; Casassa,~S.; Baima,~J.; Salustro,~S.; Kirtman,~B. Quantum-mechanical condensed matter simulations with {CRYSTAL}. \emph{WIREs Computational Molecular Science} \textbf{2018}, \emph{8}, e1360, \_eprint: https://onlinelibrary.wiley.com/doi/pdf/10.1002/wcms.1360\relax
\mciteBstWouldAddEndPuncttrue
\mciteSetBstMidEndSepPunct{\mcitedefaultmidpunct}
{\mcitedefaultendpunct}{\mcitedefaultseppunct}\relax
\EndOfBibitem
\bibitem[Dovesi \latin{et~al.}(2020)Dovesi, Pascale, Civalleri, Doll, Harrison, Bush, D’Arco, Noël, Rérat, Carbonnière, Causà, Salustro, Lacivita, Kirtman, Ferrari, Gentile, Baima, Ferrero, Demichelis, and De~La~Pierre]{dovesi_crystal_2020}
Dovesi,~R.; Pascale,~F.; Civalleri,~B.; Doll,~K.; Harrison,~N.~M.; Bush,~I.; D’Arco,~P.; Noël,~Y.; Rérat,~M.; Carbonnière,~P.; Causà,~M.; Salustro,~S.; Lacivita,~V.; Kirtman,~B.; Ferrari,~A.~M.; Gentile,~F.~S.; Baima,~J.; Ferrero,~M.; Demichelis,~R.; De~La~Pierre,~M. The {CRYSTAL} code, 1976–2020 and beyond, a long story. \emph{The Journal of Chemical Physics} \textbf{2020}, \emph{152}, 204111\relax
\mciteBstWouldAddEndPuncttrue
\mciteSetBstMidEndSepPunct{\mcitedefaultmidpunct}
{\mcitedefaultendpunct}{\mcitedefaultseppunct}\relax
\EndOfBibitem
\bibitem[Patterson(2010)]{10Charles-IRQuads}
Patterson,~C.~H. Exciton: a code for excitations in materials. \emph{Molecular Physics} \textbf{2010}, \emph{108}, 3181--3188\relax
\mciteBstWouldAddEndPuncttrue
\mciteSetBstMidEndSepPunct{\mcitedefaultmidpunct}
{\mcitedefaultendpunct}{\mcitedefaultseppunct}\relax
\EndOfBibitem
\bibitem[Csizmadia \latin{et~al.}(1966)Csizmadia, Harrison, Moskowitz, and Sutcliffe]{csizmadia_non-empirical_1966}
Csizmadia,~I.~G.; Harrison,~M.~C.; Moskowitz,~J.~W.; Sutcliffe,~B.~T. Non-empirical {LCAO}-{MO}-{SCF}-{CI} calculations on organic molecules with {Gaussian} type functions. \emph{Theoretica chimica acta} \textbf{1966}, \emph{6}, 191--216\relax
\mciteBstWouldAddEndPuncttrue
\mciteSetBstMidEndSepPunct{\mcitedefaultmidpunct}
{\mcitedefaultendpunct}{\mcitedefaultseppunct}\relax
\EndOfBibitem
\bibitem[Elder(1973)]{elder_use_1973}
Elder,~M. Use of molecular symmetry in {SCF} calculations. \emph{International Journal of Quantum Chemistry} \textbf{1973}, \emph{7}, 75--85, \_eprint: https://onlinelibrary.wiley.com/doi/pdf/10.1002/qua.560070109\relax
\mciteBstWouldAddEndPuncttrue
\mciteSetBstMidEndSepPunct{\mcitedefaultmidpunct}
{\mcitedefaultendpunct}{\mcitedefaultseppunct}\relax
\EndOfBibitem
\bibitem[{van Wüllen}(1994)]{94CPL-Wullen}
{van Wüllen},~C. An implementation of a Kohn—Sham density functional program using a Gaussian-type basis set. Application to the equilibrium geometry of C60 and C70. \emph{Chemical Physics Letters} \textbf{1994}, \emph{219}, 8--14\relax
\mciteBstWouldAddEndPuncttrue
\mciteSetBstMidEndSepPunct{\mcitedefaultmidpunct}
{\mcitedefaultendpunct}{\mcitedefaultseppunct}\relax
\EndOfBibitem
\bibitem[Rusakov \latin{et~al.}(2013)Rusakov, Frisch, and Scuseria]{rusakov_space_2013}
Rusakov,~A.~A.; Frisch,~M.~J.; Scuseria,~G.~E. Space group symmetry applied to {SCF} calculations with periodic boundary conditions and {Gaussian} orbitals. \emph{The Journal of Chemical Physics} \textbf{2013}, \emph{139}, 114110\relax
\mciteBstWouldAddEndPuncttrue
\mciteSetBstMidEndSepPunct{\mcitedefaultmidpunct}
{\mcitedefaultendpunct}{\mcitedefaultseppunct}\relax
\EndOfBibitem
\bibitem[Weigend(2002)]{02PCCP-Weigend}
Weigend,~F. A fully direct RI-HF algorithm: Implementation{,} optimised auxiliary basis sets{,} demonstration of accuracy and efficiency. \emph{Phys. Chem. Chem. Phys.} \textbf{2002}, \emph{4}, 4285--4291\relax
\mciteBstWouldAddEndPuncttrue
\mciteSetBstMidEndSepPunct{\mcitedefaultmidpunct}
{\mcitedefaultendpunct}{\mcitedefaultseppunct}\relax
\EndOfBibitem
\bibitem[Weigend(2008)]{08JCC-Weigend}
Weigend,~F. Hartree–Fock exchange fitting basis sets for H to Rn. \emph{Journal of Computational Chemistry} \textbf{2008}, \emph{29}, 167--175\relax
\mciteBstWouldAddEndPuncttrue
\mciteSetBstMidEndSepPunct{\mcitedefaultmidpunct}
{\mcitedefaultendpunct}{\mcitedefaultseppunct}\relax
\EndOfBibitem
\bibitem[Zicovich-Wilson and Dovesi(1998)Zicovich-Wilson, and Dovesi]{zicovich-wilson_use_1998}
Zicovich-Wilson,~C.~M.; Dovesi,~R. On the use of symmetry-adapted crystalline orbitals in {SCF}-{LCAO} periodic calculations. {I}. {The} construction of the symmetrized orbitals. \emph{International Journal of Quantum Chemistry} \textbf{1998}, \emph{67}, 299--309, \_eprint: https://onlinelibrary.wiley.com/doi/pdf/10.1002/\%28SICI\%291097-461X\%281998\%2967\%3A5\%3C299\%3A\%3AAID-QUA3\%3E3.0.CO\%3B2-Q\relax
\mciteBstWouldAddEndPuncttrue
\mciteSetBstMidEndSepPunct{\mcitedefaultmidpunct}
{\mcitedefaultendpunct}{\mcitedefaultseppunct}\relax
\EndOfBibitem
\bibitem[Zicovich-Wilson and Dovesi(1998)Zicovich-Wilson, and Dovesi]{zicovich-wilson_use_1998-1}
Zicovich-Wilson,~C.~M.; Dovesi,~R. On the use of symmetry-adapted crystalline orbitals in {SCF}-{LCAO} periodic calculations. {II}. {Implementation} of the self-consistent-field scheme and examples. \emph{International Journal of Quantum Chemistry} \textbf{1998}, \emph{67}, 311--320, \_eprint: https://onlinelibrary.wiley.com/doi/pdf/10.1002/\%28SICI\%291097-461X\%281998\%2967\%3A5\%3C311\%3A\%3AAID-QUA4\%3E3.0.CO\%3B2-Y\relax
\mciteBstWouldAddEndPuncttrue
\mciteSetBstMidEndSepPunct{\mcitedefaultmidpunct}
{\mcitedefaultendpunct}{\mcitedefaultseppunct}\relax
\EndOfBibitem
\bibitem[Dong and Gull(2025)Dong, and Gull]{dongxinyang-2025}
Dong,~X.; Gull,~E. Symmetry adaptation for self-consistent many-body calculations. \emph{Computer Physics Communications} \textbf{2025}, \emph{307}, 109401\relax
\mciteBstWouldAddEndPuncttrue
\mciteSetBstMidEndSepPunct{\mcitedefaultmidpunct}
{\mcitedefaultendpunct}{\mcitedefaultseppunct}\relax
\EndOfBibitem
\bibitem[Yeh and Morales(2024)Yeh, and Morales]{yeh_low-scaling_2024-1}
Yeh,~C.-N.; Morales,~M.~A. Low-{Scaling} {Algorithms} for {GW} and {Constrained} {Random} {Phase} {Approximation} {Using} {Symmetry}-{Adapted} {Interpolative} {Separable} {Density} {Fitting}. \emph{Journal of Chemical Theory and Computation} \textbf{2024}, \emph{20}, 3184--3198, Publisher: American Chemical Society\relax
\mciteBstWouldAddEndPuncttrue
\mciteSetBstMidEndSepPunct{\mcitedefaultmidpunct}
{\mcitedefaultendpunct}{\mcitedefaultseppunct}\relax
\EndOfBibitem
\bibitem[Häser(1991)]{haser_molecular_1991}
Häser,~M. Molecular point-group symmetry in electronic structure calculations. \emph{The Journal of Chemical Physics} \textbf{1991}, \emph{95}, 8259--8265\relax
\mciteBstWouldAddEndPuncttrue
\mciteSetBstMidEndSepPunct{\mcitedefaultmidpunct}
{\mcitedefaultendpunct}{\mcitedefaultseppunct}\relax
\EndOfBibitem
\bibitem[Häser \latin{et~al.}()Häser, Almlöf, and Feyereisen]{hser_exploiting_1991}
Häser,~M.; Almlöf,~J.; Feyereisen,~M.~W. Exploiting non-abelian point group symmetry in direct two-electron integral transformations. \emph{79}, 115--122\relax
\mciteBstWouldAddEndPuncttrue
\mciteSetBstMidEndSepPunct{\mcitedefaultmidpunct}
{\mcitedefaultendpunct}{\mcitedefaultseppunct}\relax
\EndOfBibitem
\bibitem[Furche \latin{et~al.}(2014)Furche, Ahlrichs, Hättig, Klopper, Sierka, and Weigend]{furche_turbomole_2014}
Furche,~F.; Ahlrichs,~R.; Hättig,~C.; Klopper,~W.; Sierka,~M.; Weigend,~F. Turbomole. \emph{WIREs Computational Molecular Science} \textbf{2014}, \emph{4}, 91--100, \_eprint: https://onlinelibrary.wiley.com/doi/pdf/10.1002/wcms.1162\relax
\mciteBstWouldAddEndPuncttrue
\mciteSetBstMidEndSepPunct{\mcitedefaultmidpunct}
{\mcitedefaultendpunct}{\mcitedefaultseppunct}\relax
\EndOfBibitem
\bibitem[Sierka \latin{et~al.}(2003)Sierka, Hogekamp, and Ahlrichs]{sierka_fast_2003}
Sierka,~M.; Hogekamp,~A.; Ahlrichs,~R. Fast evaluation of the {Coulomb} potential for electron densities using multipole accelerated resolution of identity approximation. \emph{The Journal of Chemical Physics} \textbf{2003}, \emph{118}, 9136--9148\relax
\mciteBstWouldAddEndPuncttrue
\mciteSetBstMidEndSepPunct{\mcitedefaultmidpunct}
{\mcitedefaultendpunct}{\mcitedefaultseppunct}\relax
\EndOfBibitem
\bibitem[Gao and Chelikowsky(2020)Gao, and Chelikowsky]{gao_accelerating_2020}
Gao,~W.; Chelikowsky,~J.~R. Accelerating {Time}-{Dependent} {Density} {Functional} {Theory} and {GW} {Calculations} for {Molecules} and {Nanoclusters} with {Symmetry} {Adapted} {Interpolative} {Separable} {Density} {Fitting}. \emph{Journal of Chemical Theory and Computation} \textbf{2020}, \emph{16}, 2216--2223, Publisher: American Chemical Society\relax
\mciteBstWouldAddEndPuncttrue
\mciteSetBstMidEndSepPunct{\mcitedefaultmidpunct}
{\mcitedefaultendpunct}{\mcitedefaultseppunct}\relax
\EndOfBibitem
\bibitem[Soler \latin{et~al.}(2002)Soler, Artacho, Gale, García, Junquera, Ordejón, and Sánchez-Portal]{soler_siesta_2002}
Soler,~J.~M.; Artacho,~E.; Gale,~J.~D.; García,~A.; Junquera,~J.; Ordejón,~P.; Sánchez-Portal,~D. The {SIESTA} method for ab initio order-{N} materials simulation. \emph{Journal of Physics: Condensed Matter} \textbf{2002}, \emph{14}, 2745\relax
\mciteBstWouldAddEndPuncttrue
\mciteSetBstMidEndSepPunct{\mcitedefaultmidpunct}
{\mcitedefaultendpunct}{\mcitedefaultseppunct}\relax
\EndOfBibitem
\bibitem[Dovesi(1986)]{86Dovesi}
Dovesi,~R. On the role of symmetry in the ab initio hartree-fock linear-combination-of-atomic-orbitals treatment of periodic systems. \emph{International Journal of Quantum Chemistry} \textbf{1986}, \emph{29}, 1755--1774\relax
\mciteBstWouldAddEndPuncttrue
\mciteSetBstMidEndSepPunct{\mcitedefaultmidpunct}
{\mcitedefaultendpunct}{\mcitedefaultseppunct}\relax
\EndOfBibitem
\bibitem[Ihrig \latin{et~al.}(2015)Ihrig, Wieferink, Zhang, Ropo, Ren, Rinke, Scheffler, and Blum]{15Ihrig-LRI}
Ihrig,~A.~C.; Wieferink,~J.; Zhang,~I.~Y.; Ropo,~M.; Ren,~X.; Rinke,~P.; Scheffler,~M.; Blum,~V. Accurate localized resolution of identity approach for linear-scaling hybrid density functionals and for many-body perturbation theory. \emph{New Journal of Physics} \textbf{2015}, \emph{17}, 093020\relax
\mciteBstWouldAddEndPuncttrue
\mciteSetBstMidEndSepPunct{\mcitedefaultmidpunct}
{\mcitedefaultendpunct}{\mcitedefaultseppunct}\relax
\EndOfBibitem
\bibitem[Lin \latin{et~al.}(2025)Lin, Ji, He, Ren, and He]{lin_force_2025}
Lin,~P.; Ji,~Y.; He,~L.; Ren,~X.; He Efficient hybrid-functional-based force and stress calculations for periodic systems with thousands of atoms. \emph{J. Chem. Theory Comput.} \textbf{2025}, Publisher: American Chemical Society\relax
\mciteBstWouldAddEndPuncttrue
\mciteSetBstMidEndSepPunct{\mcitedefaultmidpunct}
{\mcitedefaultendpunct}{\mcitedefaultseppunct}\relax
\EndOfBibitem
\bibitem[et~al.()]{LibRI-paper}
et~al.,~L.~P. in preparation\relax
\mciteBstWouldAddEndPuncttrue
\mciteSetBstMidEndSepPunct{\mcitedefaultmidpunct}
{\mcitedefaultendpunct}{\mcitedefaultseppunct}\relax
\EndOfBibitem
\bibitem[Hamann \latin{et~al.}(1979)Hamann, Schlüter, and Chiang]{hamann_norm-conserving_1979}
Hamann,~D.~R.; Schlüter,~M.; Chiang,~C. Norm-{Conserving} {Pseudopotentials}. \emph{Physical Review Letters} \textbf{1979}, \emph{43}, 1494--1497, Publisher: American Physical Society\relax
\mciteBstWouldAddEndPuncttrue
\mciteSetBstMidEndSepPunct{\mcitedefaultmidpunct}
{\mcitedefaultendpunct}{\mcitedefaultseppunct}\relax
\EndOfBibitem
\bibitem[Hamann(2013)]{SG15}
Hamann,~D.~R. Optimized norm-conserving Vanderbilt pseudopotentials. \emph{Phys. Rev. B} \textbf{2013}, \emph{88}, 085117\relax
\mciteBstWouldAddEndPuncttrue
\mciteSetBstMidEndSepPunct{\mcitedefaultmidpunct}
{\mcitedefaultendpunct}{\mcitedefaultseppunct}\relax
\EndOfBibitem
\bibitem[Heyd \latin{et~al.}(2003)Heyd, Scuseria, and Ernzerhof]{heyd_hybrid_2003}
Heyd,~J.; Scuseria,~G.~E.; Ernzerhof,~M. Hybrid functionals based on a screened {Coulomb} potential. \emph{The Journal of Chemical Physics} \textbf{2003}, \emph{118}, 8207--8215\relax
\mciteBstWouldAddEndPuncttrue
\mciteSetBstMidEndSepPunct{\mcitedefaultmidpunct}
{\mcitedefaultendpunct}{\mcitedefaultseppunct}\relax
\EndOfBibitem
\bibitem[Rose(1957)]{57Rose}
Rose,~M. \emph{Elementary Theory of Angular Momentum}; Structure of matter series; Wiley, 1957\relax
\mciteBstWouldAddEndPuncttrue
\mciteSetBstMidEndSepPunct{\mcitedefaultmidpunct}
{\mcitedefaultendpunct}{\mcitedefaultseppunct}\relax
\EndOfBibitem
\bibitem[Blanco \latin{et~al.}(1997)Blanco, Flórez, and Bermejo]{97Blanco-RealSphereHarm}
Blanco,~M.~A.; Flórez,~M.; Bermejo,~M. Evaluation of the rotation matrices in the basis of real spherical harmonics. \emph{Journal of Molecular Structure: THEOCHEM} \textbf{1997}, \emph{419}, 19--27\relax
\mciteBstWouldAddEndPuncttrue
\mciteSetBstMidEndSepPunct{\mcitedefaultmidpunct}
{\mcitedefaultendpunct}{\mcitedefaultseppunct}\relax
\EndOfBibitem
\end{mcitethebibliography}

\end{document}